\documentclass[twocolumn,showpacs,amsmath,prb,superscriptaddress]{revtex4-2}
\usepackage{epsfig,psfrag}
\usepackage{amsbsy}
\usepackage{amsfonts}
\usepackage{bbm}
\usepackage{tabularx}
\usepackage{euscript}
\usepackage{amsfonts}
\usepackage{exscale}
 \usepackage{slashed}
\usepackage{comment}

\usepackage{amsmath,amssymb}
\usepackage{graphicx}
\usepackage{dcolumn}
\usepackage{bm}
\usepackage{braket}
\usepackage{verbatim}
\usepackage{float}
\usepackage{bbold}
\usepackage{color}
\usepackage{xcolor}
\usepackage{relsize}
\usepackage{amsthm}
\usepackage{enumerate}
\usepackage{soul,xcolor}
\usepackage[normalem]{ulem}
\usepackage[T1]{fontenc}
\usepackage[colorlinks=true ,urlcolor=blue,urlbordercolor={0 1 1}]{hyperref}

\newcommand{\beq}{\begin{eqnarray}}
\newcommand{\eeq}{\end{eqnarray}}

\newcommand{\la}{\langle}
\newcommand{\ra}{\rangle}
\newcommand{\lla}{\langle\!\langle}
\newcommand{\rra}{\rangle\!\rangle}

\newcommand{\bsp}{\begin{split}}
\newcommand{\esp}{\end{split}}
\newcommand{\const}{{\rm const}}

\newcommand{\be}{\begin{equation}}
\newcommand{\ee}{\end{equation}}

\newcommand{\jyz}[1]{\textcolor{teal}{#1}}

\begin{document}
\graphicspath{{figures/}}

\setstcolor{red}

\title{Pseudogap and Non-Fermi-liquid criticality in double Kondo model for bilayer nickelates}

\author{Jing-Yu Zhao}
\affiliation{Department of Physics and Astronomy, Johns Hopkins University, Baltimore, Maryland 21218, USA}
\author{Ya-Hui Zhang}
\affiliation{Department of Physics and Astronomy, Johns Hopkins University, Baltimore, Maryland 21218, USA}

\date{\today}

\begin{abstract}
Motivated by recent experimental progress on high-temperature superconductivity in bilayer nickelates, we investigate the phase diagram of the normal state in a bilayer Kondo lattice model using single-site dynamical mean-field theory (DMFT). When the interlayer tunneling $t_\perp$ is absent, we identify a non-Fermi-liquid (NFL) critical point tuned by the interlayer spin coupling $J_\perp$ or hole doping $x$, which separates a standard Fermi liquid in the overdoped region from a distinct pseudogap (PG) metal in the underdoped regime. This PG phase, which we term the `second Fermi liquid' (sFL), exhibits small hole pockets and violates the perturbative Luttinger theorem despite the absence of symmetry breaking or fractionalization. The PG metal behaves like a heavy Fermi liquid, with small quasi-particle residue and large effective mass. We also provide an intuitive analytical description of the pseudogap and the ground-state wave function based on an ancilla-fermion framework. Inside the PG phase, we interpret the ancilla fermion as a spin-polaron and demonstrate a Kondo-like resonance peak in the spectral function of this composite fermion directly in DMFT calculation. Extending the analysis to finite $t_\perp$, we apply this framework to the bilayer nickelate $\mathrm{La}_3\mathrm{Ni}_2\mathrm{O}_7$. We propose that current experimental samples ($x \approx 0.5$) reside in the overdoped FL regime, suggesting that the pseudogap phase and the NFL criticality may be accessed via electron doping. 
\end{abstract}
\maketitle{}


\textit{Introduction---}
The unconventional pseudogap state emerging from doped Mott insulators remains a central problem in condensed matter physics, particularly in the underdoped cuprates \cite{leeMott2006,keimerHighTc2015a}. 
Existing theories include preformed Cooper pairs \cite{andersonRVB1987,emerySuper1995}, Fermi surface reconstruction from symmetry breaking orders \cite{SchmalianPines2,abanov_sdw_2000,Tremblay04,Chubukov23}, 
and fractionalized Fermi liquids (FL*) \cite{senthil_flp_2003,yang_yrz_2006,Punk15,zhang_pseudogap_2020}. 
Among them, one scenario is that the Fermi arcs observed in the pseudogap regime are part of small pockets with volume scaling with the hole density \cite{badouxCarrierDensity2016} instead of the total density. 
The small Fermi surface violates the Luttinger theorem \cite{luttingerWard1960,luttinger1960,oshikawa_topological_2000}, but is allowed if there is translation symmetry breaking or additional `topological order' \cite{oshikawa_topological_2000,senthil_flp_2003,paramekanti2004extending,mei2012luttinger}. 
Despite these possibilities, the nature of the cuprate pseudogap remains unsettled, hindering our understanding of strange-metal behavior that may arise near a quantum critical point (QCP) separating the pseudogap and conventional Fermi liquid phases.

The challenge in cuprates has two major origins. 
First, experiments often detect multiple intertwined symmetry-breaking orders at low temperature \cite{ fradkinIntertwined2015a, keimerHighTc2015a, cominRIX2016}, whereas in the higher-temperature pseudogap regime such orders are much less clear. 
This makes it difficult to determine whether symmetry breaking is essential to pseudogap physics. 
Second, there are limited unbiased numerical methods for the fermionic Hubbard model. 
Although density-matrix renormalization group (DMRG) and Monte Carlo approaches have identified competing orders such as stripe order \cite{whiteDMRG1998,zhengStripes2017, huangStripe2017,jiangttpJ2021, jiangStripe2022,xuStripe2024}, 
they have not provided a clear characterization of Fermi arcs or Fermi-surface reconstruction, partly because obtaining high-precision electron spectral functions is difficult. 
Dynamical mean-field theory (DMFT) \cite{georges_dynamical_1996, maierQCT2005, gullContinuoustimeMonteCarlo2011,maierQCT2005}, 
especially when combined with a numerical renormalization group (NRG) impurity solver \cite{Wilson1975,Bulla1999, Bulla2008, Weichselbaum2012}, 
is a powerful framework for Mott physics and can yield accurate Green's functions and spectral functions within its approximation of momentum independent self energy. 
However, standard single-site DMFT does not capture the pseudogap phase of the single-band Hubbard model, where intersite spin correlations are important. 
It is therefore desirable to identify a system  that suppresses competing orders while still hosting a pseudogap phase accessible to simple methods such as single-site DMFT.

The recent discovery of superconductivity in bilayer nickelates $\mathrm{La_3Ni_2O_7}$ provides a timely platform to revisit these questions \cite{sun_2023_first,hou_2023_nick,zhang_2024_nick, wangPr2024, wangPoly2024, zhang123_2024, dongVisual2024, puphalCrystal2024, yangOrbital2024, khasanovSplittingDensity2025, koAmbient2025, zhouAmbient2025, liuFilm2025, haoSrdopedFilm025}. Experiments report transition temperatures up to $T_c \approx 80-100\,\mathrm{K}$ under high pressure \cite{sun_2023_first,hou_2023_nick,zhang_2024_nick} and $T_c \approx 40-60\,\mathrm{K}$ in thin films \cite{liuFilm2025,haoSrdopedFilm025, liSCfilm2026,wangSuperconductingDome32026, liuHalfdome2026} with strain. While many theoretical works already discussed  pairing mechanisms 
\cite{lu2023interlayer,Yang2024ESD,lange2023pairing,luo2023bilayer,zhang2023electronic,huang2023impurity,Zhang2024,Geisler2024,PhysRevMaterials.8.044801,PhysRevB.109.045151,sakakibara2023possible,tian2024correlation,qin2023high,
yang2023minimal,zhan2024cooperation,chen2024non,yang2023possible,gu2023effective,liu2023s,shen2023effective,PhysRevB.109.104508,PhysRevB.109.205156,PhysRevB.109.L201124,oh2023type,zhang2023strong,zhu2025quantum,pan2023effect,maCPL2025,PhysRevB.110.024514,PhysRevB.110.L060510,PhysRevB.110.104507,PhysRevB.110.094509,lu2023superconductivity,
Luo2024,PhysRevB.109.045127,PhysRevB.109.045154,PhysRevLett.133.096002,Ouyang2024,PhysRevB.108.125105,lange2023pairing,cao2023flat,qu2023bilayer,PhysRevB.108.214522,zhang2023trends,PhysRevB.111.014515,PhysRevB.109.115114,
PhysRevB.110.205122,PhysRevB.109.L180502,tian2025spin,liu2025origin,liao2024orbital,PhysRevLett.132.126503,yin2025s,PhysRevB.111.104505,kaneko2025t,ji2025strong,Wang_2025,haque2025dft,shi2025theoretical,gao2025robust,
le2025landscape,hu2025electronic,shao2024possible,rm9g-8lm1,ushio2025theoretical,duan2025orbital,qiu2025pairing,cao2025strain,shao2025pairing,PhysRevB.111.L020504,xue2024magnetism,Oh2024ESD,wu_deconfined_2024}, the normal-state phase diagram---especially the possible pseudogap regime---remains much less explored. Current measurements are concentrated near hole doping $x\approx 0.5$ (defined relative to the Ni $d^8$ configuration), making this an opportune setting to make predictions across the full doping range to be tested in future experiments. Moreover, experiments have already reported strange-metal signatures \cite{zhang_2024_nick,zhouAmbient2025}, reminiscent of optimally doped cuprates, suggesting proximity to a QCP separating two distinct normal states. In this work, we therefore focus on predicting a pseudogap phase in the as-yet unexplored underdoped regime and on characterizing the non-Fermi-liquid (NFL) behavior in the quantum-critical region.

\begin{figure}[t]
    \centering
    \includegraphics[width=0.96\linewidth]{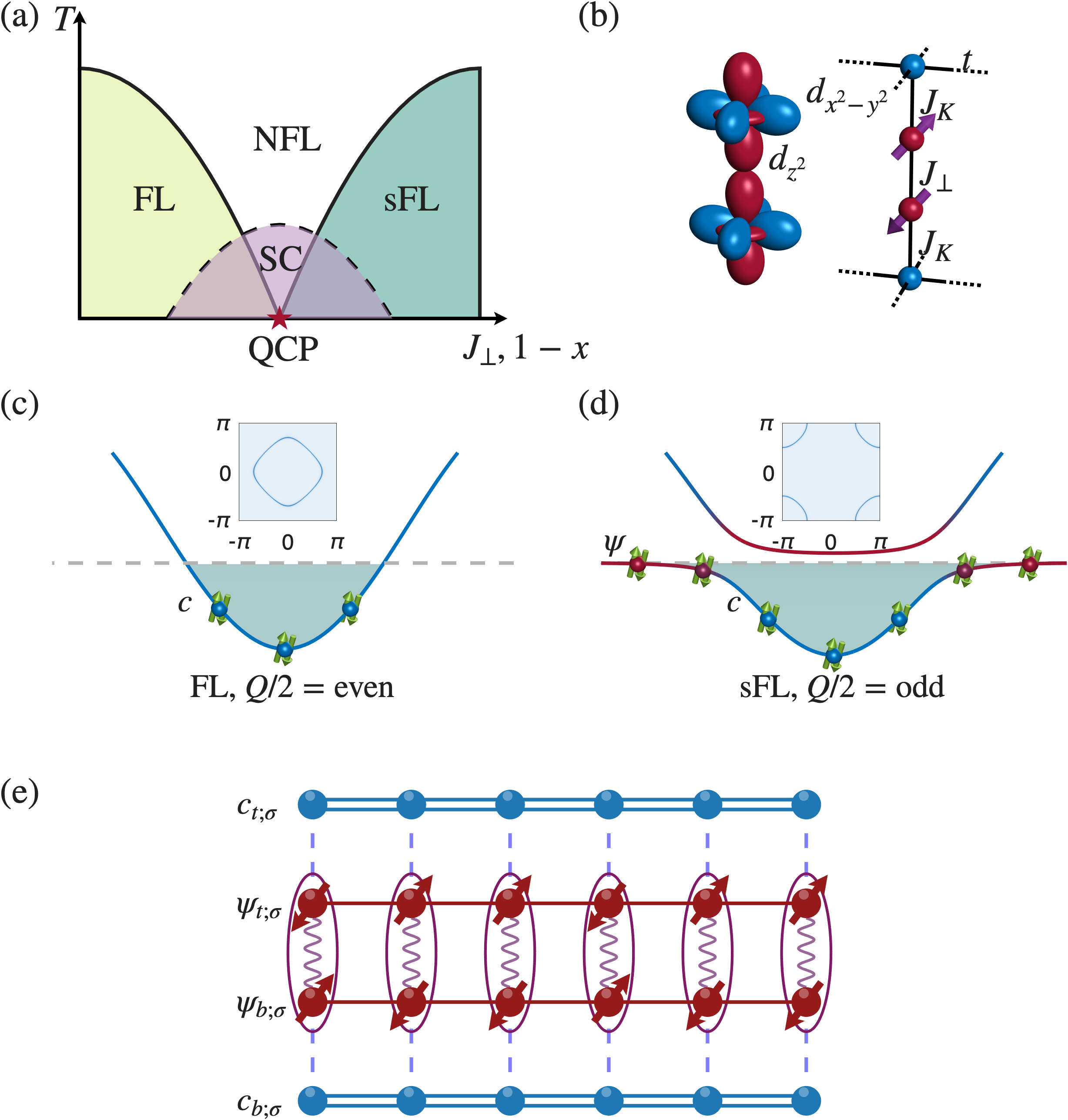}
    \caption{
    (a) Schematic phase diagram of the double Kondo model as a function of temperature $T$ and interlayer coupling $J_\perp$ or doping $1-x$. A quantum critical point is expected at an intermediate value of $J_\perp$ or $1-x$, giving rise to a critical regime with non-Fermi-liquid (NFL) behavior. 
    (b) Illustration of the double Kondo lattice model for bilayer nickelates defined in Eq.~\eqref{eqn:double_kondo}. Each layer consists of itinerant electrons from the $d_{x^2-y^2}$ orbital and localized spin-$1/2$ moments from the $d_{z^2}$ orbital, coupled via a ferromagnetic Kondo coupling $J_K=-2J_H<0$, where $J_H$ is the inter-orbital Hund's coupling. The local moments in the two layers are further coupled by an interlayer superexchange $J_\perp>0$. 
    (c) and (d) illustrate the distinction between the FL and sFL phases using a $\mathbb{Z}_2$ index $C=Q/2 \,(\mathrm{mod}\,2)$ and their corresponding Fermi-surface volumes. 
    (e) A model wave function for the sFL phase within the ancilla framework. The blue layers represent physical electrons from the $d_{x^2-y^2}$ orbital, while the red layers denote hidden ancillary fermions $\psi_{l;\sigma}$. In the final wave function, the ancilla fermions are projected into rung singlets, yielding a state in the physical Hilbert space beyond the Slater determinant framework. The pseudogap arises from hybridization between the electrons and the ancilla fermions, leading to a small hole-like Fermi pocket around $(\pi,\pi)$.}
    \label{fig:illu}
\end{figure}

The bilayer structure of nickelates allows two distinct Fermi liquids, as can be seen directly from symmetry analysis \cite{wu_deconfined_2024}.
In the absence of interlayer hopping ($t_\perp=0$), the system has an enlarged global symmetry $U(1)_t \times U(1)_b \times SU(2)_S/\mathbb{Z}_2$, together with a layer-exchange symmetry $\mathbb{Z}_2^{\text{mirror}}$, where $U(1)_{t,b}$ represent separate charge conservation in each layer and $SU(2)_S$ denotes the total spin symmetry.
For hole doping $x$ per layer relative to half filling, Oshikawa's theorem allows two possible Fermi-surface areas per flavor, $A_{FS}=(1-x)/2$ or $-x/2$, in units of the Brillouin-zone area \cite{oshikawa_topological_2000,wu_deconfined_2024}.
The latter corresponds to small hole pockets and may give rise to a pseudogap metal without symmetry breaking or topological order.
This raises the question of whether such a pseudogap Fermi liquid can be realized in a microscopic model and how it is connected to the conventional Fermi liquid.

Several effective models have been proposed for bilayer nickelates. 
Here we focus on the more realistic double Kondo model\cite{Yang2024ESD,yang2025strong,oh2026doping}. 
Each layer consists of itinerant electrons on $d_{x^2-y^2}$ and localized moments on $d_{z^2}$, as shown in Fig.~\ref{fig:illu}.
The $d_{x^2-y^2}$ and $d_{z^2}$ orbitals are coupled through a Hund’s coupling $J_K=-2J_H<0$, and $d_{z^2}$ orbitals on the two layers are coupled with an antiferromagnetic exchange $J_\perp$. 
The latter favors interlayer spin singlets and can suppress both magnetic orders and fractionalized spin-liquids. 

Importantly, all interactions are within a single unit cell, making the model particularly suitable for single-site DMFT.
Combined with the NRG impurity solver, this approach provides nonperturbative access to the low-energy physics, including the Fermi surface reconstruction across a NFL QCP.
Using DMFT+NRG, we systematically map out the phase diagram as a function of $J_\perp$, $x$, and $J_K=-2J_H$. 
For sufficiently strong Hund's coupling $J_H\gtrsim 2.5t$, we find both Fermi liquids allowed by the generalized Luttinger constraints.
For small $J_\perp$ and large $x$, the system forms a conventional Fermi liquid (FL), with a large Fermi surface of area $A_{\mathrm{FL}}=(1-x)/2$.
For large $J_\perp$ and small $x$, the system instead realizes a small Fermi surface of area $A_{\mathrm{sFL}}=-x/2$, which we refer to as a \emph{second Fermi liquid} (sFL). 
The sFL remains a Fermi liquid, but at an intrinsically strongly correlated fixed point, separated from the FL phase by a phase transition if interlayer hopping $t_\perp$ vanishes. 
Within single-site DMFT framework, the transition is continuous and appears to be in the universality class of the two-channel-two-spin Kondo impurity model 
\cite{Affleck1992,Affleck1995,Sela2011,Mitchell2012}.

In the critical regime, there are signatures of NFL behaviors, including a log divergence of susceptibility for various bosonic operators such as the total density and the layer-alternating spin.  When $t_\perp$ is finite as in the real materials, NFL regime still exists in a large energy window 
 between $T_{\mathrm{coh}}$ and $T_{\mathrm{NFL}}$, 
with $T_{\mathrm{NFL}}$ at order of $0.1 t$ 
and  $T_{\mathrm{coh}}$ as small as $10^{-3}t$ even with $t_\perp\sim 0.1 t$.  Here $t$ is the intra-layer hopping and is typically at order of a few thousands Kelvin. Given that $T_{\mathrm{coh}}$ is likely smaller than $T_c$ of superconductivity, the normal state above $T_c$ is still controlled by the NFL critical regime.  We propose that this NFL criticality is responsible for the strange metal behavior \cite{zhang_2024_nick,zhouAmbient2025} observed in the bilayer nickelates and the pseudogap side may be reached by electron doping.

\textit{Double kondo model---}
We study the following effective double Kondo model,
\begin{equation}\label{eqn:double_kondo}
\begin{aligned}
    H_{\mathrm{DK}} =& \sum_{\mathbf{k},a,\sigma} 
    \left(\epsilon(\mathbf{k})-\mu\right)c^\dagger_{\mathbf{k};a;\sigma} c_{\mathbf{k};a;\sigma} 
    \\&+ \sum_{\mathbf{k};\sigma}\left(t_\perp(\mathbf{k}) c^\dagger_{\mathbf{k};t;\sigma} c^{}_{\mathbf{k};b;\sigma}+\mathrm{h.c.}\right)
    + \sum_i H_{i;\mathrm{int}}~, \\
    H_{i;\mathrm{int}} =& 
    J_K \sum_{a} \mathbf{S}_{i;a} \cdot \mathbf{s}_{i;a}
    + U \sum_{a} n_{i;a;\uparrow} n_{i;a;\downarrow}
    \\&+ J_{\perp} \mathbf{S}_{i;t} \cdot \mathbf{S}_{i;b}~ + V_\perp n_{i;t}n_{i;b}~,
\end{aligned}
\end{equation}
where $a=t,b$ labels the top and bottom layers and $\sigma$ the spin.
The dispersion of the itinerate $c_{i;a;\sigma}$ in $d_{x^2-y^2}$ orbital is chosen as that of a square lattice  $\epsilon(\mathbf{k})=-2t \cos(k_x)-2t\cos(k_y)$. 
$t=1$ is set as the unit throughout the paper. 
$t_\perp(\mathbf{k}) = t_{\perp;1}(\cos k_x-\cos k_y)^2/4$ is the interlayer hopping\cite{oh2026doping,fan2025minimal}. 
Within every site, the itinerant $d_{x^2-y^2}$ electron is coupled to the localized $d_{z^2}$ moment $\mathbf{S}_{i;a}$ through a ferromagnetic coupling $J_K=-2J_H$, where $\mathbf{s}_{i;a}=\frac{1}{2}c^\dagger_{i;a;\rho}\boldsymbol{\sigma}_{\rho\lambda}c_{i;a;\lambda}$. Here $J_H$ is the Hund's coupling.
The two $d_{z^2}$ orbitals are coupled by an interlayer antiferromagnetic exchange $J_\perp>0$.
An intralayer Hubbard interaction $U$ and interlayer repulsion $V_\perp$ are also included for the itinerant orbital.
$V_\perp$ is found to have little effect on the phase diagram of the normal state and is taken as 0 unless otherwise mentioned. 

With a  strong Hund’s coupling, there is an effective inter-layer antiferromagnetic spin-spin coupling $\tilde{J}_\perp\propto J_\perp>0$ between the itinerant electrons in the two layers. 
In the weak inter-layer coupling limit $\tilde J_\perp \ll t$, the two layers remain effectively decoupled.
Each layer forms a FL with a large electron-like Fermi surface. 
For $\tilde J_\perp \gg t$, spins prefer to form interlayer singlets and leads to an insulator at half filling ($x=0$). 
Upon finite hole doping, the doped carriers propagate in the singlet background and form a sFL phase with small hole pockets.  
We emphasize that sFL is still a Fermi liquid, but is beyond weak coupling theory.

\begin{figure}[t]
    \centering
    \includegraphics[width=0.95\linewidth]{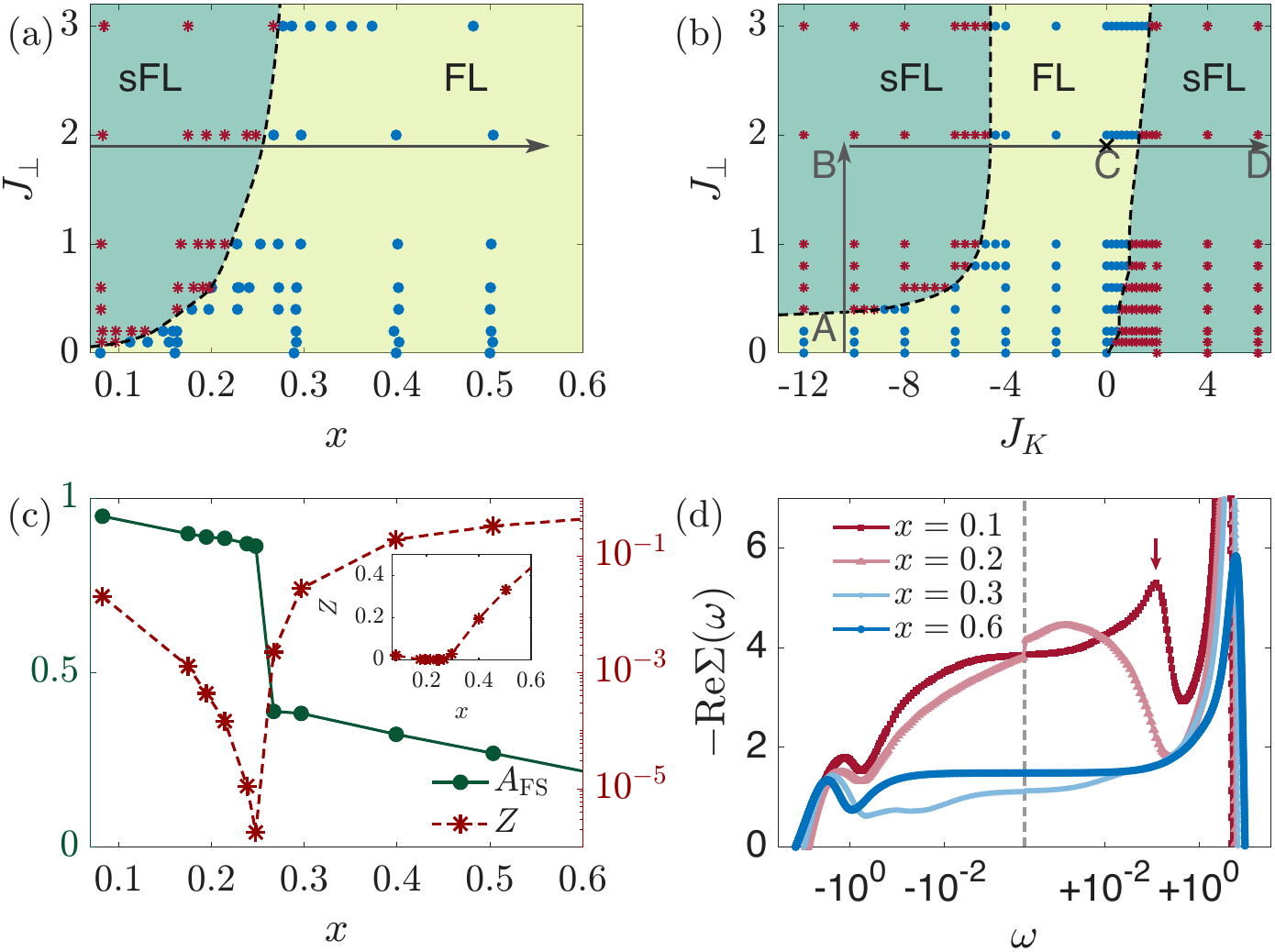}
    \caption{Phase diagrams of the double Kondo model in Eq.~\eqref{eqn:double_kondo}, obtained from self-consistent DMFT calculations with $U = 8t$ and $t_\perp=0$, with $t=1$ as the unit. 
    (a) Phase diagram as a function of $J_\perp$ and doping $x$ at fixed Hund’s coupling $J_K = -12t$.  
    Blue dots and red stars denote even and odd values of the $\mathbb{Z}_2$ charge $C$, characterizing the FL phase (yellow region) and sFL phase (green region), respectively.
    (b) Phase diagram as a function of $J_K$ and $J_\perp$ at fixed doping $x = 0.2$. 
    (c) Fermi surface volume $A_{\mathrm{FS}}$ and quasiparticle weight $Z=m_0/m_{\mathrm{eff}}=(1-\partial_\omega \mathrm{Re}\Sigma(\omega)\vert_{\omega=0})^{-1}$ along the line cut in (a) at fixed $J_\perp=2t$.
    The main panel shows $Z$ on a logarithmic scale, while the linear-scale behavior is presented in the inset.
    (d) $-\mathrm{Re}\Sigma(\omega)$ for different doping levels $x$ along the same line cut.
    Red (blue) curves correspond to the sFL (FL) phase.
    The red arrow indicates the point where the self-energy develops a divergence. 
    The large peak at energy $\sim 10t$ comes from the Hubbard $U$. 
    }
    \label{fig:phase_diagram}
\end{figure}

\textit{Phase diagram---}
In Fig.~\ref{fig:phase_diagram} (a-b), we present the self-consistent DMFT+NRG phase diagrams at $t_\perp=0$. 
The phase boundary between FL and sFL is identified by a $\mathbb{Z}_2$ index extracted from the NRG Wilson chain, in agreement with the discontinuity in the Fermi volume $\Delta A_{\mathrm{FS}}\simeq \pm 1/2$.  
As we focus on the normal states here, superconducting instabilities are artificially suppressed and will be addressed in future work. 
More technical details of the numeric can be found in the supplementary and Refs.~\cite{Wilson1975, georges_dynamical_1996, Bulla2008, 
Weichselbaum2007,Weichselbaum2012, Kugler2022, 
MitchellInter2014,StadlerInterleave2016,Gleis2024X}. 

For strong enough ferromagnetic Hund’s coupling ($J_H\gtrsim 2.5t$), the sFL is stabilized by stronger $J_\perp$ and smaller doping $x$. 
The critical doping $x_c$ increases with increasing $J_\perp$ and saturates at large $J_\perp$, reaching a maximal value $x_c \sim 0.27$. 
For weaker Hund’s coupling, the system always flows to the FL phase.
For completeness we also examine the antiferromagnetic Hund’s coupling ($J_K>0$). 
In this regime, the sFL phase can be viewed as the familiar heavy Fermi liquid phase from antiferromagnetic Kondo coupling $J_K>0$. 

Along all phase boundaries, the system exhibits QCP behavior with a continuously vanishing quasiparticle weight $Z=(1-\partial_\omega \mathrm{Re}\Sigma(\omega)\vert_{\omega=0})^{-1}$ [Fig.~\ref{fig:phase_diagram} (c)].  
The QCP is further accompanied by a jump in the Fermi surface volume per flavor, $\Delta A_{\mathrm{FS}}\simeq \pm 1/2$.

\begin{figure}[t]
    \centering
    \includegraphics[width=0.97\linewidth]{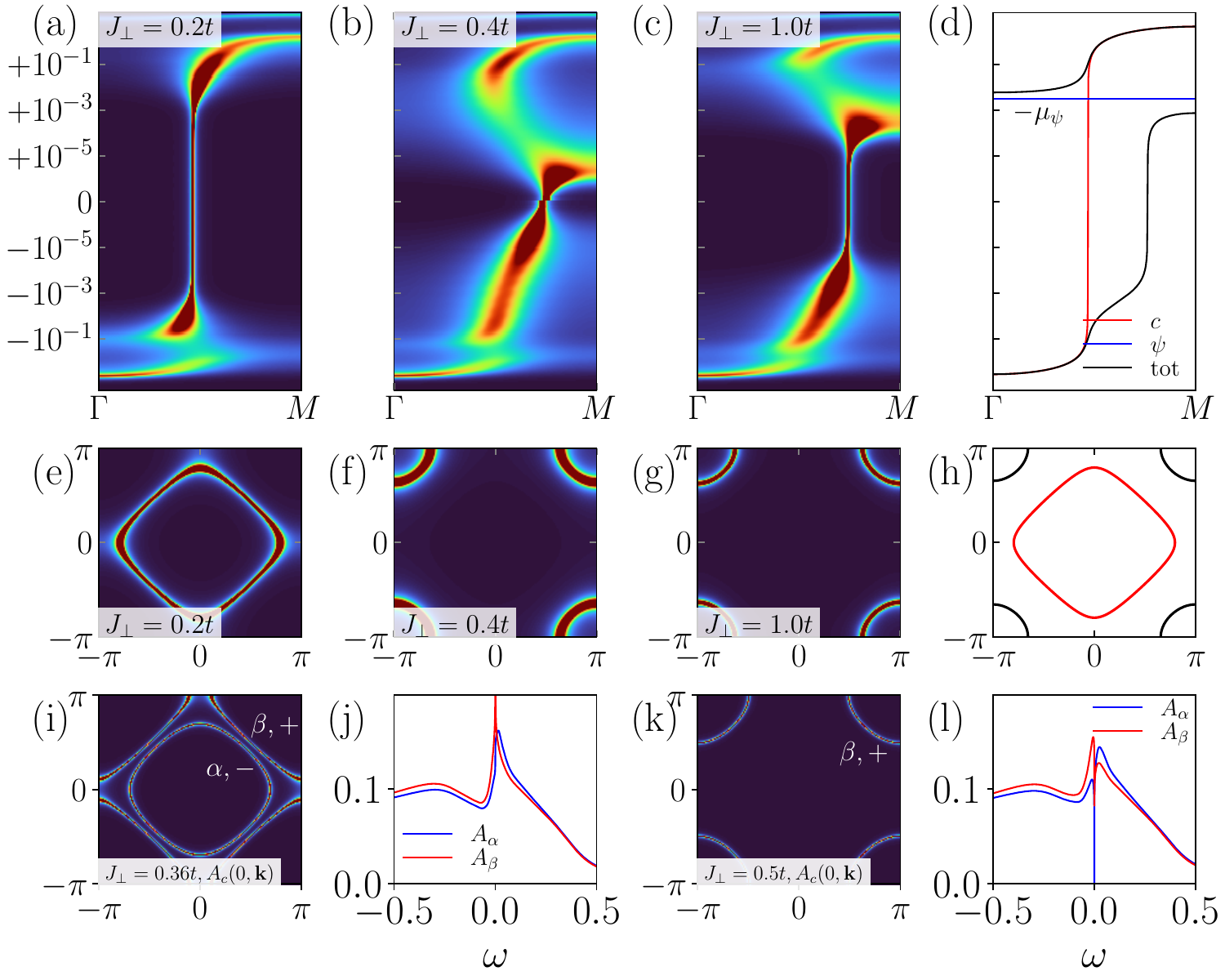}
    \caption{
    (a)-(c) Reconstructed momentum-resolved spectral function $A_c(\omega,\mathbf{k})$ 
    along the momentum line cut $(0,0)$–$(\pi,\pi)$ for $J_\perp = 0.2t$, $0.4t$, and $1.0t$, respectively, at fixed $J_K = -10t$, $x=0.2$, $U = 8t$ and $t_\perp=0$. 
    (e)-(g) Corresponding zero-frequency spectral function $A_c(\omega=0,\mathbf{k})$. 
    (d) Simulated band structure obtained from the ancilla-fermion description Eq.~\eqref{eqn:ancilla}, with $\Phi=0.1t$ and $\mu_\psi=-0.003t$. 
    The red, blue, and black lines represent the bare $c$ band, the $\psi$ band, and the hybridized bands, respectively. 
    (h) Corresponding Fermi surfaces in the ancilla-fermion theory. 
    The Fermi surfaces of the bare $c$ band and the hybridized bands are shown as red and black curves, respectively.
    (i),(j) Fermi surface and spectrum functions for a finite $t_{\perp;1}=0.2t$ at $J_\perp=0.36t$.  
    (k),(l) Fermi surface and spectrum functions for a finite $t_{\perp;1}=0.2t$ at $J_\perp=0.5t$.  
    }
    \label{fig:momentum}
\end{figure}

\textit{Pseudogap of sFL---}
The pseudogap behavior in sFL is manifested as a divergent structure in the self-energy, as indicated by the red arrow in Fig.~\ref{fig:phase_diagram}(d). 
In addition to a large kink contributed by the Hubbard $U$, 
as approaching the sFL, the self-energy progressively develops another divergent kink at smaller frequency, approximated by the form $\Sigma(\omega)\sim\Phi^2/(\omega+\mu_\psi+i\eta)$.
Such a structure can be reproduced within an ancilla-fermion framework as shown later. 
In realistic calculations, however, the broadening parameter $\eta$ can be relatively large. 

To gain further insight, we reconstruct the momentum-resolved spectrum from the DMFT self-energy:
\begin{equation}
    A_c(\omega,\mathbf{k}) = -\frac{1}{\pi}\mathrm{Im}\frac{1}{\omega+\mu-\epsilon(\mathbf{k})-\Sigma(\omega)}~,
\end{equation}
which is periodized onto the two-dimensional square lattice. 
Here $\Sigma(\omega)$ is the DMFT self-energy.  

Fig.~\ref{fig:momentum} shows $A_c(\omega,\mathbf{k})$ along the $(0,0)$–$(\pi,\pi)$ momentum cut for different $J_\perp$. 
For small $J_\perp$, the system is in FL phase with a single low-energy quasiparticle band. 
The high-energy branches at $\sim \pm 10t$ correspond to Hubbard bands generated by $U$. 
The Fermi surface is a large electron-like surface around $\Gamma$ with volume $(1-x)/2$.
For large $J_\perp$ in the sFL phase, the original quasiparticle band splits into two branches. Especially the lower band becomes very flat and lies just above $E=0$ (at $-\mu_\psi \sim 10^{-2} t$) close to $(\pi,\pi)$.  A small hole pocket forms when $E=0$ crosses this emergent flat band, with $A_{\mathrm{sFL}}=-x/2$. 
Within both the FL and the sFL phase, the imaginary self energy (the scattering rate)  vanishes at  the $\omega \rightarrow 0$ limit, hence the spectral function becomes quite sharp at $E=0$, indicating coherent quasi-particles. But the quasi-particle residue $Z$ is  much smaller in the sFL phase  and  later we will show that the quasi-particle in the sFL phase is dominated by an emergent composite fermion. In this sense the sFL phase is quite similar to the heavy Fermi liquid (HFL) in standard Kondo lattice model.  Close to the critical point ($J_\perp \approx 0.4t$), the  spectral function becomes less sharp even at $\omega \rightarrow 0$ region, indicating strong scattering in the NFL region.

Here we show that both the pseudogap bands structure deep in the sFL phase and the divergent self-energy can be naturally understood within an effective ancilla fermion theory (AFT) ~\cite{zhang_pseudogap_2020}. 
We consider a model in which the physical electrons hybridize with an auxiliary fermion $\psi_{i;a;\sigma}$ through
\begin{equation}\label{eqn:ancilla}
\begin{aligned}
    H_{\mathrm{AFT}} =& 
    \sum_{\mathbf{k},a,\sigma} 
    \left(\epsilon(\mathbf{k})-\mu\right)c^\dagger_{\mathbf{k};a;\sigma} c_{\mathbf{k};a;\sigma} \\
    &+ \Phi \sum_{i,a,\sigma} c^\dagger_{i;a;\sigma}\psi_{i;a;\sigma} +\mathrm{h.c.} - \mu_\psi \sum_{i,a} n_{i;a}^\psi~,
\end{aligned}
\end{equation}
where $\Phi$ is the hybridization between $c$ and $\psi$. 
The ancilla fermions $\psi$ are fixed at half filling by tuning $\mu_\psi$.

Already at the mean-field level this model reproduces the characteristic sFL band structure. 
First, the Fermi volume is reconstructed according to $A_{\mathrm{sFL}} = n_c + n_\psi \; \mathrm{mod}\; 1 = -x/2$, 
with $n_c=(1-x)/2$ and $n_\psi=1/2$. 
Furthermore, the resulting hybridized band structure closely mimics that of the sFL, as shown in Fig.~\ref{fig:momentum}(d,h). 
Integrating out the ancilla fermion yields an effective self-energy 
$\Sigma^{\mathrm{AFT}}(\omega)=\frac{\Phi^2}{\omega+\mu_\psi}$,
whose pole structure naturally explains the finite-energy pseudogap in the sFL phase.

The ancilla framework also provides a model wavefunction with tunable pseudogap for the sFL phase by projecting the ancilla fermions into inter-layer rung singlet in the end, as illustrated in Fig.~\ref{fig:illu}(e).  The final wavefunction is in the physical Hilbert space of a bilayer Hubbard model with only itinerant electron $c_{a;\sigma}$.
In the supplementary, we justify the wavefunction analytically in the infinite $\tilde J_\perp$ limit. 
We also interpret the ancilla fermion as a spin polaron composite operator $\psi_{i;t;\lambda}\sim 
\sum_\alpha  c_{i;t;\rho}  (\boldsymbol{\sigma}\cdot \boldsymbol{s}_{i;b})_{\rho\lambda}~$ and we demonstrate that the spectral function of this composite operator indeed has a sharp Kondo-resonance peak at $\omega=0$ inside the sFL phase from the DMFT result, which confirms that the composite fermion dominates at low energy. 

We also consider the effect of a finite $t_{\perp;1}$ in Fig.~\ref{fig:momentum}(i–l). 
Although the interlayer hopping reduces the symmetry to a single charge $U(1)$, we still observe a phase transition between the FL and sFL, indicating that these two Fermi liquids remain distinct.
The transition is weakly first-order, as discussed in the Supplemental Material. 
In the FL phase, a finite $t_{\perp;1}$ splits the degenerate band into the experimentally observed $\alpha$ and $\beta$ bands, 
with the splitting increasing with $J_\perp$.
In the sFL phase, the $\alpha$ band develops a small Mott-like gap, while the $\beta$ band exhibits a pseudogap near the Fermi energy and forms a single hole pocket with area $-x$, as shown in Fig.~\ref{fig:momentum}(l). 
We note that even if $t_{\perp;1}$ has a $(\cos k_x-\cos k_y)^2$ factor, the interaction generated self energy is momentum independent. 
It remains to see whether this is an artifact of DMFT or is intrinsic. 

\begin{figure}
    \centering
    \includegraphics[width=0.97\linewidth]{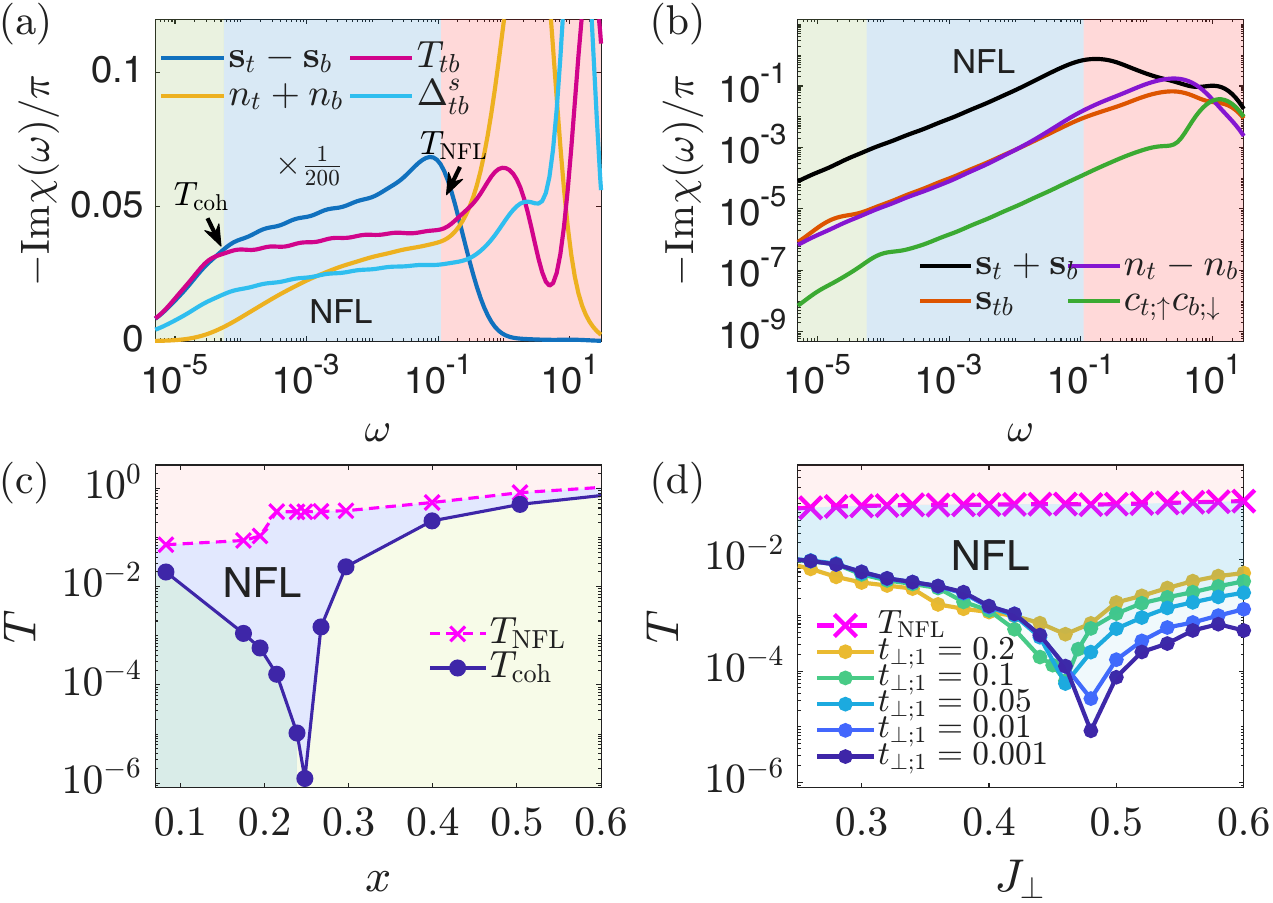}
    \caption{NFL behavior near the QCP, with parameters chosen as $J_\perp = 0.6t$, $J_K = -10t$, $x = 0.2$, and $U = 8.0t$.
    (a) The imaginary part of the bosonic correlation function, $-\mathrm{Im}\chi(\omega)/\pi$, which exhibits a frequency-independent (constant) behavior.
    The shaded blue region marks the NFL regime between $T_{\mathrm{coh}}$ and $T_{\mathrm{NFL}}$.
    (b) $-\mathrm{Im}\chi(\omega)/\pi$ showing a linear decay with frequency $\omega$.
    (c) Upper and lower bounds of the NFL regime $T_{\mathrm{NFL}}$ and $T_{\mathrm{coh}}$ along the line cut indicated in Fig.~\ref{fig:phase_diagram} (a). 
    (d) $T_{\mathrm{NFL}}$ and $T_{\mathrm{coh}}$ as functions of $J_\perp$ for different values of the interlayer hopping $t_{\perp;1}$. 
    The minimum $T_{\mathrm{coh}}$ is of order $\sim 10^{-3}t$ for $t_{\perp;1}=0.2t$. 
    The couplings are fixed at $J_K = -10t$, $U = 8t$, and doping $x = 0.2$. 
    $T_{\mathrm{NFL}}$ depends only weakly on $t_{\perp;1}$ and is shown only for $t_{\perp;1}=0.2t$.
    }
    \label{fig:correlators}
\end{figure}

\textit{Non-Fermi-liquid behavior at QCP---}
To probe the QCP regime, we compute a set of bosonic correlation functions
\begin{equation}
    \chi_O(\omega) \equiv \int dt\theta(t)\langle [O(t),O(0)]\rangle e^{i\omega t}~,
\end{equation}
in the vicinity of the critical point, as shown in Fig.~\ref{fig:correlators}.

According to their low-energy scaling, these correlators fall into two distinct classes. 
The first class exhibits a nearly constant imaginary susceptibility
$-\mathrm{Im}\chi(\omega)\sim \mathrm{const}$, 
corresponding to a logarithmically divergent real part
$\mathrm{Re}\chi(\omega)\sim -\log \omega$
and an imaginary time decay
$\langle O(\tau)O(0)\rangle \sim 1/\tau$. 
These operators represent nearly critical bosonic modes that strongly scatter the conduction electrons and are responsible for the NFL phenomenology. 
Operators in this class include the layer-alternating spin $\mathbf{s}_{i;t}-\mathbf{s}_{i;b}$, the total charge density $n_{i;t}+n_{i;b}$, the interlayer singlet pairing 
$\Delta_{tb}^s=c_{i;t;\uparrow}c_{i;b;\downarrow}-c_{i;t;\downarrow}c_{i;b;\uparrow}$, 
and the interlayer hopping
$T_{tb}=c_{i;t;\sigma}^\dagger c_{i;b;\sigma}$.

The second class instead shows regular Fermi-liquid scaling with $-\mathrm{Im}\chi(\omega)\sim\omega$, 
$\mathrm{Re}\chi(\omega)\sim\mathrm{const}$, 
and $\langle O(\tau)O(0)\rangle\sim1/\tau^2$. 
Examples include the total spin $\mathbf{s}_{i;t}+\mathbf{s}_{i;b}$, the relative charge $n_{i;t}-n_{i;b}$, intra-layer pairing $c_{i;t;\uparrow}c_{i;t;\downarrow}$, and spin-transfer operator 
$\mathbf{S}_{tb}=c_{i;t;\rho}^\dagger\boldsymbol{\sigma}_{\rho\lambda} c_{i;b;\lambda}$. 

From the divergent correlations, we extract two characteristic temperature scales, $T_{\mathrm{NFL}}$ and $T_{\mathrm{coh}}$, as the upper and lower bonds of NFL behavior. 
In Fig.~\ref{fig:correlators} (c) we show $T_{\mathrm{NFL}}$ and $T_{\mathrm{coh}}$ cross the QCP. 
At the FL–sFL boundary, $T_{\mathrm{coh}}$ vanishes continuously when $t_\perp=0$, consistent with a QCP behavior where NFL extends to zero temperature. 
In contrast, $T_{\mathrm{NFL}}$ changes only weakly with doping and stays at a value of order $\sim 0.1 t$.

In realistic materials, the interlayer charge hopping is generally finite. 
We therefore examine the stability of the NFL regime in the presence of a finite $t_{\perp;1}$. 
In Fig.~\ref{fig:correlators}(d) we show $T_{\mathrm{coh}}$ and $T_{\mathrm{NFL}}$ for different values of $t_{\perp;1}$. 
The minimum $T_{\mathrm{coh}}$ increases with $t_{\perp;1}$, and the QCP likely evolves into a crossover for finite $t_{\perp;1}$. 
However, for $t_{\perp;1}=0.2t$, the minimum $T_{\mathrm{coh}}$ can still reach values as low as $\sim 10^{-3}t$, while $T_{\mathrm{NFL}}$ is almost unaffected by $t_{\perp;1}$.  Given that $T_{\mathrm{coh}} \sim 10^{-3} t$ is likely smaller than $T_c \sim 40-100$ K of superconductivity, the NFL regime controlled by the nearby QCP should still dominate the normal state above $T_c$.  In the Supplemental Material we further examine the case of isotropic interlayer hopping.

\textit{Discussion---}
In summary, we study the bilayer nickelate system using DMFT+NRG applied to a double Kondo lattice model for bilayer nickelates.
By tuning inter-layer spin coupling $J_\perp$ and hole doping $x$, we identify both a conventional FL in overdoped region and a pseudogap metal sFL in underdoped regime.
Between these two phases, we find a quantum critical region accompanied by divergent bosonic susceptibility.
Our results suggest that the strange-metal behavior observed in bilayer nickelates may originate from this QCP between the pseudogap metal and the Fermi liquid. While the current samples are likely in the FL side, we suggest future experiments to search for the PG phase by electron doping. We plan to study the pairing mechanism from the NFL normal state in a subsequent work.

Although our discussion mainly focuses on the double Kondo lattice model relevant to bilayer nickelates, 
the existence of a distinct FL phase with small Fermi surface and the QCP is universal in systems with the $U(1)_t\times U(1)_b\times SU(2)_S/\mathbb{Z}_2$ symmetry, such as in the twisted bilayer graphene (TBG).  Therefore there may be a deep connection between  bilayer nickelates and TBG, which was not appreciated previously. Especially a similar sFL phase as studied in this work may control the normal state of TBG\cite{zhaoTopoMott2025a, zhaoRVB2025}. We hope to extend our analysis here directly to TBG in future.

\textit{Acknowledge---} 
The work is supported by the Alfred P. Sloan Foundation through a Sloan Research Fellowship (Y.-H.Z.).

\bigskip
\bigskip
\begin{center}
\textbf{\large End Matter}
\end{center}
\bigskip
\vspace{-0.2cm}

\textit{Method---}
Within the DMFT framework \cite{georges_dynamical_1996}, the interacting lattice problem is mapped onto an effective single-impurity model subject to a self-consistently determined hybridization function $\Delta_0(\omega)$.

The self-consistent procedure is implemented as follows. 
Starting from an initial guess of $\Delta_0(\omega)$, the impurity self-energy $\Sigma_{\mathrm{imp}}(\omega)$ is obtained using an impurity solver (here, the numerical renormalization group, NRG). 
Within DMFT, the lattice self-energy is assumed to be momentum-independent and identified with the impurity self-energy, $\Sigma(\omega)=\Sigma_{\mathrm{imp}}(\omega)$. The local lattice Green’s function is then computed as
\begin{equation}
    G_{ii}(\omega) = \int \frac{\mathrm{d}^2k}{(2\pi)^2} \frac{1}{\omega+\mu - \Sigma(\omega) - \epsilon(\mathbf{k})}~,
\end{equation}
from which an updated hybridization function is obtained via the DMFT self-consistency condition~\cite{georges_dynamical_1996},
\begin{equation}
    \Delta_{0}(\omega) = - \omega - \mu + \Sigma(\omega) + G^{-1}_{ii}(\omega)~.
\end{equation}
These steps are iterated until convergence of $\Delta_0(\omega)$, $\Sigma(\omega)$, and $G_{ii}(\omega)$ is achieved. 

The quantum impurity problem is solved using the Wilson NRG \cite{Wilson1975,Bulla2008}. 
The hybridization function $\Delta_0(\omega)$ is logarithmically discretized into intervals $\pm[D\Lambda^{-n-1},D\Lambda^{-n}]$ and mapped onto a semi-infinite Wilson chain, 
which is diagonalized iteratively while retaining a finite number of low-energy states at each step. 
The method becomes exact in the limits $\Lambda \to 1$ and infinite retained states. 
In this work, we use $\Lambda=3$ and retain up to $10{,}000$ states during the NRG iterations.

Dynamical correlation functions are computed using the density-matrix NRG (DM-NRG) formalism \cite{Weichselbaum2007,Weichselbaum2012}, ensuring proper spectral weight normalization. 
The discrete spectra are broadened using log-Gaussian kernels to obtain smooth spectral functions. 
The impurity self-energy is extracted using a second-order equation-of-motion approach \cite{Kugler2022} in terms of higher-order correlation functions, which improves numerical stability and guarantees a positive-definite imaginary part.

\textit{Bilayer Hubbard model---}
In the strong Hund's coupling limit $J_H\gg t$, the phase diagram of the double kondo model is similar to a bilayer Hubbard model within DMFT:
\begin{equation}\label{eqn:dtJ}
\begin{aligned}
    H=& \sum_{\mathbf{k},a,\sigma} 
    \left(\epsilon(\mathbf{k})-\mu\right)c^\dagger_{\mathbf{k};a;\sigma} c_{\mathbf{k};a;\sigma}
    + \sum_iH_{i;\mathrm{int}}~,\\
    H_{i;\mathrm{int}}=&\sum_aUn_{i;a;\uparrow}n_{i;a;\downarrow}
    +J_{\perp} \mathbf s_{i;t} \cdot \mathbf s_{i;b} + V_\perp n_{i;t}n_{i;b} ~,
\end{aligned}
\end{equation}
where the system consists of two layers of itinerant electrons $c_{i;\alpha}$.
The spin operator is defined as $\mathbf s_{i;a} = \frac{1}{2}c^\dagger_{i;a;\rho}\boldsymbol{\sigma}_{\rho\lambda}c^{}_{i;a;\lambda}$ 
and spins on the two layers are coupled through an interlayer exchange interaction $J_\perp$.
\jyz{ The model Eq.~\eqref{eqn:dtJ} shares the same $U(1)_t\times U(1)_b\times SU(2)_S$ symmetry and therefore also allows two different kinds of Fermi liquids. }
In the supplementary, we show that this model gives a phase diagram similar to the double Kondo model in DMFT.   The bilayer Hubbard model therefore serves as the minimal model supporting the FL-sFL transition.  But we emphasize that in the small $J_\perp$ limit, the double Kondo model hosts ferromagnetism or a spin-density-wave (SDW) phase, which are absent in the single-orbital bilayer Hubbard model\cite{yang2025magnetism}. 
We note that the bilayer Hubbard model is relevant to several physical systems, including multiorbital molecular solids such as $A_3C_{60}$\cite{caponeC60_2009} and moiré materials such as twisted bilayer graphene \cite{zhaoTopoMott2025a, zhaoRVB2025}.

\begin{figure}[t]
    \centering
    \includegraphics[width=0.99\linewidth]{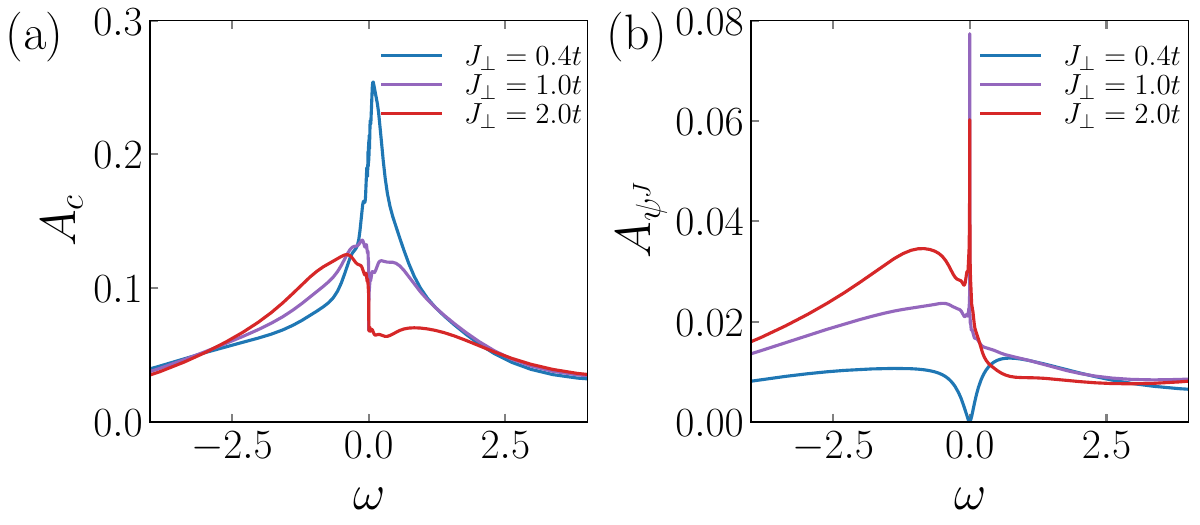}
    \caption{
    Spectral functions obtained from the double Hubbard model [Eq.~\eqref{eqn:dtJ}] with $U=8t$ and $V_\perp=0$ for $J_\perp=0.4t$, $1.0t$, and $2.0t$.
    The case $J_\perp=0.4t$ corresponds to the FL phase, $J_\perp=2.0t$ to the sFL phase, and $J_\perp=1.0t$ lies near the QCP on the sFL side.
    (a) The spectral functions $A_c(\omega)$ for the physical electron $c_{i;a;\sigma}$. (b) The specturm function $A_{\psi^J}(\omega)$ for the trion operator $\Psi^J_{i;a;\lambda}\sim \sum_\rho c_{i;t;\rho} (\boldsymbol{\sigma}\cdot \boldsymbol{s}_{i;b})_{\rho\lambda}$.
    The latter vanishes as $\omega \to 0$ in the FL phase, but instead develops a pronounced peak in the sFL phase.
    }
    \label{fig:ApsiJ}
\end{figure}

\textit{Ancilla Fermion theory---}
Within the ancilla theory Eq.~\eqref{eqn:ancilla}, the Green's functions can be obtained by solving
\begin{equation}
    \left(\begin{matrix}
    \omega + \mu -\epsilon(\mathbf{k})  & -\Phi \\ -\Phi & \omega + \mu_\psi
    \end{matrix}\right) 
    \left(\begin{matrix}G^{\mathrm{AFT}}(\omega,\mathbf{k}) \\ F^{\mathrm{AFT}}(\omega,\mathbf{k}) \end{matrix}\right)
    = \left(\begin{matrix} 1 \\ 0 \end{matrix}\right)~,
\end{equation}
where $F^\mathrm{AFT} (\omega,\mathbf{k})= \lla \psi_{\mathbf{k};\alpha};c^\dagger_{\mathbf{k};\alpha}\rra (\omega)$ is the retarded Green's function of $\psi_{\mathbf{k};\alpha}$ and $c^\dagger_{\mathbf{k};\alpha}$. 
Solving this equation gives the electron Green's function
\begin{equation}\label{eqn:ancilla_green0}
    G^{\mathrm{AFT}}(\omega,\mathbf{k}) = \frac{1}{\omega+\mu-\epsilon(\mathbf{k}) - \Phi^2/(\omega+\mu_\psi)}~. 
\end{equation}
We therefore obtain the electron self-energy within the ancilla framework as: 
\begin{equation}
    \Sigma^{\mathrm{AFT}}(\omega) = \Phi^2/(\omega+\mu_\psi). 
\end{equation}
For $\omega+i0^+ $, the self energy would show a peak at $-\mu_\psi$ in the imaginary part and a kink at $-\mu_\psi$ in the real part, 
consistent with our earlier observation.

To connect this effective description to the microscopic model, we derive one step of equation-of-motion for the electron Green’s function in bilayer Hubbard model as:
\begin{equation}
    (\omega + \mu -\epsilon(\mathbf{k})) G(\omega,\mathbf{k}) -  \lla [c_{\mathbf{k};\alpha}^{},H_{\mathrm{int}}];c^\dagger_{\mathbf{k};\alpha}\rra(\omega) = 1~,
\end{equation}
where the composite operator $[c_{\mathbf{k}\alpha},H_{\mathrm{int}}]$  can be identified as the fermionic mode $\psi$ underlying the ancilla construction in Eq.~\eqref{eqn:ancilla_green0}.  
In the bilayer Hubbard model Eq.~\eqref{eqn:dtJ}, 
$[c_{\mathbf{k}\alpha},H_{\mathrm{int}}]$ decomposes into both a Hubbard-like charge fluctuation contribution $\Psi^U_{i;a;\uparrow}\sim c_{i;a;\uparrow}n_{i;a;\downarrow}$ and a spin-polaron contribution 
$\Psi^J_{i;t;\lambda}  \sim\sum_\rho 
c_{i;t;\rho}  (\boldsymbol{\sigma}\cdot \boldsymbol{s}_{i;b})_{\rho\lambda}$
associated with interlayer spin exchange, as detailed in the Supplementary. 
While $\Psi^U$ mainly affects the high-energy physics, $\Psi^J$ exhibits a sharp change across the FS and sFL phases. 

This change is reflected in the real-frequency spectral functions $A_c(\omega)$ and $A_{\Psi^J}(\omega)$ obtained for the impurity site of DMFT, shown in Fig.~\ref{fig:ApsiJ}.
In the FL phase, $A_c(\omega)$ remains featureless, while $A_{\Psi^J}(\omega)$ vanishes as $\omega \to 0$.
Upon entering the sFL phase, a dip (pseudogap) develops in $A_c(\omega)$, with the spectral weight suppressed near $\omega=0$.
In contrast, $A_{\Psi^J}(\omega)$ develops a pronounced Kondo-like peak at $\omega=0$.
This indicates that the composite fermion $\Psi^J$ emerges as an elementary quasi-particle in the sFL phase.

\textit{Model wavefunction} Finally, a variational wave function for the bilayer Hubbard model can be constructed by projecting the ancilla wave function onto the physical Hilbert space,
\begin{equation}
    |\Psi_\mathrm{anc}\rangle = P_s |\mathrm{Slater}[c,\psi]\ra~,
\end{equation}
where $P_s$ projects the ancilla fermions $\psi_{i;t}$ and $\psi_{i;b}$ on each site $i$ onto an interlayer spin-singlet state.
Here, $|\mathrm{Slater}[c,\psi]\rangle$ denotes the ground-state wave function of the ancilla Hamiltonian in Eq.~\eqref{eqn:ancilla}.
In the limit $J_\perp\rightarrow \infty$  of Eq.~\eqref{eqn:dtJ}, this construction reproduces the exact rung-singlet ground state at half-filling. 
In the Supplementary, we further show that these results remain robust upon introducing a finite kinetic energy $t$.
Upon doping, the projected wave function naturally yields a small Fermi surface with a volume consistent with the sFL phase.

\bibliography{refs}

@article{Gleis2024X,
    archivePrefix = {arXiv},
    author = {Gleis, Andreas and Lee, Seung-Sup B. and Kotliar, Gabriel and von Delft, Jan},
    doi = {10.1103/PhysRevX.14.041036},
    eprint = {2310.12672},
    issn = {2160-3308},
    journal = {Physical Review X},
    month = {nov},
    number = {4},
    pages = {041036},
    publisher = {American Physical Society},
    title = {{Emergent Properties of the Periodic Anderson Model: A High-Resolution, Real-Frequency Study of Heavy-Fermion Quantum Criticality}},
    volume = {14},
    year = {2024}
}

@article{paramekanti2004extending,
  title={Extending Luttinger’s theorem to $Z_2$ fractionalized phases of matter},
  author={Paramekanti, Arun and Vishwanath, Ashvin},
  journal={Physical Review B},
  volume={70},
  number={24},
  pages={245118},
  year={2004},
  doi={10.1103/PhysRevB.70.245118},
  publisher={APS}
}

@article{mei2012luttinger,
  title={Luttinger-volume violating Fermi liquid in the pseudogap phase of the cuprate superconductors},
  author={Mei, Jia-Wei and Kawasaki, Shinji and Zheng, Guo-Qing and Weng, Zheng-Yu and Wen, Xiao-Gang},
  journal={Physical Review B},
  volume={85},
  number={13},
  pages={134519},
  year={2012},
  doi={10.1103/PhysRevB.85.134519},
  publisher={APS}
}

@article{oh2026doping,
  title={Doping a spin-one Mott insulator: possible application to bilayer nickelate},
  author={Oh, Hanbit and Yang, Hui and Zhang, Ya-Hui},
  journal={New Journal of Physics},
  volume={28},
  number={2},
  pages={021201},
  year={2026},
  doi={10.1088/1367-2630/ae3d5b},
  publisher={IOP Publishing}
}

@misc{yang2025magnetism,
  title={Magnetism and superconductivity in bilayer nickelate},
  author={Yang, Hui and Zhang, Ya-Hui},
  archivePrefix={arXiv},
  eprint={2512.13793},
  year={2025}
}

@misc{fan2025minimal,
  title={{Minimal two band model and experimental proposals to distinguish pairing mechanisms of the high-T$_c$ superconductor La$_3$Ni$_2$O$_7$}},
  author={Fan, Zheng-Duo and Vishwanath, Ashvin},
  archivePrefix={arXiv},
  eprint={2512.05956},
  year={2025}
}

@article{yang2025strong,
  title={Strong pairing and symmetric pseudogap metal in a double Kondo lattice model: From a nickelate superconductor to a tetralayer optical lattice},
  author={Yang, Hui and Oh, Hanbit and Zhang, Ya-Hui},
  journal={Physical Review B},
  volume={111},
  number={24},
  pages={L241102},
  year={2025},
  doi={10.1103/PhysRevB.111.L241102},
  publisher={APS}
}

@article{sun_2023_first,
    title = {Signatures of superconductivity near 80 {K} in a nickelate under high pressure},
    volume = {621},
    issn = {0028-0836, 1476-4687},
    doi = {10.1038/s41586-023-06408-7},
    number = {7979},
    urldate = {2025-04-12},
    journal = {Nature},
    author = {Sun, Hualei and Huo, Mengwu and Hu, Xunwu and Li, Jingyuan and Liu, Zengjia and Han, Yifeng and Tang, Lingyun and Mao, Zhongquan and Yang, Pengtao and Wang, Bosen and Cheng, Jinguang and Yao, Dao-Xin and Zhang, Guang-Ming and Wang, Meng},
    month = sep,
    year = {2023},
    pages = {493--498}
}

@article{hou_2023_nick,
    title = {Emergence of {High}-{Temperature} {Superconducting} {Phase} in {Pressurized} $\mathrm{La}_{3}\mathrm{Ni}_{2}\mathrm{O}_{7}$ {Crystals}},
    volume = {40},
    issn = {0256-307X, 1741-3540},
    doi = {10.1088/0256-307X/40/11/117302},
    number = {11},
    journal = {Chinese Physics Letters},
    author = {Hou , Jun  and Yang , Peng-Tao  and Liu , Zi-Yi  and Li , Jing-Yuan  and Shan , Peng-Fei  and Ma , Liang  and Wang , Gang  and Wang , Ning-Ning  and Guo , Hai-Zhong  and Sun , Jian-Ping  and Uwatoko, Yoshiya and Wang , Meng and Zhang , Guang-Ming  and Wang , Bo-Sen  and Cheng , Jin-Guang },
    month = oct,
    year = {2023},
    pages = {117302},
}

@article{zhang_2024_nick,
    title = {High-temperature superconductivity with zero resistance and strange-metal behaviour in $\mathrm{La}_3\mathrm{Ni}_2\mathrm{O}_{7-\delta}$},
    volume = {20},
    issn = {1745-2473, 1745-2481},
    doi = {10.1038/s41567-024-02515-y},
    number = {8},
    journal = {Nature Physics},
    author = {Zhang, Yanan and Su, Dajun and Huang, Yanen and Shan, Zhaoyang and Sun, Hualei and Huo, Mengwu and Ye, Kaixin and Zhang, Jiawen and Yang, Zihan and Xu, Yongkang and Su, Yi and Li, Rui and Smidman, Michael and Wang, Meng and Jiao, Lin and Yuan, Huiqiu},
    month = aug,
    year = {2024},
    pages = {1269--1273}
}

@article{wangPr2024,
  title = {{Bulk High-Temperature Superconductivity in Pressurized Tetragonal $\mathrm{La}_2\mathrm{PrNi}_2\mathrm{O}_7$}},
  author = {Wang, Ningning and Wang, Gang and Shen, Xiaoling and Hou, Jun and Luo, Jun and Ma, Xiaoping and Yang, Huaixin and Shi, Lifen and Dou, Jie and Feng, Jie and Yang, Jie and Shi, Yunqing and Ren, Zhian and Ma, Hanming and Yang, Pengtao and Liu, Ziyi and Liu, Yue and Zhang, Hua and Dong, Xiaoli and Wang, Yuxin and Jiang, Kun and Hu, Jiangping and Nagasaki, Shoko and Kitagawa, Kentaro and Calder, Stuart and Yan, Jiaqiang and Sun, Jianping and Wang, Bosen and Zhou, Rui and Uwatoko, Yoshiya and Cheng, Jinguang},
  year = 2024,
  month = oct,
  journal = {Nature},
  volume = {634},
  number = {8034},
  pages = {579--584},
  issn = {0028-0836, 1476-4687},
  doi = {10.1038/s41586-024-07996-8}
}

@article{wangPoly2024,
  title = {{Pressure-Induced Superconductivity In Polycrystalline $\mathrm{La}_3\mathrm{Ni}_2\mathrm{O}_{7-\delta}$}},
  author = {Wang, G. and Wang, N. N. and Shen, X. L. and Hou, J. and Ma, L. and Shi, L. F. and Ren, Z. A. and Gu, Y. D. and Ma, H. M. and Yang, P. T. and Liu, Z. Y. and Guo, H. Z. and Sun, J. P. and Zhang, G. M. and Calder, S. and Yan, J.-Q. and Wang, B. S. and Uwatoko, Y. and Cheng, J.-G.},
  date = {2024-03-07},
  journal = {Physical Review X},
  year = 2024,
  month = mar,
  volume = {14},
  number = {1},
  pages = {011040},
  issn = {2160-3308},
  doi = {10.1103/PhysRevX.14.011040}
}

@article{zhang123_2024,
  title = {{Effects of Pressure and Doping on Ruddlesden-Popper Phases $\mathrm{La}_{n+1}\mathrm{Ni}_{n}\mathrm{O}_{3n+1}$}},
  author = {Zhang, Mingxin and Pei, Cuiying and Wang, Qi and Zhao, Yi and Li, Changhua and Cao, Weizheng and Zhu, Shihao and Wu, Juefei and Qi, Yanpeng},
  date = {2024-06-20},
  journal = {Journal of Materials Science \& Technology},
  volume = {185},
  pages = {147--154},
  issn = {1005-0302},
  doi = {10.1016/j.jmst.2023.11.011},
  url = {https://www.sciencedirect.com/science/article/pii/S1005030223009829},
  year = 2026, 
  mon = jun
}

@article{puphalCrystal2024,
  title = {{Unconventional Crystal Structure of the High-Pressure Superconductor $\mathrm{La}_ 3\mathrm{Ni}_2\mathrm{O}_7$}},
  author = {Puphal, P. and Reiss, P. and Enderlein, N. and Wu, Y.-M. and Khaliullin, G. and Sundaramurthy, V. and Priessnitz, T. and Knauft, M. and Suthar, A. and Richter, L. and Isobe, M. and {van Aken}, P. A. and Takagi, H. and Keimer, B. and Suyolcu, Y. E. and Wehinger, B. and Hansmann, P. and Hepting, M.},
  year = 2024,
  month = oct,
  journal = {Physical Review Letters},
  volume = {133},
  number = {14},
  pages = {146002},
  publisher = {American Physical Society},
  doi = {10.1103/PhysRevLett.133.146002},
  urldate = {2026-03-12}
}

@article{dongVisual2024,
  title = {{Visualization of Oxygen Vacancies and Self-Doped Ligand Holes in $\mathrm{La}_3\mathrm{Ni}_2\mathrm{O}_{7-\delta}$}},
  author = {Dong, Zehao and Huo, Mengwu and Li, Jie and Li, Jingyuan and Li, Pengcheng and Sun, Hualei and Gu, Lin and Lu, Yi and Wang, Meng and Wang, Yayu and Chen, Zhen},
  date = {2024-06},
  journal = {Nature},
  volume = {630},
  number = {8018},
  pages = {847--852},
  publisher = {Nature Publishing Group},
  issn = {1476-4687},
  doi = {10.1038/s41586-024-07482-1},
  year = 2024,
  mon = jun
}

@article{yangOrbital2024,
  title = {{Orbital-Dependent Electron Correlation in Double-Layer Nickelate $\mathrm{La}_3\mathrm{Ni}_2\mathrm{O}_7$}},
  author = {Yang, Jiangang and Sun, Hualei and Hu, Xunwu and Xie, Yuyang and Miao, Taimin and Luo, Hailan and Chen, Hao and Liang, Bo and Zhu, Wenpei and Qu, Gexing and Chen, Cui-Qun and Huo, Mengwu and Huang, Yaobo and Zhang, Shenjin and Zhang, Fengfeng and Yang, Feng and Wang, Zhimin and Peng, Qinjun and Mao, Hanqing and Liu, Guodong and Xu, Zuyan and Qian, Tian and Yao, Dao-Xin and Wang, Meng and Zhao, Lin and Zhou, X. J.},
  year = 2024,
  month = may,
  journal = {Nature Communications},
  volume = {15},
  number = {1},
  pages = {4373},
  publisher = {Nature Publishing Group},
  issn = {2041-1723},
  doi = {10.1038/s41467-024-48701-7},
  urldate = {2026-03-12}
}

@article{khasanovSplittingDensity2025,
  title = {{Pressure-Enhanced Splitting of Density Wave Transitions in $\mathrm{La}_3\mathrm{Ni}_2\mathrm{O}_{7-\delta}$}},
  author = {Khasanov, Rustem and Hicken, Thomas J. and Gawryluk, Dariusz J. and Sazgari, Vahid and Plokhikh, Igor and Sorel, Lo{\"i}c Pierre and Bartkowiak, Marek and B{\"o}tzel, Steffen and Lechermann, Frank and Eremin, Ilya M. and Luetkens, Hubertus and Guguchia, Zurab},
  year = 2025,
  month = mar,
  journal = {Nature Physics},
  volume = {21},
  number = {3},
  pages = {430--436},
  publisher = {Nature Publishing Group},
  issn = {1745-2481},
  doi = {10.1038/s41567-024-02754-z},
  urldate = {2026-03-12}
}

@article{koAmbient2025,
  title = {{Signatures of Ambient Pressure Superconductivity in Thin Film $\mathrm{La}_3\mathrm{Ni}_2\mathrm{O7}$}},
  author = {Ko, Eun Kyo and Yu, Yijun and Liu, Yidi and Bhatt, Lopa and Li, Jiarui and Thampy, Vivek and Kuo, Cheng-Tai and Wang, Bai Yang and Lee, Yonghun and Lee, Kyuho and Lee, Jun-Sik and Goodge, Berit H. and Muller, David A. and Hwang, Harold Y.},
  year = 2025,
  month = feb,
  journal = {Nature},
  volume = {638},
  number = {8052},
  pages = {935--940},
  publisher = {Nature Publishing Group},
  issn = {1476-4687},
  doi = {10.1038/s41586-024-08525-3},
  urldate = {2026-02-19}
}

@article{zhouAmbient2025,
  title = {{Ambient-Pressure Superconductivity Onset above 40 {{K}} in $(\mathrm{La},\mathrm{Pr})_3\mathrm{Ni}_2\mathrm{O}_7$ Films}},
  author = {Zhou, Guangdi and Lv, Wei and Wang, Heng and Nie, Zihao and Chen, Yaqi and Li, Yueying and Huang, Haoliang and Chen, Wei-Qiang and Sun, Yu-Jie and Xue, Qi-Kun and Chen, Zhuoyu},
  year = 2025,
  month = apr,
  journal = {Nature},
  volume = {640},
  number = {8059},
  pages = {641--646},
  publisher = {Nature Publishing Group},
  issn = {1476-4687},
  doi = {10.1038/s41586-025-08755-z},
  urldate = {2026-03-12}
}

@article{liuFilm2025,
  title = {{Superconductivity and Normal-State Transport in Compressively Strained $\mathrm{La}_2\mathrm{PrNi}_2\mathrm{O}_7$ Thin Films}},
  author = {Liu, Yidi and Ko, Eun Kyo and Tarn, Yaoju and Bhatt, Lopa and Li, Jiarui and Thampy, Vivek and Goodge, Berit H. and Muller, David A. and Raghu, Srinivas and Yu, Yijun and Hwang, Harold Y.},
  year = 2025,
  month = aug,
  journal = {Nature Materials},
  volume = {24},
  number = {8},
  pages = {1221--1227},
  publisher = {Nature Publishing Group},
  issn = {1476-4660},
  doi = {10.1038/s41563-025-02258-y},
  urldate = {2026-03-12}
}

@article{haoSrdopedFilm025,
  title = {{Superconductivity in Sr-doped $\mathrm{La}_3\mathrm{Ni}_2\mathrm{O}_7$ Thin Films}},
  author = {Hao, Bo and Wang, Maosen and Sun, Wenjie and Yang, Yang and Mao, Zhangwen and Yan, Shengjun and Sun, Haoying and Zhang, Hongyi and Han, Lu and Gu, Zhengbin and Zhou, Jian and Ji, Dianxiang and Nie, Yuefeng},
  year = 2025,
  month = nov,
  journal = {Nature Materials},
  volume = {24},
  number = {11},
  pages = {1756--1762},
  publisher = {Nature Publishing Group},
  issn = {1476-4660},
  doi = {10.1038/s41563-025-02327-2},
  urldate = {2026-03-12}
}

@article{liSCfilm2026,
  title = {Enhanced Superconductivity in the Compressively Strained Bilayer Nickelate Thin Films by Pressure},
  author = {Li, Qing and Sun, Jianping and B{\"o}tzel, Steffen and Ou, Mengjun and Xiang, Zhe-Ning and Lechermann, Frank and Wang, Bosen and Wang, Yi and Zhang, Ying-Jie and Cheng, Jinguang and Eremin, Ilya M. and Wen, Hai-Hu},
  year = 2026,
  month = feb,
  journal = {Nature Communications},
  publisher = {Nature Publishing Group},
  issn = {2041-1723},
  doi = {10.1038/s41467-026-69660-1},
  urldate = {2026-03-24},
  copyright = {2026 The Author(s)},
  langid = {english}
}

@article{wangSuperconductingDome32026,
  title = {{Superconducting Dome in $\mathrm{La}_{3-x}\mathrm{Sr}_x\mathrm{Ni}_2 \mathrm{O}_{7-\delta}$ Thin Films}},
  author = {Wang, Maosen and Hao, Bo and Sun, Wenjie and Yan, Shengjun and Sun, Shengwang and Zhang, Hongyi and Gu, Zhengbin and Nie, Yuefeng},
  year = 2026,
  month = feb,
  journal = {Physical Review Letters},
  volume = {136},
  number = {6},
  pages = {066002},
  issn = {0031-9007, 1079-7114},
  doi = {10.1103/qrkk-l2ng},
  urldate = {2026-03-24},
  langid = {english}
}

@misc{liuHalfdome2026,
  title = {A Superconducting Half-Dome in Bilayer Nickelates},
  author = {Liu, Yidi and Wang, Bai Yang and Li, Jiarui and Tarn, Yaoju and Bhatt, Lopa and Colletta, Michael and Wu, Yi-Ming and Kuo, Cheng-Tai and Lee, Jun-Sik and Goodge, Berit H. and Muller, David A. and Shen, Zhi-Xun and Raghu, Srinivas and Hwang, Harold Y. and Yu, Yijun},
  year = 2026,
  month = mar,
  number = {arXiv:2603.12196},
  eprint = {2603.12196},
  publisher = {arXiv},
  urldate = {2026-04-06},
  archiveprefix = {arXiv}
}

@article{lu2023interlayer,
  title = {Interlayer-Coupling-Driven High-Temperature Superconductivity in $\mathrm{La}_{3}\mathrm{Ni}_{2}\mathrm{O}_{7}$ under Pressure},
  author = {Lu, Chen and Pan, Zhiming and Yang, Fan and Wu, Congjun},
  journal = {Physical Review Letter},
  volume = {132},
  issue = {14},
  pages = {146002},
  numpages = {6},
  year = {2024},
  month = {Apr},
  publisher = {American Physical Society},
  doi = {10.1103/PhysRevLett.132.146002}
}

@article{Yang2024ESD,
    archivePrefix = {arXiv},
    arxivId = {2309.15095},
    author = {Yang, Hui and Oh, Hanbit and Zhang, Ya-hui},
    doi = {10.1103/PhysRevB.110.104517},
    eprint = {2309.15095},
    issn = {2469-9950},
    journal = {Physical Review B},
    month = {sep},
    number = {10},
    pages = {104517},
    title = {{Strong pairing from a small Fermi surface beyond weak coupling: Application to $\mathrm{La}_3\mathrm{Ni}_2\mathrm{O}_7$}},
    volume = {110},
    year = {2024}
}

@misc{Oh2024ESD,
    archivePrefix = {arXiv},
    arxivId = {2411.07292},
    author = {Oh, Hanbit and Yang, Hui and Zhang, Ya-Hui},
    journal = {},
    eprint = {2411.07292},
    title = {{High-temperature superconductivity from kinetic energy}},
    pages = {1},
    year = {2024}
}

@article{wu_deconfined_2024,
    title = {Deconfined {Fermi} liquid to {Fermi} liquid transition and superconducting instability},
    volume = {110},
    doi = {10.1103/PhysRevB.110.125122},
    number = {12},
    urldate = {2025-04-20},
    journal = {Physical Review B},
    author = {Wu, Xiaofan and Yang, Hui and Zhang, Ya-Hui},
    month = sep,
    year = {2024},
    pages = {125122}
}

@article{lange2023pairing,
  title = {Pairing dome from an emergent Feshbach resonance in a strongly repulsive bilayer model},
  author = {Lange, Hannah and Homeier, Lukas and Demler, Eugene and Schollw\"ock, Ulrich and Bohrdt, Annabelle and Grusdt, Fabian},
  journal = {Physical Review B},
  volume = {110},
  issue = {8},
  pages = {L081113},
  numpages = {7},
  year = {2024},
  month = {Aug},
  publisher = {American Physical Society},
  doi = {10.1103/PhysRevB.110.L081113}
}

@article{luo2023bilayer,
  title = {Bilayer Two-Orbital Model of $\mathrm{L}{\mathrm{a}}_{3}\mathrm{N}{\mathrm{i}}_{2}\mathrm{O}_{7}$ under Pressure},
  author = {Luo, Zhihui and Hu, Xunwu and Wang, Meng and W\'u, W\'ei and Yao, Dao-Xin},
  journal = {Physical Review Lettter},
  volume = {131},
  issue = {12},
  pages = {126001},
  numpages = {6},
  year = {2023},
  month = {Sep},
  publisher = {American Physical Society},
  doi = {10.1103/PhysRevLett.131.126001}
}

@article{zhang2023electronic,
  title = {Electronic structure, dimer physics, orbital-selective behavior, and magnetic tendencies in the bilayer nickelate superconductor $\mathrm{L}{\mathrm{a}}_{3}\mathrm{N}{\mathrm{i}}_{2}\mathrm{O}_{7}$ under pressure},
  author = {Zhang, Yang and Lin, Ling-Fang and Moreo, Adriana and Dagotto, Elbio},
  journal = {Physical Review B},
  volume = {108},
  issue = {18},
  pages = {L180510},
  numpages = {5},
  year = {2023},
  month = {Nov},
  publisher = {American Physical Society},
  doi = {10.1103/PhysRevB.108.L180510},
}

@article{huang2023impurity,
  title={Impurity and vortex states in the bilayer high-temperature superconductor $\mathrm{La}_3\mathrm{Ni}_2\mathrm{O}_7$},
  author={Huang, Junkang and Wang, ZD and Zhou, Tao},
  journal={Physical Review B},
  volume={108},
  number={17},
  pages={174501},
  year={2023},
  doi={10.1103/PhysRevB.108.174501},
  publisher={APS}
}

@Article{Zhang2024,
    author={Zhang, Yang and Lin, Ling-Fang and Moreo, Adriana and Maier, Thomas A. and Dagotto, Elbio},
    title={Structural phase transition, $s_{\pm}$-wave pairing, and magnetic stripe order in bilayered superconductor $\mathrm{La}_3\mathrm{Ni}_2\mathrm{O}_7$ under pressure},
    journal={Nature Communications},
    year={2024},
    month={Mar},
    day={19},
    volume={15},
    number={1},
    pages={2470},
    doi={10.1038/s41467-024-46622-z}
}

@Article{Geisler2024,
    author={Geisler, Benjamin and Hamlin, James J. and Stewart, Gregory R. and Hennig, Richard G. and Hirschfeld, P. J.},
    title={Structural transitions, octahedral rotations, and electronic properties of $\mathrm{A}_3\mathrm{Ni}_2\mathrm{O}_7$ rare-earth nickelates under high pressure},
    journal={npj Quantum Materials},
    year={2024},
    month={Apr},
    day={26},
    volume={9},
    number={1},
    pages={38},
    issn={2397-4648},
    doi={10.1038/s41535-024-00648-0}
}

@article{PhysRevMaterials.8.044801,
  title = {Structural routes to stabilize superconducting $\mathrm{La}_{3}\mathrm{Ni}_{2}\mathrm{O}_{7}$ at ambient pressure},
  author = {Rhodes, Luke C. and Wahl, Peter},
  journal = {Physical Review Materials},
  volume = {8},
  issue = {4},
  pages = {044801},
  numpages = {9},
  year = {2024},
  month = {Apr},
  publisher = {American Physical Society},
  doi = {10.1103/PhysRevMaterials.8.044801}
}

@article{PhysRevB.109.045151,
  title = {Electronic structure, magnetic correlations, and superconducting pairing in the reduced Ruddlesden-Popper bilayer $\mathrm{La}_{3}\mathrm{Ni}_{2}\mathrm{O}_{6}$ under pressure: Different role of ${d}_{3{z}^{2}\ensuremath{-}{r}^{2}}$ orbital compared with $\mathrm{La}_{3}\mathrm{Ni}_{2}\mathrm{O}_{7}$},
  author = {Zhang, Yang and Lin, Ling-Fang and Moreo, Adriana and Maier, Thomas A. and Dagotto, Elbio},
  journal = {Physical Review B},
  volume = {109},
  issue = {4},
  pages = {045151},
  numpages = {10},
  year = {2024},
  month = {Jan},
  publisher = {American Physical Society},
  doi = {10.1103/PhysRevB.109.045151},
  url = {https://link.aps.org/doi/10.1103/PhysRevB.109.045151}
}

@article{sakakibara2023possible,
  title = {{Possible High ${T}_{c}$ Superconductivity in {La}$_3${Ni}$_2${O}$_7$ under High Pressure through Manifestation of a Nearly Half-Filled Bilayer Hubbard Model}},
  author = {Sakakibara, Hirofumi and Kitamine, Naoya and Ochi, Masayuki and Kuroki, Kazuhiko},
  journal = {Physical Review Letter},
  volume = {132},
  issue = {10},
  pages = {106002},
  numpages = {6},
  year = {2024},
  month = {Mar},
  publisher = {American Physical Society},
  doi = {10.1103/PhysRevLett.132.106002},
  url = {https://link.aps.org/doi/10.1103/PhysRevLett.132.106002}
}

@article{tian2024correlation,
  title={Correlation effects and concomitant two-orbital $s_\pm$-wave superconductivity in $\mathrm{La}_{3}\mathrm{Ni}_{2}\mathrm{O}_{7}$ under high pressure},
  author={Tian, Yi-Heng and Chen, Yin and Wang, Jia-Ming and He, Rong-Qiang and Lu, Zhong-Yi},
  journal={Physical Review B},
  volume={109},
  number={16},
  pages={165154},
  year={2024},
  doi={10.1103/PhysRevB.109.165154},
  publisher={APS}
}

@article{qin2023high,
  title = {High-${T}_{c}$ superconductivity by mobilizing local spin singlets and possible route to higher ${T}_{c}$ in pressurized $\mathrm{La}_{3}\mathrm{Ni}_{2}\mathrm{O}_{7}$},
  author = {Qin, Qiong and Yang, Yi-feng},
  journal = {Physical Review B},
  volume = {108},
  issue = {14},
  pages = {L140504},
  numpages = {6},
  year = {2023},
  month = {Oct},
  publisher = {American Physical Society},
  doi = {10.1103/PhysRevB.108.L140504},
  url = {https://link.aps.org/doi/10.1103/PhysRevB.108.L140504}
}

@article{yang2023minimal,
  title = {Interlayer valence bonds and two-component theory for high-${T}_{c}$ superconductivity of $\mathrm{La}_{3}\mathrm{Ni}_{2}\mathrm{O}_{7}$ under pressure},
  author = {Yang, Yi-feng and Zhang, Guang-Ming and Zhang, Fu-Chun},
  journal = {Physical Review B},
  volume = {108},
  issue = {20},
  pages = {L201108},
  numpages = {6},
  year = {2023},
  month = {Nov},
  publisher = {American Physical Society},
  doi = {10.1103/PhysRevB.108.L201108},
  url = {https://link.aps.org/doi/10.1103/PhysRevB.108.L201108}
}

@article{zhan2024cooperation,
  title = {Cooperation between Electron-Phonon Coupling and Electronic Interaction in Bilayer Nickelates $\mathrm{La}_{3}\mathrm{Ni}_{2}\mathrm{O}_{7}$},
  author = {Zhan, Jun and Gu, Yuhao and Wu, Xianxin and Hu, Jiangping},
  journal = {Physical Review Letter},
  volume = {134},
  issue = {13},
  pages = {136002},
  numpages = {7},
  year = {2025},
  month = {Mar},
  publisher = {American Physical Society},
  doi = {10.1103/PhysRevLett.134.136002},
  url = {https://link.aps.org/doi/10.1103/PhysRevLett.134.136002}
}

@article{Chen2024non,
  title = {Non-Fermi liquid and antiferromagnetic correlations with hole doping in the bilayer two-orbital Hubbard model of $\mathrm{La}_{3}\mathrm{Ni}_{2}\mathrm{O}_{7}$ at zero temperature},
  author = {Chen, Yin and Tian, Yi-Heng and Wang, Jia-Ming and He, Rong-Qiang and Lu, Zhong-Yi},
  journal = {Physical Review B},
  volume = {110},
  issue = {23},
  pages = {235119},
  numpages = {7},
  year = {2024},
  month = {Dec},
  publisher = {American Physical Society},
  doi = {10.1103/PhysRevB.110.235119},
  url = {https://link.aps.org/doi/10.1103/PhysRevB.110.235119}
}

@article{yang2023possible,
  title = {Possible {s}$_{\ifmmode\pm\else\textpm\fi{}}$-wave superconductivity in $\mathrm{L}{\mathrm{a}}_{3}\mathrm{N}{\mathrm{i}}_{2}\mathrm{O}_{7}$},
  author = {Yang, Qing-Geng and Wang, Da and Wang, Qiang-Hua},
  journal = {Physical Review B},
  volume = {108},
  issue = {14},
  pages = {L140505},
  numpages = {5},
  year = {2023},
  month = {Oct},
  publisher = {American Physical Society},
  doi = {10.1103/PhysRevB.108.L140505},
  url = {https://link.aps.org/doi/10.1103/PhysRevB.108.L140505}
}

@article{gu2023effective,
  title = {{Effective model and pairing tendency in the bilayer Ni-based superconductor ${\mathrm{La}}_{3}{\mathrm{Ni}}_{2}{\mathrm{O}}_{7}$}},
  author = {Gu, Yuhao and Le, Congcong and Yang, Zhesen and Wu, Xianxin and Hu, Jiangping},
  journal = {Physical Review B},
  volume = {111},
  issue = {17},
  pages = {174506},
  numpages = {7},
  year = {2025},
  month = {May},
  publisher = {American Physical Society},
  doi = {10.1103/PhysRevB.111.174506},
  url = {https://link.aps.org/doi/10.1103/PhysRevB.111.174506}
}

@article{liu2023s,
  title = {${s}^{\ifmmode\pm\else\textpm\fi{}}$-Wave Pairing and the Destructive Role of Apical-Oxygen Deficiencies in $\mathrm{La}_{3}\mathrm{Ni}_{2}\mathrm{O}_{7}$ under Pressure},
  author = {Liu, Yu-Bo and Mei, Jia-Wei and Ye, Fei and Chen, Wei-Qiang and Yang, Fan},
  journal = {Physical Review Letter},
  volume = {131},
  issue = {23},
  pages = {236002},
  numpages = {6},
  year = {2023},
  month = {Dec},
  publisher = {American Physical Society},
  doi = {10.1103/PhysRevLett.131.236002},
  url = {https://link.aps.org/doi/10.1103/PhysRevLett.131.236002}
}

@article{shen2023effective,
    doi = {10.1088/0256-307X/40/12/127401},
    year = {2023},
    month = {nov},
    publisher = {Chinese Physical Society and IOP Publishing Ltd},
    volume = {40},
    number = {12},
    pages = {127401},
    author = {Yang Shen and Mingpu Qin and Guang-Ming Zhang},
    title = {{Effective Bi-Layer Model Hamiltonian and Density-Matrix Renormalization Group Study for the High-Tc Superconductivity in $\mathrm{La}_3\mathrm{Ni}_2\mathrm{O}_7$ under High Pressure}},
    journal = {Chinese Physics Letters}
}

@article{PhysRevB.109.104508,
  title = {{Competing ${d}_{xy}$ and ${s}_{\ifmmode\pm\else\textpm\fi{}}$ pairing symmetries in superconducting ${\mathrm{La}}_{3}{\mathrm{Ni}}_{2}{\mathrm{O}}_{7}$: $\mathrm{LDA}+\mathrm{FLEX}$ calculations}},
  author = {Heier, Griffin and Park, Kyungwha and Savrasov, Sergey Y.},
  journal = {Physical Review B},
  volume = {109},
  issue = {10},
  pages = {104508},
  numpages = {9},
  year = {2024},
  month = {Mar},
  publisher = {American Physical Society},
  doi = {10.1103/PhysRevB.109.104508},
  url = {https://link.aps.org/doi/10.1103/PhysRevB.109.104508}
}

@article{PhysRevB.109.205156,
  title = {{Electronic properties of the bilayer nickelates ${R}_{3}\mathrm{N}{\mathrm{i}}_{2}{\mathrm{O}}_{7}$ with oxygen vacancies ($R=\mathrm{La}$ or Ce)}},
  author = {Sui, Xuelei and Han, Xiangru and Jin, Heng and Chen, Xiaojun and Qiao, Liang and Shao, Xiaohong and Huang, Bing},
  journal = {Physical Review B},
  volume = {109},
  issue = {20},
  pages = {205156},
  numpages = {12},
  year = {2024},
  month = {May},
  publisher = {American Physical Society},
  doi = {10.1103/PhysRevB.109.205156},
  url = {https://link.aps.org/doi/10.1103/PhysRevB.109.205156}
}

@article{PhysRevB.109.L201124,
  title = {{Pair correlations of the hybridized orbitals in a ladder model for the bilayer nickelate ${\mathrm{La}}_{3}{\mathrm{Ni}}_{2}{\mathrm{O}}_{7}$}},
  author = {Kakoi, Masataka and Kaneko, Tatsuya and Sakakibara, Hirofumi and Ochi, Masayuki and Kuroki, Kazuhiko},
  journal = {Physical Review B},
  volume = {109},
  issue = {20},
  pages = {L201124},
  numpages = {6},
  year = {2024},
  month = {May},
  publisher = {American Physical Society},
  doi = {10.1103/PhysRevB.109.L201124},
  url = {https://link.aps.org/doi/10.1103/PhysRevB.109.L201124}
}

@article{oh2023type,
  title = {Type-{II} ${t}\ensuremath{-}{J}$ model and shared superexchange coupling from Hund's rule in superconducting $\mathrm{La}_{3}\mathrm{Ni}_{2}\mathrm{O}_{7}$},
  author = {Oh, Hanbit and Zhang, Ya-Hui},
  journal = {Physical Review B},
  volume = {108},
  issue = {17},
  pages = {174511},
  numpages = {8},
  year = {2023},
  month = {Nov},
  publisher = {American Physical Society},
  doi = {10.1103/PhysRevB.108.174511},
  url = {https://link.aps.org/doi/10.1103/PhysRevB.108.174511}
}

@article{zhang2023strong,
  title = {Strong Pairing Originated from an Emergent ${Z}_{2}$ Berry Phase in $\mathrm{La}_{3}\mathrm{Ni}_{2}\mathrm{O}_{7}$},
  author = {Zhang, Jia-Xin and Zhang, Hao-Kai and You, Yi-Zhuang and Weng, Zheng-Yu},
  journal = {Physical Review Letter},
  volume = {133},
  issue = {12},
  pages = {126501},
  numpages = {7},
  year = {2024},
  month = {Sep},
  publisher = {American Physical Society},
  doi = {10.1103/PhysRevLett.133.126501},
  url = {https://link.aps.org/doi/10.1103/PhysRevLett.133.126501}
}

@misc{zhu2025quantum,
  title={Quantum phase transition driven by competing intralayer and interlayer hopping in bilayer nickelates},
  author={Zhu, Xiaoyu and Qin, Wei and Cui, Ping and Zhang, Zhenyu},
  archivePrefix = {arXiv},
  arxivId = {2507.11169},
  eprint={2507.11169},
  year={2025}
}

@article{pan2023effect,
  title={{Effect of Rare-earth Element Substitution in Superconducting {R}$_3$Ni$_2$O$_7$ Under Pressure}},
  author = {Pan, Zhiming and Lu, Chen and Yang, Fan and Wu, Congjun},
  year = 2024,
  month = aug,
  journal = {Chinese Physics Letters},
  volume = {41},
  number = {8},
  pages = {087401},
  publisher = {{Chinese Physical Society and IOP Publishing Ltd}},
  issn = {0256-307X},
  doi = {10.1088/0256-307X/41/8/087401},
  urldate = {2026-03-22},
  langid = {english}
}

@article{maCPL2025,
  title = {Parameter-{{Dependent Superconducting Transition Temperature}} in a {{Sign-Problem-Free Bilayer Model}}},
  author = {Ma, Runyu and Fan, Zenghui and Ma, Tianxing and Wu, Congjun},
  year = 2025,
  month = nov,
  journal = {Chinese Physics Letters},
  volume = {42},
  number = {11},
  pages = {110705--110705},
  issn = {0256-307X},
  doi = {10.1088/0256-307X/42/11/110705},
  urldate = {2026-04-06},
}

@article{PhysRevB.110.024514,
  title = {Superconductivity in nickelate and cuprate superconductors with strong bilayer coupling},
  author = {Fan, Zhen and Zhang, Jian-Feng and Zhan, Bo and Lv, Dingshun and Jiang, Xing-Yu and Normand, Bruce and Xiang, Tao},
  journal = {Physical Review B},
  volume = {110},
  issue = {2},
  pages = {024514},
  numpages = {10},
  year = {2024},
  month = {Jul},
  publisher = {American Physical Society},
  doi = {10.1103/PhysRevB.110.024514},
  url = {https://link.aps.org/doi/10.1103/PhysRevB.110.024514}
}

@article{PhysRevB.110.L060510,
  title = {{Electronic structure, self-doping, and superconducting instability in the alternating single-layer trilayer stacking nickelates ${\mathrm{La}}_{3}{\mathrm{Ni}}_{2}{\mathrm{O}}_{7}$}},
  author = {Zhang, Yang and Lin, Ling-Fang and Moreo, Adriana and Maier, Thomas A. and Dagotto, Elbio},
  journal = {Physical Review B},
  volume = {110},
  issue = {6},
  pages = {L060510},
  numpages = {7},
  year = {2024},
  month = {Aug},
  publisher = {American Physical Society},
  doi = {10.1103/PhysRevB.110.L060510},
  url = {https://link.aps.org/doi/10.1103/PhysRevB.110.L060510}
}

@article{PhysRevB.110.104507,
  title = {Decomposition of multilayer superconductivity with interlayer pairing},
  author = {Yang, Yi-feng},
  journal = {Physical Review B},
  volume = {110},
  issue = {10},
  pages = {104507},
  numpages = {6},
  year = {2024},
  month = {Sep},
  publisher = {American Physical Society},
  doi = {10.1103/PhysRevB.110.104507},
  url = {https://link.aps.org/doi/10.1103/PhysRevB.110.104507}
}

@article{PhysRevB.110.094509,
  title = {{Interplay of two ${E}_{g}$ orbitals in superconducting ${\mathrm{La}}_{3}{\mathrm{Ni}}_{2}{\mathrm{O}}_{7}$ under pressure}},
  author = {Lu, Chen and Pan, Zhiming and Yang, Fan and Wu, Congjun},
  journal = {Physical Review B},
  volume = {110},
  issue = {9},
  pages = {094509},
  numpages = {16},
  year = {2024},
  month = {Sep},
  publisher = {American Physical Society},
  doi = {10.1103/PhysRevB.110.094509},
  url = {https://link.aps.org/doi/10.1103/PhysRevB.110.094509}
}

@misc{lu2023superconductivity,
  title={Superconductivity from Doping Symmetric Mass Generation Insulators: Application to {La}$_3${Ni}$_2${O}$_7$ under Pressure},
  author={Lu, Da-Chuan and Li, Miao and Zeng, Zhao-Yi and Hou, Wanda and Wang, Juven and Yang, Fan and You, Yi-Zhuang},
  archievePrefix={arXiv},
  eprint={2308.11195},
  year={2023}
}

@Article{Luo2024,
    author={Luo, Zhihui and Lv, Biao and Wang, Meng and W{\'u}, W{\'e}i and Yao, Dao-Xin},
    title={{High-$T_C$ superconductivity in $\mathrm{La}_3\mathrm{Ni}_2\mathrm{O}_7$ based on the bilayer two-orbital t-J model}},
    journal={npj Quantum Materials},
    year={2024},
    month={Aug},
    day={13},
    volume={9},
    number={1},
    pages={61},
    issn={2397-4648},
    doi={10.1038/s41535-024-00668-w},
    url={https://doi.org/10.1038/s41535-024-00668-w}
}

@article{PhysRevB.109.045127,
  title = {Feshbach resonance in a strongly repulsive ladder of mixed dimensionality: A possible scenario for bilayer nickelate superconductors},
  author = {Lange, Hannah and Homeier, Lukas and Demler, Eugene and Schollw\"ock, Ulrich and Grusdt, Fabian and Bohrdt, Annabelle},
  journal = {Physical Review B},
  volume = {109},
  issue = {4},
  pages = {045127},
  numpages = {16},
  year = {2024},
  month = {Jan},
  publisher = {American Physical Society},
  doi = {10.1103/PhysRevB.109.045127},
  url = {https://link.aps.org/doi/10.1103/PhysRevB.109.045127}
}

@article{PhysRevB.109.045154,
  title = {{Pair correlations in the two-orbital Hubbard ladder: Implications for superconductivity in the bilayer nickelate ${\mathrm{La}}_{3}{\mathrm{Ni}}_{2}{\mathrm{O}}_{7}$}},
  author = {Kaneko, Tatsuya and Sakakibara, Hirofumi and Ochi, Masayuki and Kuroki, Kazuhiko},
  journal = {Physical Review B},
  volume = {109},
  issue = {4},
  pages = {045154},
  numpages = {5},
  year = {2024},
  month = {Jan},
  publisher = {American Physical Society},
  doi = {10.1103/PhysRevB.109.045154},
  url = {https://link.aps.org/doi/10.1103/PhysRevB.109.045154}
}

@article{PhysRevLett.133.096002,
  title = {{Quenched Pair Breaking by Interlayer Correlations as a Key to Superconductivity in ${\mathrm{La}}_{3}{\mathrm{Ni}}_{2}{\mathrm{O}}_{7}$}},
  author = {Ryee, Siheon and Witt, Niklas and Wehling, Tim O.},
  journal = {Physical Review Letter},
  volume = {133},
  issue = {9},
  pages = {096002},
  numpages = {7},
  year = {2024},
  month = {Aug},
  publisher = {American Physical Society},
  doi = {10.1103/PhysRevLett.133.096002},
  url = {https://link.aps.org/doi/10.1103/PhysRevLett.133.096002}
}

@Article{Ouyang2024,
  author={Ouyang, Zhenfeng and Gao, Miao and Lu, Zhong-Yi},
  title={Absence of electron-phonon coupling superconductivity in the bilayer phase of $\mathrm{La}_3\mathrm{Ni}_2\mathrm{O}_7$ under pressure},
  journal={npj Quantum Materials},
  year={2024},
  month={Oct},
  day={15},
  volume={9},
  number={1},
  pages={80},
  issn={2397-4648},
  doi={10.1038/s41535-024-00689-5},
  url={https://doi.org/10.1038/s41535-024-00689-5}
}

@article{PhysRevB.108.125105,
  title = {{Correlated electronic structure, orbital-selective behavior, and magnetic correlations in double-layer ${\mathrm{La}}_{3}{\mathrm{Ni}}_{2}{\mathrm{O}}_{7}$ under pressure}},
  author = {Shilenko, D. A. and Leonov, I. V.},
  journal = {Physical Review B},
  volume = {108},
  issue = {12},
  pages = {125105},
  numpages = {9},
  year = {2023},
  month = {Sep},
  publisher = {American Physical Society},
  doi = {10.1103/PhysRevB.108.125105},
  url = {https://link.aps.org/doi/10.1103/PhysRevB.108.125105}
}

@article{cao2023flat,
  title = {Flat bands promoted by Hund's rule coupling in the candidate double-layer high-temperature superconductor {La}$_3${Ni}$_2${O}$_7$ under high pressure},
  author = {Cao, Yingying and Yang, Yi-feng},
  journal = {Physical Review B},
  volume = {109},
  issue = {8},
  pages = {L081105},
  numpages = {6},
  year = {2024},
  month = {Feb},
  publisher = {American Physical Society},
  doi = {10.1103/PhysRevB.109.L081105},
  url = {https://link.aps.org/doi/10.1103/PhysRevB.109.L081105}
}

@article{qu2023bilayer,
  title = {Bilayer ${t\text{\ensuremath{-}}J\text{\ensuremath{-}}J}_{\ensuremath{\perp}}$ Model and Magnetically Mediated Pairing in the Pressurized Nickelate $\mathrm{La}_{3}\mathrm{Ni}_{2}\mathrm{O}_{7}$},
  author = {Qu, Xing-Zhou and Qu, Dai-Wei and Chen, Jialin and Wu, Congjun and Yang, Fan and Li, Wei and Su, Gang},
  journal = {Physiew Review Letter},
  volume = {132},
  issue = {3},
  pages = {036502},
  numpages = {6},
  year = {2024},
  month = {Jan},
  publisher = {American Physical Society},
  doi = {10.1103/PhysRevLett.132.036502},
  url = {https://link.aps.org/doi/10.1103/PhysRevLett.132.036502}
}

@article{PhysRevB.108.214522,
  title = {{Electron correlations and superconductivity in ${\mathrm{La}}_{3}{\mathrm{Ni}}_{2}{\mathrm{O}}_{7}$ under pressure tuning}},
  author = {Liao, Zhiguang and Chen, Lei and Duan, Guijing and Wang, Yiming and Liu, Changle and Yu, Rong and Si, Qimiao},
  journal = {Physical Review B},
  volume = {108},
  issue = {21},
  pages = {214522},
  numpages = {9},
  year = {2023},
  month = {Dec},
  publisher = {American Physical Society},
  doi = {10.1103/PhysRevB.108.214522},
  url = {https://link.aps.org/doi/10.1103/PhysRevB.108.214522}
}

@article{zhang2023trends,
  title = {{Trends in electronic structures and ${s}_{\ifmmode\pm\else\textpm\fi{}}$-wave pairing for the rare-earth series in bilayer nickelate superconductor $R_{3}\mathrm{Ni}_{2}\mathrm{O}_{7}$}},
  author = {Zhang, Yang and Lin, Ling-Fang and Moreo, Adriana and Maier, Thomas A. and Dagotto, Elbio},
  journal = {Physical Review B},
  volume = {108},
  issue = {16},
  pages = {165141},
  numpages = {8},
  year = {2023},
  month = {Oct},
  publisher = {American Physical Society},
  doi = {10.1103/PhysRevB.108.165141},
  url = {https://link.aps.org/doi/10.1103/PhysRevB.108.165141}
}

@article{PhysRevB.111.014515,
  title = {{Charge and spin instabilities in superconducting ${\mathrm{La}}_{3}{\mathrm{Ni}}_{2}{\mathrm{O}}_{7}$}},
  author = {Chen, Xuejiao and Jiang, Peiheng and Li, Jie and Zhong, Zhicheng and Lu, Yi},
  journal = {Physical Review B},
  volume = {111},
  issue = {1},
  pages = {014515},
  numpages = {8},
  year = {2025},
  month = {Jan},
  publisher = {American Physical Society},
  doi = {10.1103/PhysRevB.111.014515},
  url = {https://link.aps.org/doi/10.1103/PhysRevB.111.014515}
}

@article{PhysRevB.109.115114,
  title = {{Hund electronic correlation in ${\mathrm{La}}_{3}{\mathrm{Ni}}_{2}{\mathrm{O}}_{7}$ under high pressure}},
  author = {Ouyang, Zhenfeng and Wang, Jia-Ming and Wang, Jing-Xuan and He, Rong-Qiang and Huang, Li and Lu, Zhong-Yi},
  journal = {Physical Review B},
  volume = {109},
  issue = {11},
  pages = {115114},
  numpages = {7},
  year = {2024},
  month = {Mar},
  publisher = {American Physical Society},
  doi = {10.1103/PhysRevB.109.115114},
  url = {https://link.aps.org/doi/10.1103/PhysRevB.109.115114}
}

@article{PhysRevB.110.205122,
  title = {{Electronic and magnetic structures of bilayer ${\text{La}}_{3}{\text{Ni}}_{2}{\text{O}}_{7}$ at ambient pressure}},
  author = {Wang, Yuxin and Jiang, Kun and Wang, Ziqiang and Zhang, Fu-Chun and Hu, Jiangping},
  journal = {Physical Review B},
  volume = {110},
  issue = {20},
  pages = {205122},
  numpages = {7},
  year = {2024},
  month = {Nov},
  publisher = {American Physical Society},
  doi = {10.1103/PhysRevB.110.205122},
  url = {https://link.aps.org/doi/10.1103/PhysRevB.110.205122}
}

@article{PhysRevB.109.L180502,
  title = {{Theory of magnetic excitations in the multilayer nickelate superconductor ${\mathrm{La}}_{3}{\mathrm{Ni}}_{2}{\mathrm{O}}_{7}$}},
  author = {B\"otzel, Steffen and Lechermann, Frank and Gondolf, Jannik and Eremin, Ilya M.},
  journal = {Physical Review B},
  volume = {109},
  issue = {18},
  pages = {L180502},
  numpages = {8},
  year = {2024},
  month = {May},
  publisher = {American Physical Society},
  doi = {10.1103/PhysRevB.109.L180502},
  url = {https://link.aps.org/doi/10.1103/PhysRevB.109.L180502}
}

@article{tian2025spin,
  title={{Spin density wave and superconductivity in the bilayer t-J model of $\mathrm{La}_3\mathrm{Ni}_2\mathrm{O}_7$ under renormalized mean-field theory}},
  author={Tian, Yang and Chen, Yan},
  journal={Physical Review B},
  volume={112},
  number={1},
  pages={014520},
  year={2025},
  publisher={APS}
}

@article{liu2025origin,
  title = {{Origin of the diagonal double-stripe spin density wave and potential superconductivity in bulk ${\mathrm{La}}_{3}{\mathrm{Ni}}_{2}{\mathrm{O}}_{7}$ at ambient pressure}},
  author = {Liu, Yu-Bo and Sun, Hongyi and Zhang, Ming and Liu, Qihang and Chen, Wei-Qiang and Yang, Fan},
  journal = {Physical Review B},
  volume = {112},
  issue = {1},
  pages = {014510},
  numpages = {13},
  year = {2025},
  month = {Jul},
  publisher = {American Physical Society},
  doi = {10.1103/24f4-349n},
  url = {https://link.aps.org/doi/10.1103/24f4-349n}
}

@misc{liao2024orbital,
  title={{Orbital-selective electron correlations in high-$T_c$ bilayer nickelates: from a global phase diagram to implications for spectroscopy}},
  author={Liao, Zhiguang and Wang, Yiming and Chen, Lei and Duan, Guijing and Yu, Rong and Si, Qimiao},
  archivePrefix={arXiv},
  eprint={2412.21019},
  year={2024}
}

@article{PhysRevLett.132.126503,
  title = {Pressure Driven Fractionalization of Ionic Spins Results in Cupratelike High-${T}_{c}$ Superconductivity in $\mathrm{La}_{3}\mathrm{Ni}_{2}\mathrm{O}_{7}$},
  author = {Jiang, Ruoshi and Hou, Jinning and Fan, Zhiyu and Lang, Zi-Jian and Ku, Wei},
  journal = {Physical Review Lett.},
  volume = {132},
  issue = {12},
  pages = {126503},
  numpages = {7},
  year = {2024},
  month = {Mar},
  publisher = {American Physical Society},
  doi = {10.1103/PhysRevLett.132.126503},
  url = {https://link.aps.org/doi/10.1103/PhysRevLett.132.126503}
}

@misc{yin2025s,
  title={The $s\pm $ pairing symmetry in the pressured $ \mathrm{La}_3 \mathrm{Ni}_2\mathrm{O}_7 $ from electron-phonon coupling},
  author={Yin, Yucong and Zhan, Jun and Liu, Boyang and Han, Xinloong},
  archivePrefix={arXiv},
  eprint={2502.21016},
  year={2025}
}

@article{PhysRevB.111.104505,
  title = {{Transition from ${s}_{\ifmmode\pm\else\textpm\fi{}}$-wave to ${d}_{{x}^{2}\ensuremath{-}{y}^{2}}$-wave superconductivity driven by interlayer interaction in the bilayer two-orbital model of ${\text{La}}_{3}{\text{Ni}}_{2}{\text{O}}_{7}$}},
  author = {Xi, Wenhan and Yu, Shun-Li and Li, Jian-Xin},
  journal = {Physical Review B},
  volume = {111},
  issue = {10},
  pages = {104505},
  numpages = {10},
  year = {2025},
  month = {Mar},
  publisher = {American Physical Society},
  doi = {10.1103/PhysRevB.111.104505},
  url = {https://link.aps.org/doi/10.1103/PhysRevB.111.104505}
}

@misc{kaneko2025t,
  title={{$t $-$ J $ model for strongly correlated two-orbital systems: Application to bilayer nickelate superconductors}},
  author={Kaneko, Tatsuya and Kakoi, Masataka and Kuroki, Kazuhiko},
  archivePrefix={arXiv},
  eprint={2504.10114},
  year={2025}
}

@misc{ji2025strong,
  title={{A Strong-Coupling-Limit Study on the Pairing Mechanism in the Pressurized ${\text{La}}_{3}{\text{Ni}}_{2}{\text{O}}_{7}$}},
  author={Ji, Jia-Heng and Lu, Chen and Shao, Zhi-Yan and Pan, Zhiming and Yang, Fan and Wu, Congjun},
  archivePrefix={arXiv},
  eprint={2504.12127},
  year={2025}
}

@article{Wang_2025,
    doi = {10.1088/1674-1056/adbacc},
    year = {2025},
    month = {apr},
    publisher = {Chinese Physical Society and IOP Publishing Ltd},
    volume = {34},
    number = {4},
    pages = {047105},
    author = {Wang, Yuxin and Zhang, Yi and Jiang, Kun},
    title = {{Electronic structure and disorder effect of ${\text{La}}_{3}{\text{Ni}}_{2}{\text{O}}_{7}$ superconductor}},
    journal = {Chinese Physics B},
}

@misc{haque2025dft,
  title={DFT exploration of pressure dependent physical properties of the recently discovered $\mathrm{La}_3\mathrm{Ni}_2\mathrm{O}_7$ superconductor},
  archivePrefix={arXiv},
  author={Haque, Md Enamul and Ali, Ruman and Masum, MA and Hassan, Jahid and Naqib, SH},
  eprint={arXiv:2504.15853},
  year={2025}
}

@misc{shi2025theoretical,
  title={Theoretical Investigation of High-Tc Superconductivity in Sr-Doped $\mathrm{La}_3\mathrm{Ni}_2\mathrm{O}_7$ at Ambient Pressure},
  archivePrefix={arXiv},
  author={Shi, Lei and Luo, Ying and Wu, Wei and Zhang, Yunwei},
  eprint={arXiv:2503.13197},
  year={2025}
}

@misc{gao2025robust,
  title={Robust $ s\pm $-wave pairing in a bilayer two-orbital model of pressurized $\mathrm{La}_3\mathrm{Ni}_2\mathrm{O}_7$ without the $\gamma $ Fermi surface},
  author={Gao, Yi},
  archivePrefix={arXiv},
  eprint={2502.19840},
  year={2025}
}

@misc{le2025landscape,
  title={Landscape of correlated orders in strained bilayer nickelate thin films},
  author={Le, Congcong and Zhan, Jun and Wu, Xianxin and Hu, Jiangping},
  eprints={2501.14665},
  archivePrefix={arXiv},
  year={2025}
}

@misc{hu2025electronic,
  title={Electronic structures and multi-orbital models of $\mathrm{La}_3\mathrm{Ni}_2\mathrm{O}_7$  thin films at ambient pressure},
  author={Hu, Xunwu and Qiu, Wenyuan and Chen, Cun-Qun and Luo, Zhihui and Yao, Dao-Xin},
  eprint={2503.17223},
  archivePrefix={arXiv},
  year={2025}
}

@misc{shao2024possible,
  title={Possible High-Temperature Superconductivity Driven by Perpendicular Electric Field in the $\mathrm{La}_3\mathrm{Ni}_2\mathrm{O}_7$ Single-Bilayer Film at Ambient Pressure},
  author={Shao, Zhi-Yan and Ji, Jia-Heng and Wu, Congjun and Yao, Dao-Xin and Yang, Fan},
  archivePrefix={arXiv},
  eprint={2411.13554},
  year={2024}
}

@article{rm9g-8lm1,
  title = {Effective perpendicular electric field as a probe for interlayer pairing in ambient-pressure superconducting ${\mathrm{La}}_{2.85}{\mathrm{Pr}}_{0.15}{\mathrm{Ni}}_{2}{\mathrm{O}}_{7}$ thin films},
  author = {Huang, Junkang and Zhou, Tao},
  journal = {Physical Review B},
  volume = {112},
  issue = {5},
  pages = {054506},
  numpages = {9},
  year = {2025},
  month = {Aug},
  publisher = {American Physical Society},
  doi = {10.1103/rm9g-8lm1},
  url = {https://link.aps.org/doi/10.1103/rm9g-8lm1}
}

@misc{ushio2025theoretical,
  title={Theoretical study on ambient pressure superconductivity in $\mathrm{La}_3\mathrm{Ni}_2\mathrm{O}_7$ thin films: structural analysis, model construction, and robustness of $ s\pm$-wave pairing},
  author={Ushio, Kensei and Kamiyama, Shu and Hoshi, Yuto and Mizuno, Ryota and Ochi, Masayuki and Kuroki, Kazuhiko and Sakakibara, Hirofumi},
  archivePrefix={arXiv},
  eprint={2506.20497},
  year={2025}
}

@misc{duan2025orbital,
  archivePrefix={arXiv},
  title={Orbital-selective correlation effects and superconducting pairing symmetry in a multiorbital $ t $-$ J $ model for bilayer nickelates},
  author={Duan, Guijing and Liao, Zhiguang and Chen, Lei and Wang, Yiming and Yu, Rong and Si, Qimiao},
  eprint={2502.09195},
  year={2025}
}

@misc{qiu2025pairing,
  title={Pairing symmetry and superconductivity in $\mathrm{La}_3\mathrm{Ni}_2\mathrm{O}_7$  thin films},
  author={Qiu, Wenyuan and Luo, Zhihui and Hu, Xunwu and Yao, Dao-Xin},
  archivePrefix={arXiv},
  eprint={2506.20727},
  year={2025}
}

@misc{cao2025strain,
  title={Strain-Engineered Electronic Structure and Superconductivity in $\mathrm{La}_3\mathrm{Ni}_2\mathrm{O}_7$ Thin Films},
  author={Cao, Yu-Han and Jiang, Kai-Yue and Lu, Hong-Yan and Wang, Da and Wang, Qiang-Hua},
  archivePrefix={arXiv},
  eprint={2507.13694},
  year={2025}
}

@misc{shao2025pairing,
  title={Pairing without $\gamma $-Pocket in the $\mathrm{La}_3\mathrm{Ni}_2\mathrm{O}_7$ Thin Film},
  author={Shao, Zhi-Yan and Lu, Chen and Liu, Min and Liu, Yu-Bo and Pan, Zhiming and Wu, Congjun and Yang, Fan},
  archivePrefix={arXiv},
  eprint={2507.20287},
  year={2025}
}

@article{PhysRevB.111.L020504,
  title = {{Type-II $t\text{\ensuremath{-}}J$ model in charge transfer regime in bilayer ${\mathrm{La}}_{3}{\mathrm{Ni}}_{2}{\mathrm{O}}_{7}$ and trilayer ${\mathrm{La}}_{4}{\mathrm{Ni}}_{3}{\mathrm{O}}_{10}$}},
  author = {Oh, Hanbit and Zhou, Boran and Zhang, Ya-Hui},
  journal = {Physical Review B},
  volume = {111},
  issue = {2},
  pages = {L020504},
  numpages = {7},
  year = {2025},
  month = {Jan},
  publisher = {American Physical Society},
  doi = {10.1103/PhysRevB.111.L020504},
  url = {https://link.aps.org/doi/10.1103/PhysRevB.111.L020504}
}

@article{xue2024magnetism,
  title={{Magnetism and Superconductivity in the t--J Model of $\mathrm{La}_3\mathrm{Ni}_2\mathrm{O}_7$ Under Multiband Gutzwiller Approximation}},
  author={Xue, Jie-Ran and Wang, Fa},
  journal={Chinese Physics Letters},
  doi={10.1088/0256-307X/41/5/057403},
  volume={41},
  number={5},
  pages={057403},
  year={2024},
  publisher={IOP Publishing}
}

@article{luttingerWard1960,
  title = {{Ground-State Energy} of a {{Many-Fermion System}}. {{II}}},
  author = {Luttinger, J. M. and Ward, J. C.},
  year = 1960,
  month = jun,
  journal = {Physical Review},
  volume = {118},
  number = {5},
  pages = {1417--1427},
  publisher = {American Physical Society},
  doi = {10.1103/PhysRev.118.1417},
  urldate = {2026-03-12}
}

@article{luttinger1960,
  title = {Fermi {{Surface}} and {{Some Simple Equilibrium Properties}} of a {{System}} of {{Interacting Fermions}}},
  author = {Luttinger, J. M.},
  year = 1960,
  month = aug,
  journal = {Physical Review},
  volume = {119},
  number = {4},
  pages = {1153--1163},
  publisher = {American Physical Society},
  doi = {10.1103/PhysRev.119.1153},
  urldate = {2026-03-12}
}

@article{oshikawa_topological_2000,
    title = {Topological approach to {Luttinger}'s theorem and the {Fermi} surface of a {Kondo} lattice},
    volume = {84},
    doi = {10.1103/PhysRevLett.84.3370},
    number = {15},
    journal = {Physical Review Letters},
    author = {Oshikawa, Masaki},
    month = apr,
    year = {2000},
    pages = {3370--3373}
}

@article{StadlerInterleave2016,
  title = {{Interleaved Numerical Renormalization Group as an Efficient Multiband Impurity Solver}},
  author = {Stadler, K. M. and Mitchell, A. K. and von Delft, J. and Weichselbaum, A.},
  date = {2016-06-01},
  journal = {Physical Review B},
  volume = {93},
  number = {23},
  year = 2016,
  pages = {235101},
  publisher = {American Physical Society},
  doi = {10.1103/PhysRevB.93.235101},
  urldate = {2026-03-03}
}

@article{MitchellInter2014,
  title = {{Generalized {{Wilson}} Chain for Solving Multichannel Quantum Impurity Problems}},
  author = {Mitchell, Andrew K. and Galpin, Martin R. and {Wilson-Fletcher}, Samuel and Logan, David E. and Bulla, Ralf},
  year = 2014,
  month = mar,
  journal = {Physical Review B},
  volume = {89},
  number = {12},
  pages = {121105},
  publisher = {American Physical Society},
  doi = {10.1103/PhysRevB.89.121105},
  urldate = {2026-03-03}
}

@article{Kugler2022, 
    archivePrefix = {arXiv},
    arxivId = {2202.04063},
    author = {Kugler, Fabian B.},
    doi = {10.1103/PhysRevB.105.245132},
    eprint = {2202.04063},
    issn = {2469-9950},
    journal = {Physical Review B},
    month = {jun},
    number = {24},
    pages = {245132},
    title = {{Improved estimator for numerical renormalization group calculations of the self-energy}},
    volume = {105},
    year = {2022}
}

@article{Weichselbaum2012,
    archivePrefix = {arXiv},
    arxivId = {1209.2062},
    author = {Weichselbaum, Andreas},
    doi = {10.1103/PhysRevB.86.245124},
    eprint = {1209.2062},
    issn = {1098-0121},
    journal = {Phys. Rev. B},
    month = {dec},
    number = {24},
    pages = {245124},
    title = {{Tensor networks and the numerical renormalization group}},
    volume = {86},
    year = {2012}
}

@article{Bulla2008,
    archivePrefix = {arXiv},
    arxivId = {cond-mat/0701105},
    author = {Bulla, Ralf and Costi, Theo A. and Pruschke, Thomas},
    doi = {10.1103/RevModPhys.80.395},
    eprint = {0701105},
    issn = {0034-6861},
    journal = {Rev. Mod. Phys.},
    month = {apr},
    number = {2},
    pages = {395--450},
    title = {{Numerical renormalization group method for quantum impurity systems}},
    volume = {80},
    year = {2008}
}

@article{Bulla1999,
   author = {R. Bulla},
   doi = {10.1103/PhysRevLett.83.136},
   issn = {0031-9007},
   issue = {1},
   journal = {Physical Review Letters},
   month = {7},
   pages = {136-139},
   title = {Zero Temperature Metal-Insulator Transition in the Infinite-Dimensional Hubbard Model},
   volume = {83},
   url = {https://link.aps.org/doi/10.1103/PhysRevLett.83.136},
   year = {1999}
}

@article{Weichselbaum2007,
    archivePrefix = {arXiv},
    arxivId = {cond-mat/0607497},
    author = {Weichselbaum, Andreas and von Delft, Jan},
    doi = {10.1103/PhysRevLett.99.076402},
    eprint = {0607497},
    issn = {0031-9007},
    journal = {Physical Review Letters},
    month = {aug},
    number = {7},
    pages = {076402},
    title = {{Sum-Rule Conserving Spectral Functions from the Numerical Renormalization Group}},
    volume = {99},
    year = {2007}
}

@article{Wilson1975,
    author = {Wilson, Kenneth G.},
    doi = {10.1103/RevModPhys.47.773},
    issn = {0034-6861},
    journal = {Reviews of Modern Physics},
    month = {oct},
    number = {4},
    pages = {773--840},
    title = {{The renormalization group: Critical phenomena and the Kondo problem}},
    volume = {47},
    year = {1975}
}

@article{georges_dynamical_1996,
    title = {Dynamical mean-field theory of strongly correlated fermion systems and the limit of infinite dimensions},
    volume = {68},
    issn = {0034-6861, 1539-0756},
    doi = {10.1103/RevModPhys.68.13},
    number = {1},
    journal = {Reviews of Modern Physics},
    author = {Georges, Antoine and Kotliar, Gabriel and Krauth, Werner and Rozenberg, Marcelo J.},
    month = jan,
    year = {1996},
    pages = {13--125}
}

@article{maierQCT2005,
  title = {Quantum Cluster Theories},
  author = {Maier, Thomas and Jarrell, Mark and Pruschke, Thomas and Hettler, Matthias H.},
  year = 2005,
  month = oct,
  journal = {Reviews of Modern Physics},
  volume = {77},
  number = {3},
  pages = {1027--1080},
  publisher = {American Physical Society},
  doi = {10.1103/RevModPhys.77.1027},
  urldate = {2026-03-24}
}

@article{gullContinuoustimeMonteCarlo2011,
  title = {Continuous-Time {{Monte Carlo}} Methods for Quantum Impurity Models},
  author = {Gull, Emanuel and Millis, Andrew J. and Lichtenstein, Alexander I. and Rubtsov, Alexey N. and Troyer, Matthias and Werner, Philipp},
  year = 2011,
  month = may,
  journal = {Reviews of Modern Physics},
  volume = {83},
  number = {2},
  pages = {349--404},
  publisher = {American Physical Society},
  doi = {10.1103/RevModPhys.83.349},
  urldate = {2026-03-24}
}

@article{Mitchell2012,
    author = {Mitchell, Andrew K. and Sela, Eran},
    doi = {10.1103/PhysRevB.85.235127},
    issn = {1098-0121},
    journal = {Phys. Rev. B},
    month = {jun},
    number = {23},
    pages = {235127},
    title = {{Universal low-temperature crossover in two-channel Kondo models}},
    volume = {85},
    year = {2012}
}

@article{Sela2011,
    archivePrefix = {arXiv},
    arxivId = {1101.3028},
    author = {Sela, Eran and Mitchell, Andrew K. and Fritz, Lars},
    doi = {10.1103/PhysRevLett.106.147202},
    eprint = {1101.3028},
    issn = {0031-9007},
    journal = {Phys. Rev. Lett.},
    month = {apr},
    number = {14},
    pages = {147202},
    title = {{Exact Crossover Green Function in the Two-Channel and Two-Impurity Kondo Models}},
    volume = {106},
    year = {2011}
}

@article{Affleck1995,
    archivePrefix = {arXiv},
    arxivId = {cond-mat/9409100},
    author = {Affleck, Ian and Ludwig, Andreas W. W. and Jones, Barbara A.},
    doi = {10.1103/PhysRevB.52.9528},
    eprint = {9409100},
    issn = {0163-1829},
    journal = {Phys. Rev. B},
    month = {oct},
    number = {13},
    pages = {9528--9546},
    primaryClass = {cond-mat},
    title = {{Conformal-field-theory approach to the two-impurity Kondo problem: Comparison with numerical renormalization-group results}},
    volume = {52},
    year = {1995}
}

@article{Affleck1992,
    author = {Affleck, Ian and Ludwig, Andreas W. W.},
    doi = {10.1103/PhysRevLett.68.1046},
    issn = {0031-9007},
    journal = {Phys. Rev. Lett.},
    month = {feb},
    number = {7},
    pages = {1046--1049},
    title = {{Exact critical theory of the two-impurity Kondo model}},
    volume = {68},
    year = {1992}
}

@article{zhang_pseudogap_2020,
    title = {From the pseudogap metal to the {Fermi} liquid using ancilla qubits},
    volume = {2},
    issn = {2643-1564},
    doi = {10.1103/PhysRevResearch.2.023172},
    number = {2},
    journal = {Physical Review Research},
    author = {Zhang, Ya-Hui and Sachdev, Subir},
    month = {may},
    year = {2020},
    pages = {023172}
}

@article{andersonRVB1987,
  title = {The {{Resonating Valence Bond State}} in {{La2CuO4}} and {{Superconductivity}}},
  author = {Anderson, P. W.},
  year = 1987,
  month = mar,
  journal = {Science},
  volume = {235},
  number = {4793},
  pages = {1196--1198},
  publisher = {American Association for the Advancement of Science},
  doi = {10.1126/science.235.4793.1196},
  urldate = {2026-03-23}
}

@article{emerySuper1995,
  title = {Importance of Phase Fluctuations in Superconductors with Small Superfluid Density},
  author = {Emery, V. J. and Kivelson, S. A.},
  year = 1995,
  month = mar,
  journal = {Nature},
  volume = {374},
  number = {6521},
  pages = {434--437},
  issn = {0028-0836, 1476-4687},
  doi = {10.1038/374434a0},
  urldate = {2026-03-23},
  copyright = {http://www.springer.com/tdm},
  langid = {english}
}

@article{leeMott2006,
  title = {Doping a {{Mott}} Insulator: {{Physics}} of High-Temperature Superconductivity},
  shorttitle = {Doping a {{Mott}} Insulator},
  author = {Lee, Patrick A. and Nagaosa, Naoto and Wen, Xiao-Gang},
  year = 2006,
  month = jan,
  journal = {Reviews of Modern Physics},
  volume = {78},
  number = {1},
  pages = {17--85},
  publisher = {American Physical Society},
  doi = {10.1103/RevModPhys.78.17},
  urldate = {2026-03-12}
}

@article{keimerHighTc2015a,
  title = {From Quantum Matter to High-Temperature Superconductivity in Copper Oxides},
  author = {Keimer, B. and Kivelson, S. A. and Norman, M. R. and Uchida, S. and Zaanen, J.},
  year = 2015,
  month = feb,
  journal = {Nature},
  volume = {518},
  number = {7538},
  pages = {179--186},
  issn = {0028-0836, 1476-4687},
  doi = {10.1038/nature14165},
  urldate = {2026-03-12},
  langid = {english}
}

@article{fradkinIntertwined2015a,
  title = {Colloquium: {{Theory}} of Intertwined Orders in High Temperature Superconductors},
  author = {Fradkin, Eduardo and Kivelson, Steven A. and Tranquada, John M.},
  year = 2015,
  month = may,
  journal = {Reviews of Modern Physics},
  volume = {87},
  number = {2},
  pages = {457--482},
  publisher = {American Physical Society},
  doi = {10.1103/RevModPhys.87.457},
  urldate = {2026-03-12}
}

@article{cominRIX2016,
  title = {Resonant {{X-Ray Scattering Studies}} of {{Charge Order}} in {{Cuprates}}},
  author = {Comin, Riccardo and Damascelli, Andrea},
  year = 2016,
  month = mar,
  journal = {Annual Review of Condensed Matter Physics},
  volume = {7},
  number = {Volume 7, 2016},
  pages = {369--405},
  publisher = {Annual Reviews},
  issn = {1947-5454, 1947-5462},
  doi = {10.1146/annurev-conmatphys-031115-011401}
}

@article{senthil_flp_2003,
    author = {{Senthil}, T. and {Sachdev}, S. and {Vojta}, M.},
    title = "{Fractionalized Fermi Liquids}",
    journal = {Phys. Rev. Lett.},
    eprint = {cond-mat/0209144},
    year = 2003,
    month = may,
    volume = 90,
    number = 21,
    eid = {216403},
    pages = {216403},
    doi = {10.1103/PhysRevLett.90.216403},
}

@article{yang_yrz_2006,
    author = {{Yang}, Kai-Yu and {Rice}, T.~M. and {Zhang}, Fu-Chun},
    title = "{Phenomenological theory of the pseudogap state}",
    journal = {Phys. Rev. B},
    year = 2006,
    month = may,
    volume = {73},
    number = {17},
    eid = {174501},
    pages = {174501},
    doi = {10.1103/PhysRevB.73.174501},
    archivePrefix = {arXiv},
    eprint = {cond-mat/0602164}
}

@article{Punk15,
    author = "Punk, Matthias and Allais, Andrea and Sachdev, Subir",
    title = "{A quantum dimer model for the pseudogap metal}",
    journal = "Proceedings of the National Academy of Sciences", 
    volume = "112",
    year = "2015",
    pages = "9552",
    doi = "10.1073/pnas.1512206112",
    eprint = "1501.00978",
    archivePrefix = "arXiv"
}

@article{badouxCarrierDensity2016,
  title = {Change of Carrier Density at the Pseudogap Critical Point of a Cuprate Superconductor},
  author = {Badoux, S. and Tabis, W. and Lalibert{\'e}, F. and Grissonnanche, G. and Vignolle, B. and Vignolles, D. and B{\'e}ard, J. and Bonn, D. A. and Hardy, W. N. and Liang, R. and {Doiron-Leyraud}, N. and Taillefer, Louis and Proust, Cyril},
  year = 2016,
  month = mar,
  journal = {Nature},
  volume = {531},
  number = {7593},
  pages = {210--214},
  issn = {0028-0836, 1476-4687},
  doi = {10.1038/nature16983},
  urldate = {2026-03-12}
}

@article{whiteDMRG1998,
  title = {Density {{Matrix Renormalization Group Study}} of the {{Striped Phase}} in the {{2D}} $t$-$J$ {{Model}}},
  author = {White, Steven R. and Scalapino, D. J.},
  year = 1998,
  month = feb,
  journal = {Physical Review Letters},
  volume = {80},
  number = {6},
  pages = {1272--1275},
  publisher = {American Physical Society},
  doi = {10.1103/PhysRevLett.80.1272}
}

@article{zhengStripes2017,
  title = {Stripe Order in the Underdoped Region of the Two-Dimensional {{Hubbard}} Model},
  author = {Zheng, Bo-Xiao and Chung, Chia-Min and Corboz, Philippe and Ehlers, Georg and Qin, Ming-Pu and Noack, Reinhard M. and Shi, Hao and White, Steven R. and Zhang, Shiwei and Chan, Garnet Kin-Lic},
  year = 2017,
  month = dec,
  journal = {Science},
  volume = {358},
  number = {6367},
  pages = {1155--1160},
  publisher = {American Association for the Advancement of Science},
  doi = {10.1126/science.aam7127}
}

@article{jiangttpJ2021,
  title = {Ground-State Phase Diagram of the t-T{$\prime$}-{{J}} Model},
  author = {Jiang, Shengtao and Scalapino, Douglas J. and White, Steven R.},
  year = 2021,
  month = nov,
  journal = {Proceedings of the National Academy of Sciences},
  volume = {118},
  number = {44},
  pages = {e2109978118},
  publisher = {Proceedings of the National Academy of Sciences},
  doi = {10.1073/pnas.2109978118},
  urldate = {2026-03-24}
}

@article{jiangStripe2022,
  title = {Stripe Order Enhanced Superconductivity in the {{Hubbard}} Model},
  author = {Jiang, Hong-Chen and Kivelson, Steven A.},
  year = 2022,
  month = jan,
  journal = {Proceedings of the National Academy of Sciences},
  volume = {119},
  number = {1},
  pages = {e2109406119},
  publisher = {Proceedings of the National Academy of Sciences},
  doi = {10.1073/pnas.2109406119}
}

@article{xuStripe2024,
  title = {Coexistence of Superconductivity with Partially Filled Stripes in the {{Hubbard}} Model},
  author = {Xu, Hao and Chung, Chia-Min and Qin, Mingpu and Schollw{\"o}ck, Ulrich and White, Steven R. and Zhang, Shiwei},
  year = 2024,
  month = may,
  journal = {Science},
  volume = {384},
  number = {6696},
  pages = {eadh7691},
  publisher = {American Association for the Advancement of Science},
  doi = {10.1126/science.adh7691},
  urldate = {2026-03-24}
}

@article{huangStripe2017,
  title = {Numerical Evidence of Fluctuating Stripes in the Normal State of High-{{Tc}} Cuprate Superconductors},
  author = {Huang, Edwin W. and Mendl, Christian B. and Liu, Shenxiu and Johnston, Steve and Jiang, Hong-Chen and Moritz, Brian and Devereaux, Thomas P.},
  year = 2017,
  month = dec,
  journal = {Science},
  volume = {358},
  number = {6367},
  pages = {1161--1164},
  publisher = {American Association for the Advancement of Science},
  doi = {10.1126/science.aak9546}
}

@ARTICLE{SchmalianPines2,
    author = {{Schmalian}, J{\"o}rg and {Pines}, David and {Stojkovi{\'c}}, Branko},
    title = "{Microscopic theory of weak pseudogap behavior in the underdoped cuprate superconductors: General theory and quasiparticle properties}",
    journal = {Physical Review B},
    year = 1999,
    month = jul,
    volume = {60},
    number = {1},
    pages = {667-686},
    doi = {10.1103/PhysRevB.60.667},
    archivePrefix = {arXiv},
    eprint = {cond-mat/9804129}
}

@article{abanov_sdw_2000,
    author = {{Abanov}, Ar. and {Chubukov}, Andrey V.},
    title = "{Spin-Fermion Model near the Quantum Critical Point: One-Loop Renormalization Group Results}",
    journal = {Physical Review Letter},
    year = 2000,
    month = jun,
    volume = {84},
    number = {24},
    pages = {5608-5611},
    doi = {10.1103/PhysRevLett.84.5608},
    archivePrefix = {arXiv},
    eprint = {cond-mat/0002122}
}

@ARTICLE{Tremblay04,
    author = {{Kyung}, B. and {Hankevych}, V. and {Dar{\'e}}, A. -M. and {Tremblay}, A. -M.~S.},
    title = "{Pseudogap and Spin Fluctuations in the Normal State of the Electron-Doped Cuprates}",
    journal = {Physical Review Letter},
    year = 2004,
    month = sep,
    volume = {93},
    number = {14},
    eid = {147004},
    pages = {147004},
    doi = {10.1103/PhysRevLett.93.147004},
    archivePrefix = {arXiv},
    eprint = {cond-mat/0312499}
}

@article{Chubukov23,
    author = {{Ye}, Mengxing and {Chubukov}, Andrey V.},
    title = "{Crucial role of thermal fluctuations and vertex corrections for the magnetic pseudogap}",
    journal = {Physical Review B},
    year = 2023,
    month = aug,
    volume = {108},
    number = {8},
    eid = {L081118},
    pages = {L081118},
    doi = {10.1103/PhysRevB.108.L081118},
    archivePrefix = {arXiv},
    eprint = {2306.05489}
}

@article{caponeC60_2009,
  title = {{\emph{Colloquium}} : {{Modeling}} the Unconventional Superconducting Properties of Expanded $\mathrm{A}_3\mathrm{C}_{60}$ Fullerides},
  shorttitle = {{\emph{Colloquium}}},
  author = {Capone, Massimo and Fabrizio, Michele and Castellani, Claudio and Tosatti, Erio},
  year = 2009,
  month = jun,
  journal = {Reviews of Modern Physics},
  volume = {81},
  number = {2},
  pages = {943--958},
  issn = {0034-6861, 1539-0756},
  doi = {10.1103/RevModPhys.81.943},
  urldate = {2025-12-06},
  copyright = {http://link.aps.org/licenses/aps-default-license},
  langid = {english}
}

@article{zhaoTopoMott2025a,
  title = {Topological {{Mott}} Localization and Pseudogap Metal in Twisted Bilayer Graphene},
  author = {Zhao, Jing-Yu and Zhou, Boran and Zhang, Ya-Hui},
  year = 2025,
  month = aug,
  journal = {Physical Review B},
  volume = {112},
  number = {8},
  pages = {085107},
  publisher = {American Physical Society},
  doi = {10.1103/9n8v-7rx2},
  urldate = {2026-03-13}
}

@misc{zhaoRVB2025,
  title = {Resonating-Valence-Bond Superconductor from Small {{Fermi}} Surface in Twisted Bilayer Graphene},
  author = {Zhao, Jing-Yu and Zhang, Ya-Hui},
  year = 2025,
  month = nov,
  eprint = {2510.26801},
  publisher = {arXiv},
  urldate = {2026-03-13},
  archiveprefix = {arXiv}
}

@misc{ledwithTrion2025,
  title = {Exotic {{Carriers}} from {{Concentrated Topology}}: {{Dirac Trions}} as the {{Origin}} of the {{Missing Spectral Weight}} in {{Twisted Bilayer Graphene}}},
  shorttitle = {Exotic {{Carriers}} from {{Concentrated Topology}}},
  author = {Ledwith, Patrick J. and Vishwanath, Ashvin and Khalaf, Eslam},
  year = 2025,
  month = may,
  number = {arXiv:2505.08779},
  eprint = {2505.08779},
  primaryclass = {cond-mat},
  publisher = {arXiv},
  urldate = {2025-05-15},
  archiveprefix = {arXiv},
  langid = {american}
}

\appendix 
\onecolumngrid

\section{Dynamical Mean Field Theory and NRG solver}\label{app:dmft}

In this appendix, we provide a brief review of the DMFT methods and application of NRG solver. 

\subsection{DMFT self-consistence.}

Consider a general fermionic system with local onsite interactions, 
\begin{equation}
    H = - \sum_{ij,\alpha}t_{ij}c^{\dagger}_{i;\alpha}c_{j;\alpha}^{}
    -\mu \sum_{i,\alpha}c^\dagger_{i;\alpha}c^{}_{i;\alpha} 
    + \sum_iH_{i;\mathrm{int}}~,
\end{equation}
where $\alpha$ denotes a generic flavor index, and $H_{i;\mathrm{int}}$ represents a local interaction acting on site $i$.
For example, in the Hubbard model, $\alpha = \uparrow,\downarrow$, and $H_{i;\mathrm{int}} = Un_{i;\uparrow}n_{i;\downarrow}$. 
In the double Kondo model considered in Eq.~\eqref{eqn:double_kondo}, $\alpha= (a;\sigma)$ includes both the layer index $a$ and the spin index $\sigma$, 
and $H_{i;\mathrm{int}} = J_K \sum_{a}\mathbf s_{i;a} \cdot \mathbf S_{i;a} + J_\perp \mathbf S_{i;t}\cdot \mathbf S_{i;b} + U\sum_a n_{i;a;\uparrow}n_{i;a;\downarrow} + V_\perp n_{i;t}n_{i;b}$. 

In the strong-coupling limit where the local interaction $H_{i;\mathrm{int}}$ dominates, it is natural to start from the atomic limit and treat the hopping term perturbatively.
From this perspective, the lattice problem can be mapped onto an effective impurity problem.
Specifically, we select a reference site $o$ as the impurity and treat all remaining sites as an effective thermal bath.
The action can then be decomposed into three parts:
\begin{equation}
    S =  S^{(o)} + S^o + \Delta S~,
\end{equation}
where $S^{o}$ denotes the action of the selected impurity site $o$, 
\begin{equation}
    S^o = \int _0^\beta d\tau\left(\sum_{\sigma}c^\dagger_{o\sigma}(\partial_\tau-\mu) c^{}_{o\sigma} 
    +H_{o,\mathrm{int}}\right)~,
\end{equation}
$\Delta S$ is the action contributed by the hopping terms between the $o$ site and all other sites, 
\begin{equation}
    \Delta S = -\int_0^\beta d\tau \sum_{i,\alpha}t_{io}\left(c^{\dagger}_{i;\alpha}c_{o;\alpha}^{} + c^\dagger_{o;\alpha}c^{}_{i;\alpha}
    \right)~,
\end{equation}
Finally, $S^{(o)}$ denotes the remaining action of the environment, i.e., the lattice with site $o$ removed.

We now integrate out all fermionic degrees of freedom except those on site $o$, obtaining an effective impurity action,
\begin{equation}
    e^{-S_{\mathrm{eff}}[c^\dagger_{o;\alpha},c^{}_{o;\alpha}]}\equiv \int \prod_{i\neq o; \alpha}Dc^\dagger_{i;\alpha}Dc^{}_{i;\alpha}e^{-S^o-S^{(o)}-\Delta S}~. 
\end{equation}
The resulting effective action takes the form
\begin{equation} \label{eqn:dmft_effact}
    \begin{aligned}
    S_{\mathrm {eff}} = 
    \int_0^\beta\mathrm d\tau \sum_\alpha c^\dagger_{o;\alpha}(\tau) (\partial_\tau - \mu) c_{o;\alpha} + \int_0^\beta\mathrm H_{o;\mathrm{int}}(\tau)
    -\int_0^\beta \mathrm d\tau_1 \mathrm d\tau_2 \sum_\alpha c^\dagger_{o;\alpha}(\tau_1)\Delta_{0;\alpha}(\tau_1-\tau_2)c_{o;\alpha}^{}(\tau_2) ~.
    \end{aligned}
\end{equation}
Here $\Delta_{0;\alpha}(\tau)$ is the hybridization function describing the dynamical coupling between the impurity and the conduction bath.
It is generally nonlocal in imaginary time and encodes the full influence of the lattice environment.
Determining $\Delta_{0;\alpha}(\tau)$ self-consistently is the central step in DMFT calculation.

When the spatial dimension $d$ is high, only the second-order contribution $(\Delta S)^2$ survives in the cumulant expansion of the effective action.
Under this condition, the hybridization function satisfies the self-consistency relation
\cite{georges_dynamical_1996}: 
\begin{equation}\label{eqn:self_cons}
    \begin{aligned}
        \Delta_{0;\alpha} (\omega) = - \omega - \mu + \Sigma_\alpha(\omega) + G^{-1}_{oo;\alpha}(\omega)~,
    \end{aligned}
\end{equation}
where all quantities have been Fourier transformed to real-frequency space $\omega$, 
$\Sigma_\alpha(\omega)$ denotes the electron self-energy, which is assumed to be momentum-independent within the DMFT framework.
It is determined self-consistently from the effective action in Eq.~\eqref{eqn:dmft_effact}.
Solving for the self-energy constitutes the central task of the impurity solver and will be discussed in detail in the next subsection.
The local Green’s function at site $o$, $G_{oo;\alpha}(\omega)$, is related to the self-energy through the lattice dispersion,
\begin{equation}\label{eqn:greens}
    G_{oo;\alpha}(\omega) = \int \frac{\mathrm{d}^2k}{(2\pi)^2} \frac{1}{\omega+\mu - \Sigma_\alpha(\omega) - \epsilon_\alpha(\mathbf{k})}~,
\end{equation}
where $\epsilon_\alpha(\mathbf{k})$ is the dispersion of non-interacting conduction band. 
In this work, we choose 
\begin{equation}
    \epsilon_\alpha(\mathbf{k}) = -2t(\cos(k_x)+\cos(k_y))~
\end{equation}
as the dispersion of a 2 dimensional square lattice tight binding model. 
Throughout this paper, we choose $t=1$ as unit of the energy. 

With the relations established above, the lattice problem is reduced to solving the effective single-site action in Eq.~\eqref{eqn:dmft_effact}.
The central difficulty lies in determining the self-energy $\Sigma_\alpha(\omega)$ and the local Green’s function $G_{oo;\alpha}(\omega)$ for a given hybridization function $\Delta_{0;\alpha}(\omega)$.
Various impurity solvers have been developed for this purpose.
A widely used approach is quantum Monte Carlo (QMC), which is particularly effective for accessing high-energy properties and finite-temperature behavior.
However, it is less suitable for resolving subtle critical properties at very low temperatures.
In the present problem, our primary interest is the QCP separating two distinct Fermi-liquid phases, which is intrinsically a zero-temperature phenomenon.
For this reason, we employ the Wilson numerical renormalization group method as the impurity solver, which is especially well suited for capturing low-energy fixed points and quantum critical behavior.

To summarize, the problem reduces to solving the following self-consistent loop:
\begin{equation}
    \Delta_{0;\alpha}
    \xrightarrow{\text{NRG}} \Sigma_\alpha(\omega) 
    \xrightarrow{\text{Eq.~\eqref{eqn:greens}}} G_{oo;\alpha}(\omega) \xrightarrow{\text{Eq.~\eqref{eqn:self_cons}}} \Delta_{0;\alpha}~,
\end{equation}
which is iterated until convergence is achieved.

Finally, we note that in solving the self-consistent equations, two definitions of the local impurity Green’s function $G_{oo;\alpha}(\omega)$ naturally arise.
The first is obtained directly from the NRG solver, which exhibits a pronounced even–odd effect associated with the Wilson-chain discretization.
The second is reconstructed from the lattice self-consistency relation in Eq.~\eqref{eqn:greens}.
The two Green’s functions are qualitatively consistent with each other.
However, when the interaction strength is large, the doping level $x$ extracted from the two definitions shows a small quantitative discrepancy.
This difference originates from the sensitivity of the lattice reconstruction to high-frequency details of the self-energy. 
As the NRG result directly reflects the impurity occupancy and is therefore numerically more stable, 
in all phase diagrams presented in the main text, we adopt the doping level $x$ calculated directly from the NRG results.

\begin{figure}[t]
    \centering
    \includegraphics[width=0.5\linewidth]{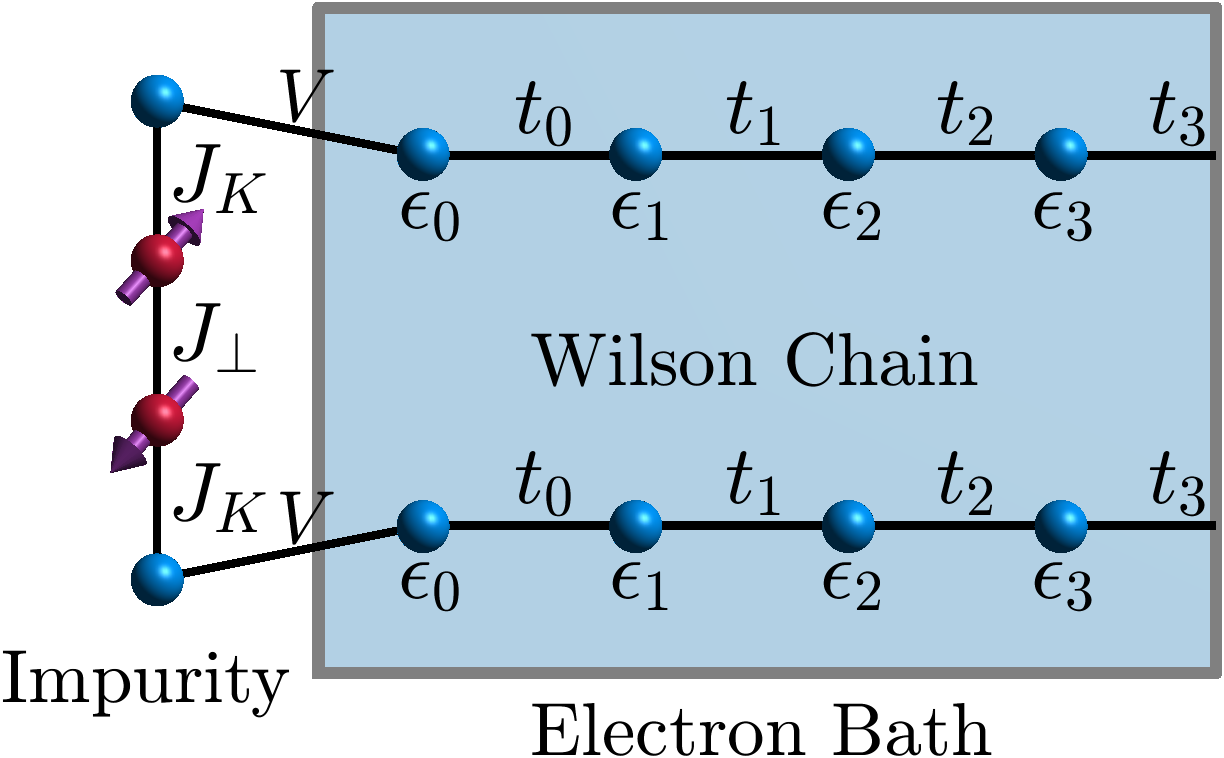}
    \caption{Illustration of the Wilson chain in the NRG calculation.
    The system is mapped onto an impurity coupled to a conduction electron bath via a hybridization $V$.
    The parameters $t_n$ and $\epsilon_n$ denote the hopping amplitudes and on-site energies of the Wilson chain, which are extracted from the hybridization function $\Delta(\omega)$.
    Their asymptotic behavior follows $t_n, \epsilon_n \sim \Lambda^{-n/2}$.}
    \label{fig:wilson_chain}
\end{figure}

\subsection{Numerical Renormalization Group}\label{subapp:nrg}

In this appendix, we briefly review the Numerical Renormalization Group method and specify the parameters used in our calculations. 
Consider an impurity fermion $c_{o;\alpha}$ interacting with a bath of electrons $d_{k;\alpha}$, 
\begin{equation}
    H_{\mathrm{imp}} = H_{o;\mathrm{int}} + \sum_{k,\alpha} V_{k;\alpha} c_{o;\alpha}^\dagger d_{k;\alpha} + \mathrm{h.c.} + 
    \sum_{k,\alpha} \epsilon_{k;\alpha}  d^\dagger_{k;\alpha} d_{k;\alpha}^{}~,
\end{equation}
where $d_{k;\alpha}$ represents bath fermions with eigenenergies $\epsilon_{k;\alpha}$, 
$V_{k;\alpha}$ denotes the hybridization amplitude between the impurity and the bath. 
Integrating out the conduction electrons leads to an effective action of the same form as Eq.~\eqref{eqn:dmft_effact}, with hybridization function
\begin{equation}
    \Delta_{0;\alpha}(\omega) =  - \sum_{k} \frac{V_{k;\alpha}^2}{\omega - \epsilon_{k;\alpha}}~,
\end{equation}
We define the hybridization spectral density as
\begin{equation}\label{eqn:Delta_NRG}
    \Delta_\alpha(\omega) \equiv \pi\sum_n V_{k;\alpha}^2\delta(\omega-\epsilon_{k;\alpha}) = \mathrm{Im}\Delta_{0;\alpha}(\omega+i0^+)~. 
\end{equation}

Following Wilson's original idea \cite{Wilson1975}, the continuous spectrum $\Delta_\alpha(\omega)$ is discretized logarithmically into energy intervals 
$\pm [D\Lambda^{-n-1}, D\Lambda^{-n}]$. 
Here $\Lambda > 1$ is the discretization parameter, which is the central approximation made in the NRG calculation. 
The exact result recovers in the $\Lambda\rightarrow 1$ limit. 
In this work, we choose $\Lambda=3$ without otherwise mentioned. 
After a sequence of transformations \cite{Bulla2008}, the impurity problem can be mapped onto a semi-infinite one-dimensional chain, 
as illustrated in Fig.~\ref{fig:wilson_chain}, where physically each site of the Wilson chain represents electronic states within a corresponding energy shell. 
The resulting Hamiltonian reads:
\begin{equation}
\begin{aligned}
    H = H_{o;\mathrm{int}} + \sum_\alpha V_\alpha c_{o;\alpha}^\dagger \tilde d_{0;\alpha} + \mathrm{h.c.}
    + \sum_{n=0,\alpha}^\infty t_{n;\alpha} (\tilde d^\dagger_{n;\alpha} \tilde d^{}_{n+1;\alpha}+\mathrm{h.c.}) + \sum_{n=0,\alpha}^{\infty} \epsilon_{n;\alpha} \tilde d^\dagger_{n;\alpha}\tilde d^{}_{n;\alpha}~,
\end{aligned}
\end{equation}
where $\tilde d_{n;\alpha}$ denotes the electron operator at site $n$ of the Wilson chain.
Physically, sites with smaller $n$ represent conduction electrons farther from the Fermi level, while larger $n$ correspond to those closer to the Fermi level. 
The Wilson chain parameters $t_{n;\alpha}$, $\epsilon_{n;\alpha}$, and $V_\alpha$ are uniquely determined by the hybridization function $\Delta_\alpha(\omega)$. 
In the special case where $\Delta_\alpha(\omega)$ is constant $\Delta$ over the interval $[-D, D]$, the parameters simplify to $\epsilon_{n;\alpha} = 0$, $V_\alpha = \sqrt{2D \Delta/\pi}$, and
$t_{n;\alpha}=\frac{(1+\Lambda^{-1})(1-\Lambda^{-n-1})}{2\sqrt{1-\Lambda^{-2n-1}}\sqrt{1-\Lambda^{-2n-3}}}\Lambda^{-n/2}.$ 
For more general $\Delta_\alpha(\omega)$, the parameters $t_{n;\alpha}$ and $\epsilon_{n;\alpha}$ can be determined numerically using high-precision iterative algorithms~\cite{Bulla2008}. 
Their asymptotic behavior follows $t_{n;\alpha}, \epsilon_{n;\alpha} \sim \Lambda^{-n/2}$.

To compare the spectra across iterations on the same energy scale, a rescaling of the Hamiltonian is performed at each step. 
Specifically, we define the rescaled Hamiltonian at iteration $n$ as
\begin{equation}
    H_n = \Lambda^{\frac{n-1}{2}} H,
\end{equation}
where $H$ is the truncated Hamiltonian that includes Wilson chain sites from $n = 0$ to $n = N$. 
This rescaling ensures that all terms in the Hamiltonian remain of comparable magnitude despite the exponentially decaying hoppings.
The iterative diagonalization proceeds via the recurrence relation:
\begin{equation}
    H_{n+1} = \sqrt{\Lambda} H_n + \Lambda^{n/2} \sum_\alpha t_{n;\alpha} (d^\dagger_{n+1,\alpha} d^{}_{n,\alpha} + \text{h.c.})~.
\end{equation}

The key idea of the NRG is to focus on the low-energy physics of a quantum impurity model by iteratively diagonalizing the Hamiltonian 
and retaining only the lowest-lying eigenstates at each step. 
At each iteration, high-energy states are discarded, while low-energy states are kept and used to build up the system iteratively. 
This controlled truncation enables access to exponentially small energy scales, making NRG particularly well-suited for 
studying strongly correlated systems at low temperatures.
In our calculation, we retain 10000 states for an $N=40$ Wilson chain in deriving the phase diagram. 

In the derivation above, we assume that the hybridization $V_{k;\alpha}$ is diagonal in the flavor index $\alpha$. 
In general, hybridization terms between different flavors, $V_{k;\alpha\beta}$, may also be present. 
However, in the double Kondo model Eq.~\eqref{eqn:double_kondo} with symmetry $U(1)_t \times U(1)_b \times SU(2)_S/\mathbb{Z}_2$, 
$V_{k;\alpha}$ is constrained to be diagonal and identical for each flavor $\alpha$. 
The only exception arises when an interlayer coupling $t_\perp$ is introduced, as discussed in Appendix~\ref{app:tperp}. 
In that case, the separate $U(1)_{t,b}$ symmetries of the top and bottom layers are broken down to a single global $U(1)$ symmetry, allowing hybridization between the two layers. 
Nevertheless, after performing the linear recombination 
$c_\pm = \frac{1}{\sqrt{2}}(c_t \pm c_b)$, 
both $G_\pm$ and $\Sigma_\pm$ remain diagonal in the $\pm$ basis due to the mirror symmetry between the top and bottom layers. 
Consequently, $\Delta_{\alpha}$, $V_{k;\alpha}$, $t_{n;\alpha}$, and $\epsilon_{n;\alpha}$ are still diagonal in the $\pm$ basis.
In order to accelerate the calculation, in the case with $t_\perp$ and smaller symmetry, we would use an interleaved chain \cite{MitchellInter2014,StadlerInterleave2016,Gleis2024X} of alternating $+$ and $-$ orbitals. 

\subsubsection{Measuring spectrum functions }

In the context of DMFT, we are primarily interested in dynamical correlation functions on the impurity site, such as the local Green's function $G_{oo;\alpha}(\omega)$ and self-energy $\Sigma_\alpha(\omega)$. 
To accurately compute such quantities while preserving spectral sum rules, we adopt the density-matrix NRG (DM-NRG) approach \cite{Weichselbaum2007,Weichselbaum2012}. 
This method incorporates information from discarded states across all iterations via a thermal density matrix constructed from these states.
The full thermal density matrix is expressed as:
\begin{equation}
\begin{aligned}
    \rho (T) = \sum_n \frac{d^{N-n}Z_n}{Z}\sum_s\frac{e^{-\beta E_s^n}}{Z_n}|s\rangle_n^D \phantom{ }_n^D\langle s|
    =\sum_n w_n \rho_n~,
\end{aligned}
\end{equation}
where $d$ is the local Hilbert space dimension per Wilson site, $|s\rangle_n^D$ denotes the discarded eigenstates at iteration step $n$,
$E_s^n$ is the energy of state $|s\rangle_n^D$, measured relative to the ground state at the final iteration. 
The total density matrix can be further decomposed into sum of each step density matrix $\rho_n\equiv \sum_s \frac{e^{-\beta E_s^n}}{Z_n}|s\rangle_n^D\phantom{}_n^D\langle s|$, where 
$Z_n \equiv\sum_{s \in D_n} e^{-\beta E_s^n}$ is the partial partition function for iteration $n$,
$Z = \sum_n d^{N-n} Z_n$ is the total partition function,
and $w_n = d^{N-n} Z_n / Z$ are normalized weights.
This formulation ensures that spectral functions obtained using DM-NRG satisfy the exact sum rule by construction, 
leading to physically consistent and accurate dynamical properties even at finite temperatures.
In realistic calculation, we take $\beta=+\infty$ such that only the final Wilson shell contribute to the density matrix. 

Here, we denote the retarded Green's function of operators $B$ and $C$ as 
\begin{equation}
    \langle\!\langle B; C\rangle\!\rangle(t) \equiv -i\theta(t)\langle [B(t), C(0)]_\zeta\rangle~,
\end{equation}
where $\zeta = \mp 1$ for bosonic and fermionic operators. 
The corresponding spectrum functions are measured as 
\begin{equation}
\begin{aligned}
    A_{BC}(\omega) \equiv & -\frac{1}{\pi}\mathrm{Im}\langle\!\langle B;C\rangle\!\rangle(\omega+i0^+) \\
    = & \sum_n \sum_{k',k,l}C^n_{l,k'}\rho^n_{k',k} B^n_{k,l}\delta_\zeta(\omega,E^n_l-E^n_k)
     -\zeta\sum_n\sum_{k',k,l} B^n_{l,k'}\rho^n_{k',k} C^n_{k,l}\delta_\zeta(\omega,E^n_k-E^n_l)~.
\end{aligned}
\end{equation}
Here $B^n_{k,l}$ are the corresponding matrix elements between the kept state $k$ and discarded state $l$ at step $n$. 
The same notation used for $C$ and $\rho$. 

An important approximation in NRG is the logarithmic discretization of the conduction-band continuum into finite energy intervals. 
The resulting spectrum consists of discrete $\delta$-function peaks rather than a continuous spectral function. 
To recover a smooth approximation to the original continuum spectrum, one must therefore replace each $\delta$ function by an appropriate broadening kernel. 
Here we follow Ref.~\cite{Gleis2024X} to choose the broading function as: 
\begin{equation}
    \delta_F(\omega,\omega_k) = \frac{\theta(\omega \omega_k)}{\sqrt{\pi}b |\omega_k|} e^{-(\ln|\omega/\omega_k|/b + b / 4)^2}~,
\end{equation}
for fermionic correlation functions and 
\begin{equation}
    \delta_B(\omega,\omega_k) = \frac{\theta(\omega \omega_k)}{\sqrt{\pi}b |\omega_k|} e^{-(\ln|\omega/\omega_k|/b)^2}e^{b^2 / 4}~,
\end{equation}
for bosonic correlation function. 
Throughout our calculation, we choose a smaller $b=0.1$ during the iteration of DMFT to reduce error, 
and a larger $b=0.7$ for the final measurement to avoid the artificial even-odd oscillation. 
Once $A_{BC}(\omega)$ is obtained, the full Green's function $\lla B;C\rra(\omega)$ can be derived through the Kramers–Kronig relation. 

\subsubsection{Derivation of self-energy. }

Estimating the self-energy is the central task in the NRG iteration. 
Several methods exist to extract the self-energy from NRG spectral data. 
Here we follow Ref.~\cite{Kugler2022}, which makes use of a second-order equation-of-motion (EOM) approach. 
This method has the advantage that the even–odd oscillations in the Wilson chain largely compensate each other, and the imaginary part of the resulting self-energy is guaranteed to remain positive.

For completeness, we briefly summarize the derivation of the relevant relations. 
Using the equation of motion for Green's functions,
\begin{equation}\label{eqn:eom000}
    z \lla A; B\rra(z) = \la [A,B]_\zeta\ra + \lla [A,H];B\rra(z)~, 
\end{equation}
\begin{equation}\label{eqn:eom001}
    z \lla A; B\rra(z) = \la [A,B]\ra_\zeta + \lla A;[H,B]\rra(z)~, 
\end{equation}
one obtains the following relations:  
\begin{equation}\label{eqn:eom1}
    (z + \mu + \Delta_{0;\alpha}(z)) G_\alpha^{\mathrm{loc}}(z) = 1 + F_\alpha^{\mathrm{loc}}(z)~,
\end{equation}
\begin{equation}\label{eqn:eom2}
    (z + \mu + \Delta_{0;\alpha}(z)) F_\alpha^{\mathrm{loc}}(z) = \Sigma_\alpha^H + I_\alpha^{\mathrm{loc}}(z)~,
\end{equation}
where
\begin{equation}
    G_\alpha^{\mathrm{loc}}(z) = \lla c_{o;\alpha}; c^\dagger_{o;\alpha} \rangle\!\rangle(z)~,
\end{equation}
\begin{equation}
    F_\alpha^{\mathrm{loc}}(z) = \langle\!\langle [c_{o;\alpha},H_{o;\mathrm{int}}]; c^\dagger_{o;\alpha} \rangle\!\rangle(z)~,
\end{equation}
\begin{equation}
    I_\alpha^{\mathrm{loc}}(z) = \langle\!\langle [c_{o;\alpha},H_{o;\mathrm{int}}]; [c_{o;\alpha}, H_{o;\mathrm{int}}]^\dagger \rangle\!\rangle(z)~, 
\end{equation}
and 
\begin{equation}
    \Sigma^H_\alpha = \langle\{[c_{o;\alpha}, H_{o;\mathrm{int}}], c_{o;\alpha}^\dagger\}\rangle~
\end{equation}
is the Hartree contribution.

Here the superscript ``loc'' indicates that these correlation functions are evaluated on the impurity site, 
in contrast to the lattice Green's function reconstructed in Eq.~\eqref{eqn:greens}.

Combining Eqs.~\eqref{eqn:eom1} with the Dyson equation
\begin{equation}
G_\alpha^{\mathrm{loc}}(z)
=
\frac{1}
{z+\mu+\Delta_{0;\alpha}(z)-\Sigma_\alpha^{\mathrm{loc}}(z)},
\end{equation}
we immediately obtain
\begin{equation*}
    F_\alpha^{\mathrm{loc}}(z)=\Sigma_\alpha^{\mathrm{loc}}(z)G_\alpha^{\mathrm{loc}}(z)~.
\end{equation*}
Substituting this identity into Eq.~\eqref{eqn:eom2} gives
\begin{equation*}
    \Sigma_\alpha^H + I_\alpha^{\mathrm{loc}}(z)
    =
    \left(z+\mu+\Delta_{0;\alpha}(z)\right)\Sigma_\alpha^{\mathrm{loc}}(z)G_\alpha^{\mathrm{loc}}(z)~.
\end{equation*}
Using the Dyson equation in the form
\begin{equation*}
    z+\mu+\Delta_{0;\alpha}(z)
    =
    G_\alpha^{\mathrm{loc}}(z)^{-1}+\Sigma_\alpha^{\mathrm{loc}}(z)~,
\end{equation*}
we then obtain
\begin{equation*}
    \Sigma_\alpha^H + I_\alpha^{\mathrm{loc}}(z)
    =
    \Sigma_\alpha^{\mathrm{loc}}(z)
    + \Sigma_\alpha^{\mathrm{loc}}(z)^2 G_\alpha^{\mathrm{loc}}(z)
    =
    \Sigma_\alpha^{\mathrm{loc}}(z)
    + \frac{[F_\alpha^{\mathrm{loc}}(z)]^2}{G_\alpha^{\mathrm{loc}}(z)}~.
\end{equation*}
Rearranging finally yields
\begin{equation}\label{eqn:selfenergy}
    \Sigma^{\mathrm{loc}}_\alpha(\omega) = \Sigma_\alpha^H + I^{\mathrm{loc}}_\alpha(\omega) - F^{\mathrm{loc}}_\alpha(\omega)^2 / G^{\mathrm{loc}}_\alpha(\omega). 
\end{equation}

With this formula, the self-energy can be directly obtained from the NRG solver by computing the correlation functions 
$G_\alpha^{\mathrm{loc}}(\omega)$, $F_\alpha^{\mathrm{loc}}(\omega)$, and $I_\alpha^{\mathrm{loc}}(\omega)$. 
Finally, within the DMFT framework we identify
$\Sigma_\alpha(\omega) = \Sigma_\alpha^{\mathrm{loc}}(\omega)$, 
which corresponds to the central assumption of DMFT that the self-energy is momentum independent. 

\section{Distinction between FL and sFL}

\begin{figure}[t]
    \centering
    \includegraphics[width=0.9\linewidth]{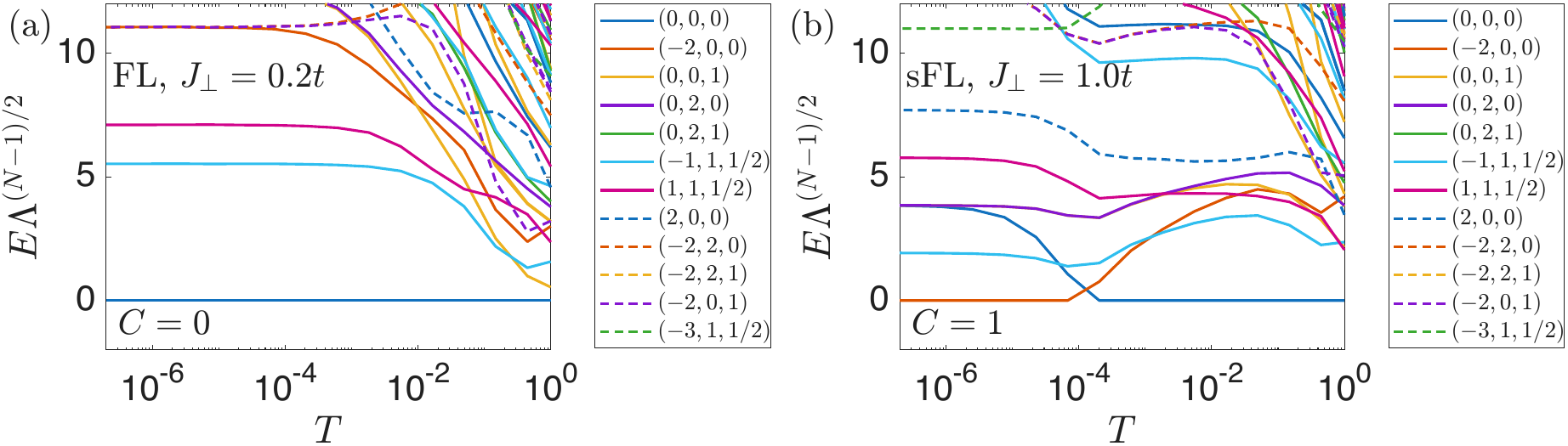}
    \caption{Spectrum flow of the even sequence Wilson chain for the double kondo model. 
    Each level is labeled by the quantum numbers $(Q_t+Q_b - 2N, Q_t - Q_b, 2S)$, where $Q_t$, $Q_b$, and $S$ denote the top-layer charge, bottom-layer charge, and total spin of the NRG impurity system at step $N$, respectively.
    The temperature scale is estimated as $T = 4t \Lambda^{-(N-1)/2}$.
    (a) Spectrum flow in the FL phase with $J_K = -10t$, $J_\perp = 0.2t$, $U = 8t$, and $x = 0.2$. The fixed-point ground state is characterized by $C = 0$.
    (b) Spectrum flow in the sFL phase with $J_K = -10t$, $J_\perp = 1.0t$, $U = 8t$, and $x = 0.2$. The fixed-point ground state is characterized by $C = 1$.
    }
    \label{fig:specflow_ferro}
\end{figure}

\begin{figure}[t]
    \centering
    \includegraphics[width=0.9\linewidth]{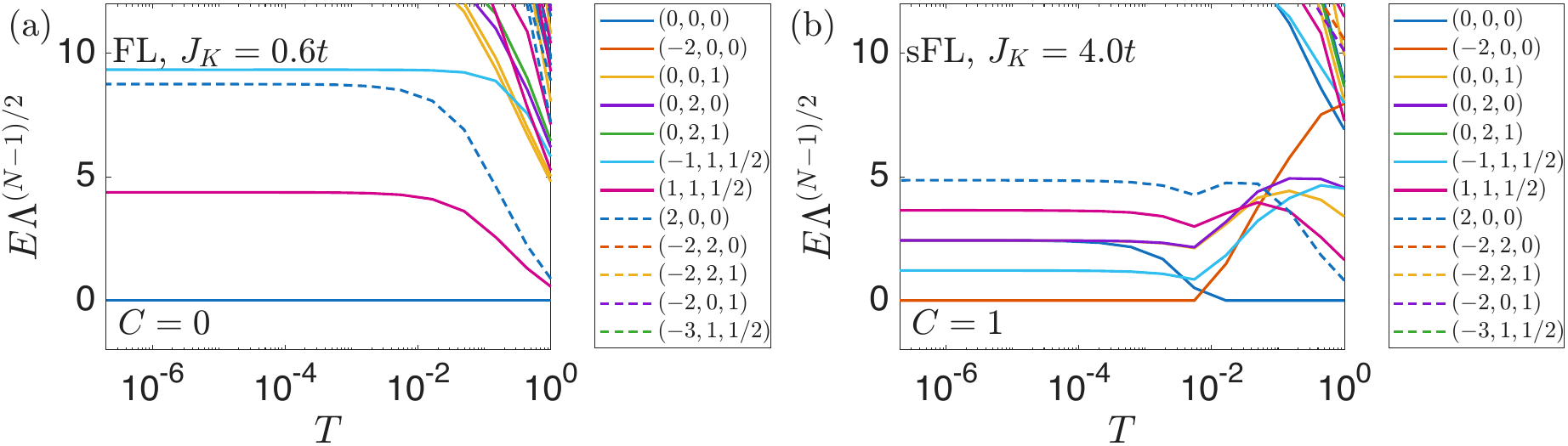}
    \caption{Spectrum flow of the even sequence Wilson chain for the double kondo model. 
    Each level is labeled by the quantum numbers $(Q_t+Q_b - 2N, Q_t - Q_b, 2S)$, where $Q_t$, $Q_b$, and $S$ denote the top-layer charge, bottom-layer charge, and total spin of the NRG impurity system at step $N$, respectively.
    The temperature scale is estimated as $T = 4t \Lambda^{-(N-1)/2}$.
    (a) Spectrum flow in the FL phase with $J_K = 0.6t$, $J_\perp = 2.0t$, $U = 8t$, and $x = 0.2$. The fixed-point ground state is characterized by $C = 0$.
    (b) Spectrum flow in the sFL phase with $J_K = 4.0t$, $J_\perp = 2.0t$, $U = 8t$, and $x = 0.2$. The fixed-point ground state is characterized by $C = 1$.}
    \label{fig:specflow_antiferro}
\end{figure}

In the main text, we show the differences between FL and sFL through the spectrum function restored in the momentum space and the Fermi surface shape. 
The sFL shows a pseudogap physics with a dip in the spectrum function at energy $\sim -\mu_\psi$ and a Fermi volume jump at zero energy. 
In this appendix, we show the difference between the FL and sFL phase from more perspectives. 

\subsection{Topological identification of sFL and FL}\label{app:topoz2}

The FL and sFL phases can already be clearly distinguished within the impurity model, without referring to the momentum-resolved spectral function.
In solving the NRG impurity model described in Appendix~\ref{app:dmft}, we obtain a sequence of many-body energy levels at each iteration step $N$.
Each energy level is labeled by a set of quantum numbers $Q_t, Q_b$ and $S$ associated with the $U(1)_t \times U(1)_b \times SU(2)_S/\mathbb{Z}_2$ symmetries of the lattice model. 
The charge number of the spin layer $d_{z^2}$ orbital is counted as zero. 
Our focus is on the quantum numbers of the ground state to which the NRG flow converges at the fixed point.

We define a $\mathbb{Z}_2$ index as 
\begin{equation}\label{eqn:Z2topo_app}
    C = (Q_{\mathrm{tot}}/2) (\mathrm{mod}\ 2),~
\end{equation}
where $Q_{\mathrm{tot}} = Q_t + Q_b$ is the total charge of the fixed-point ground state.
For a non-degenerate ground state, system symmetry requires $Q_{\mathrm{tot}}$ to be even.

We now argue that $C$ indeed defines a topological invariant.
On both sides of the QCP, the Wilson chain flows to a Fermi-liquid fixed point.
In such a fixed point, $Q_{\mathrm{tot}}$ counts the number of occupied free-fermion states below the Fermi level.
In the particle–hole asymmetric case considered here, the free-electron bands generally do not align exactly with the Fermi level such that the ground state is non-degenerate.
Taking into account the $U(1)_t \times U(1)_b \times SU(2)_S/\mathbb{Z}_2$ symmetry, the number of occupied free-fermion states must take the form $Q_{\mathrm{tot}} = 4n$, yielding $C = 0$, where the factor of $4$ arises from two spin and two layer degrees of freedom. 

On the other hand, when the interaction is strong enough, the above argument from free fermions may fail. 
Specifically, there is the possibility where $Q_{\mathrm{tot}} = 4n - 2$ and $C = 1$. 
For instance, the extra contribution of $2$ can originate from an effective hidden local singlet layer formed on the impurity [c.f. Fig.~\ref{fig:illu} (e)], 
which means parts of the fermion degrees of freedom are effectively decoupled from the Fermi sea.
This quantum number therefore effectively counts the total number of conductive electrons and thus characterizes the jump in the Fermi volume across the transition.

In Figs.~\ref{fig:specflow_ferro} and \ref{fig:specflow_antiferro}, we present typical NRG spectrum flows of the self-consistently converged DMFT solution for both $J_K<0$ and $J_K>0$.
The low-lying energies, rescaled as $ E\Lambda^{(N-1)/2} $, together with the corresponding quantum numbers $(Q_t+Q_b - 2N, Q_t-Q_b, S)$ of the NRG impurity system, are plotted as functions of temperature $T =4t \Lambda^{-(N-1)/2}$.
Here $N$ denotes the NRG iteration index, and only even iterations are shown.

The quantum numbers at the NRG fixed point differ sharply between the FL and sFL phases, while exhibiting universal behavior with respect to the sign of $J_K$.
In the FL phase, for both $J_K<0$ and $J_K>0$, the fixed-point ground state is characterized by $(Q_t+Q_b,Q_t-Q_b, S) = (2N,0,0)$,  corresponding to $C = 0$.
By contrast, in the sFL phase, the fixed-point ground state is consistently given by
$(Q_t+Q_b,Q_t-Q_b, S) = (2N-2,0,0)$,  corresponding to $C = 1$.
We note that the topological indicator $C$ remains well defined for odd Wilson-chain iterations as well, although only even iterations are displayed here for clarity.

\begin{figure}[t]
    \includegraphics[width=0.75\linewidth]{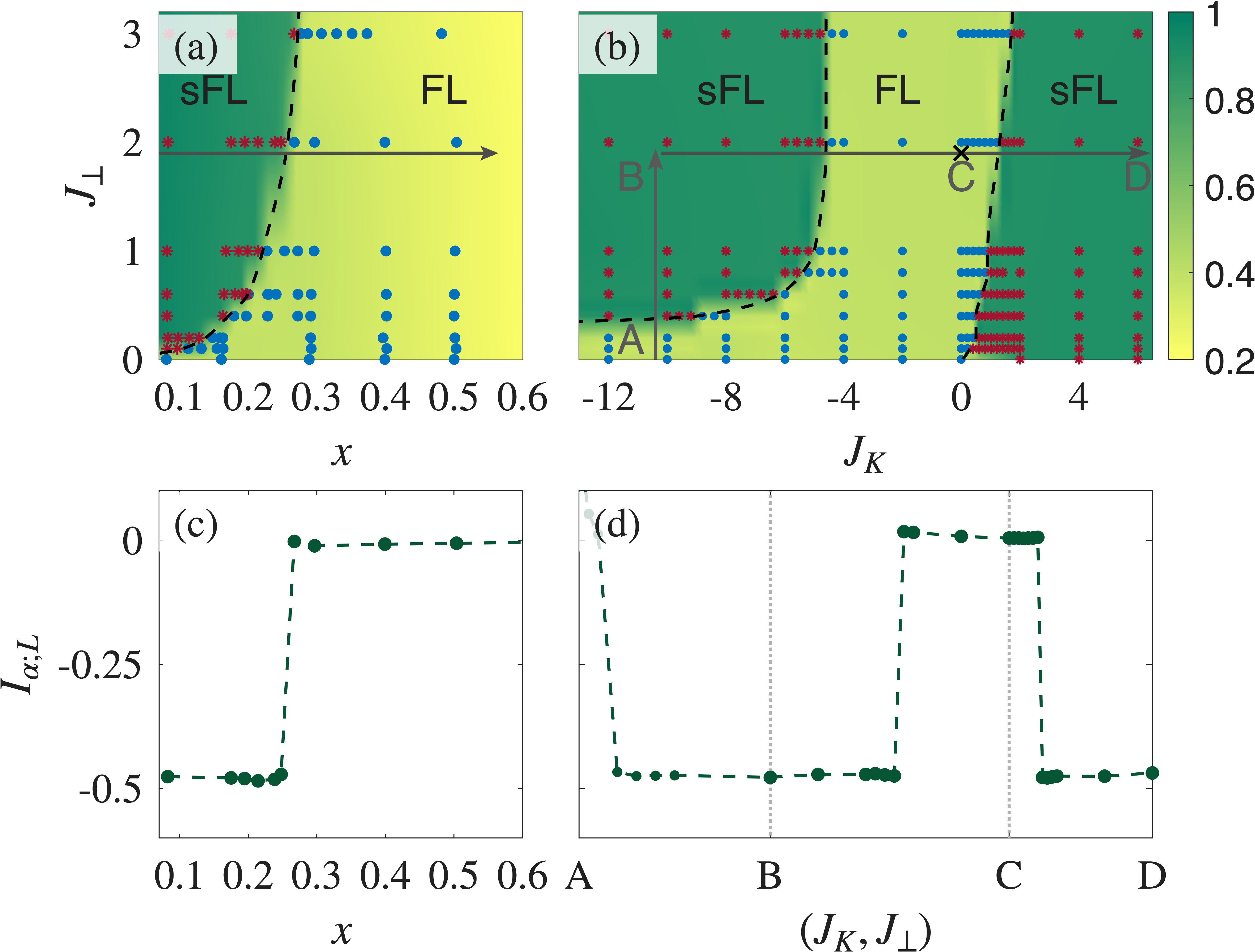}
    \caption{The Fermi surface volume as a function of $J_K$, $J_\perp$ and doping $x$ with $U=8t$, $t_\perp=0$ and (a) $J_K$ fixed at $J_K=-12t$ and (b) doping fixed at $x = 0.2$.
    Blue dots and red stars denote even and odd values of the $\mathbb{Z}_2$ charge $C$. 
    The phase diagrams are consistent with those identified by the $\mathbb{Z}_2$ indicator. 
    (c) and (d) show the Luttinger integral $I_{\alpha;L}$ defined in Eq.~\eqref{eqn:luttinger_integral} along the line cuts of (a) and (b), respectively. 
    A jumping of $\Delta I_{\alpha;L} \approx \mp 1/2$ is seen at the phase boundary. 
    }
    \label{fig:muResigma}
\end{figure}

\begin{figure}[t]
    \centering
    \includegraphics[width=0.99\linewidth]{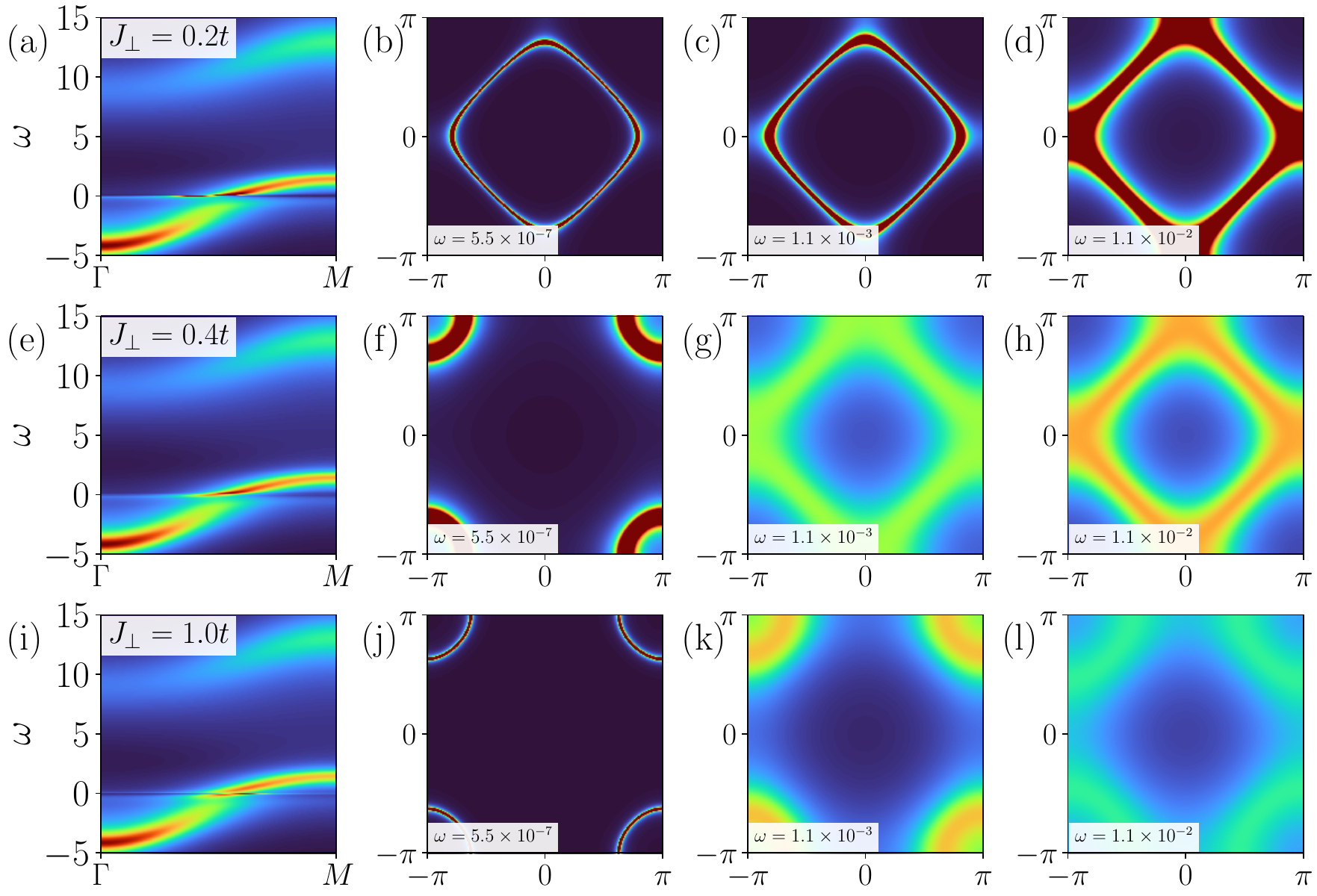}
    \caption{
    Momentum-resolved spectral function for $J_K=-10t$, $U=8t$, $t_\perp=0$, and doping $x=0.2$. 
    Panels (a)–(d), (e)–(h), and (i)–(l) correspond to $J_\perp = 0.2t$, $0.4t$, and $1.0t$, respectively. 
    In each row, the first panel shows the spectrum along the high-symmetry direction $\Gamma$–$M$, 
    while the remaining panels display constant-energy cuts of the spectral intensity in momentum space at 
    $\omega = 5.5\times10^{-7}t$, $1.1\times10^{-3}t$, and $1.1\times10^{-2}t$. 
    A small imaginary part $\eta = 10^{-3}t$ is introduced via $\omega \rightarrow \omega + i\eta$ to broaden the spectrum.
    }
    \label{fig:FS_T}
\end{figure}

\subsection{Fermi volume and the Luttinger's theorem}

To distinguish between the FL and sFL phases, a direct way is to examine the jump in the Fermi surface volume across the QCP, as shown in the main text.
This information is in fact already contained in the impurity self-energy.
To see this, we reconstruct the momentum-dependent Green’s function as
\begin{equation}
    G_\alpha(\omega,\mathbf{k}) 
    = \frac{1}{\omega + \mu - \epsilon_\alpha(\mathbf{k}) - \Sigma_\alpha(\omega)}~.
\end{equation}
The Fermi surface is determined by the poles of the Green's function at zero frequency, namely,
\begin{equation}
    \epsilon_\alpha(\mathbf{k}) - \mu + \mathrm{Re}\,\Sigma_\alpha(\omega=0) = 0~.
\end{equation}
Here, $\mu - \mathrm{Re}\,\Sigma_\alpha(\omega=0)$ plays the role of an effective chemical potential. 
The quantity $\Sigma_\alpha(\omega=0)$ refers to the self-energy evaluated at the zero-energy fixed point of the NRG flow. 
Since strictly zero energy is not accessible in practical calculations, we instead use the smallest available $\omega$.

We then compute the Fermi surface volume as 
\begin{equation}
    A_{\alpha;\mathrm{FS}} 
    = \int \frac{\mathrm{d}^2 k}{(2\pi)^2} 
    \, \theta\!\left( \mu - \epsilon_\alpha(\mathbf{k}) - \mathrm{Re}\,\Sigma_\alpha(\omega=0) \right),
\end{equation}
where $\theta(x)$ denotes the Heaviside step function. 

In Fig.~\ref{fig:muResigma}, we present the phase diagram determined from the Fermi surface volume as a function of $J_K$, $J_\perp$, and $x$, with fixed $U=8t$ and $t_\perp=0$. 
Across the transition from the FL to the sFL phase, $\mu - \mathrm{Re}\,\Sigma_\alpha(0)$ always exhibits a jump, which in turn leads to a corresponding jump of the Fermi surface areas as $\Delta A_{\mathrm{FS}} =\pm 1/2$. 
We note the phase boundaries characterized by a jump of the Fermi surface volume $\Delta A_{\mathrm{FS}} = \pm 0.5$ 
are fully consistent with those obtained from the $\mathbb{Z}_2$ indicator $C$.

\subsubsection{Luttinger's theorem}

According to the Luttinger's theorem \cite{luttinger1960}, 
\begin{equation}
    n_\mathrm{tot} = \sum_\alpha (A_{\alpha;\mathrm{FS}} +  A_{\alpha;\mathrm{LS}} + I_{\alpha;L})~,
\end{equation}
where $n_\mathrm{tot}$ is the total electron number and  $A_{\alpha;\mathrm{LS}}$ denotes the Luttinger surface where $G_\alpha(\omega,\mathbf{k}) = 0$,   
$I_{\alpha;L}$ is the Luttinger integral 
\begin{equation}\label{eqn:luttinger_integral}
    I_{\alpha;L} = -\frac{1}{\pi}\mathrm{Im}\int\ \frac{\mathrm{d}^2k}{(2\pi)^2} \int_{-\infty}^0\mathrm{d}\omega 
    G_{\alpha}(\omega,\mathbf{k}) \partial_\omega\Sigma_\alpha(\omega)~.
\end{equation}
In a conventional Fermi liquid, $I_{\alpha;L}$ is generally believed to vanish. 

In the single-layer Hubbard model, the emergence of a Luttinger surface is widely assumed to account for the Fermi volume jump. 
In our DMFT calculations, however, the self-energy $\Sigma$ does not exhibit a true divergence, and the Luttinger surface always has zero area, $A_{\alpha;\mathrm{LS}}=0$. 
Consequently, the jump of the Fermi surface area $\Delta A_{\alpha;\mathrm{FS}}=\pm 1/2$ implies a corresponding jump in the Luttinger integral, $\Delta I_{\alpha;L}=\mp 1/2$. 
The evolution of $I_{\alpha;L}$ is numerically examined along several cuts of the phase diagram in Figs.~\ref{fig:muResigma}(c) and (d). 
A clear jump from $I_{\alpha;L}=0$ in the FL phase to $I_{\alpha;L}\approx -1/2$ in the sFL phase is observed.

\subsubsection{Fermi surface evolution with temperature}

The sharp distinction between the two types of Fermi surfaces is not directly observed in experiments. 
In particular, the small Fermi pockets around the $(\pi,\pi)$ point have not been clearly detected.

In Fig.~\ref{fig:FS_T}, we present the spectral function on a linear scale together with constant-energy cuts at different energies. 
In both the FL and sFL phases, the spectra on a linear scale appear quite similar, with only subtle differences at very low energies. 
The high-energy peaks at $\sim 10t$ and $\sim -4t$ correspond to the Hubbard bands induced by the finite interaction $U$.

The distinction becomes clearer when examining the spectral function at fixed energy cuts. 
At very low energies, the difference between the FL and sFL phases is sharp. 
However, as the energy increases, the small Fermi surface gradually broadens and eventually evolves into a large Fermi-surface-like structure at energies of order $\sim 10^{-2}t$.  This energy scale corresponds to several meV and is comparable to the current energy resolution of ARPES experiments. 
Consequently, even if a small Fermi surface exists, it is difficult to observe directly in spectroscopic measurements. 
Instead, its presence is more likely to be inferred from transport measurements.

\begin{figure}[t]
    \centering
    \includegraphics[width=1.0\linewidth]{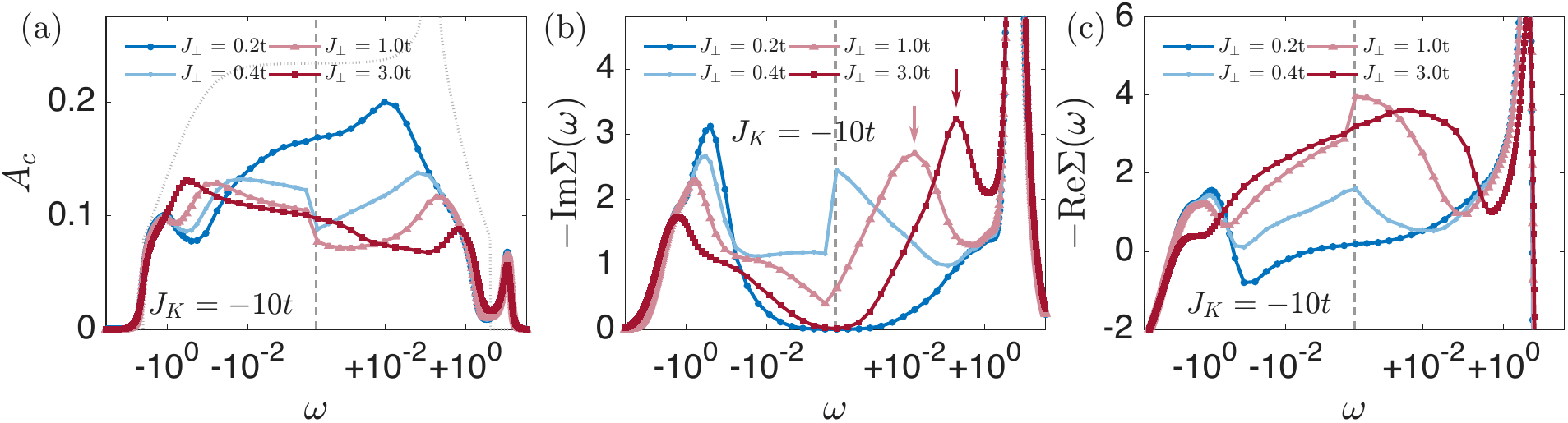}
    \caption{ 
    Spectral functions and self-energies for the Hund’s-coupling side of the double Kondo model Eq.~\eqref{eqn:double_kondo} with $J_K = -10t < 0$, $U = 8t$, and $x = 0.2$.
    Four values of $J_\perp$ are shown: $J_\perp = 0.2t$, $0.4t$, $1.0t$, and $3.0t$.
    The cases $J_\perp = 0.2t$ and $0.4t$ lie in the FL phase, while $J_\perp = 1.0t$ and $3.0t$ lie in the sFL phase.
    (a)The electron spectral function $A_c(\omega)$. 
    (b) The imaginary part of the self-energy $-\mathrm{Im}\Sigma(\omega)$. 
    The pseudogap positions in the sFL phase are identified by the peaks of $-\mathrm{Im}\Sigma(\omega)$, as indicated by the arrows.
    (c) Real part of the self-energy, $-\mathrm{Re}\Sigma(\omega)$.
    }   
    \label{fig:pseudogap_ferro}
\end{figure}

\begin{figure}[t]
    \centering
    \includegraphics[width=1.0\linewidth]{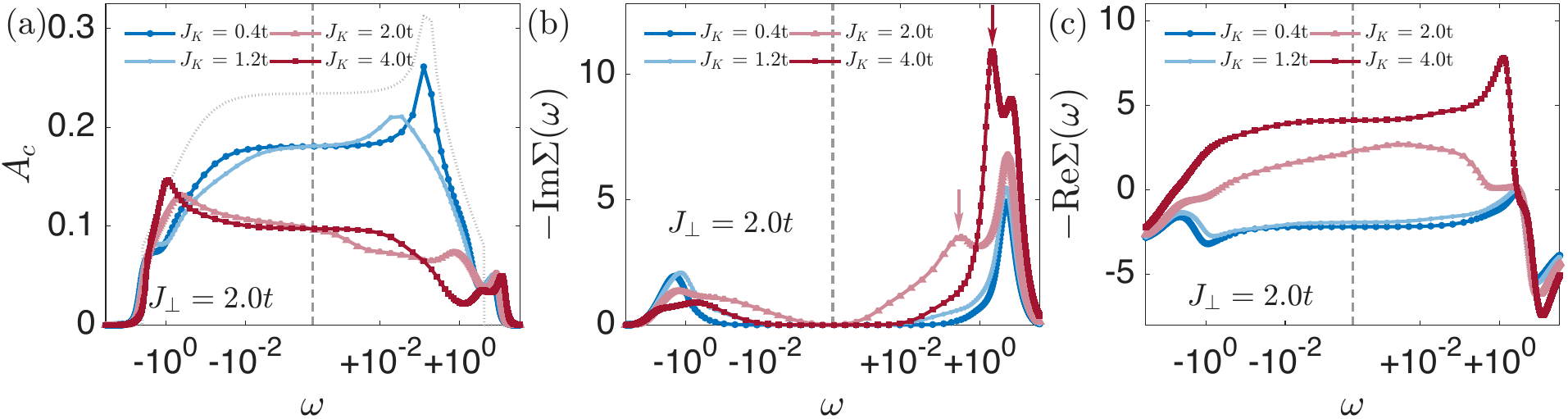}
    \caption{ 
    Spectral functions and self-energies for the Hund’s-coupling side of the double Kondo model Eq.~\eqref{eqn:double_kondo} with $J_\perp = 2.0t$, $U = 8t$, and $x = 0.2$.
    Four values of $J_K>0$ are shown: $J_K = 0.4t$, $1.2t$, $2.0t$, and $4.0t$.
    The cases $J_K = 0.4t$ and $1.2t$ lie in the FL phase, while $J_K = 2.0t$ and $4.0t$ lie in the sFL phase.
    (a)The electron spectral function $A_c(\omega)$. 
    (b) The imaginary part of the self-energy $-\mathrm{Im}\Sigma(\omega)$. 
    The pseudogap positions in the sFL phase are identified by the peaks of $-\mathrm{Im}\Sigma(\omega)$, as indicated by the arrows.
    (c) Real part of the self-energy, $-\mathrm{Re}\Sigma(\omega)$.
    }   
    \label{fig:pseudogap_anti}
\end{figure}

\subsection{Spectrum functions and self-energies of FL and sFL.}

In Figs.~\ref{fig:pseudogap_ferro} and \ref{fig:pseudogap_anti}, we present typical spectral functions and self-energies in both FL and sFL phase for both the Hund’s coupling $J_K<0$ and anti-Hund’s coupling $J_K>0$ cases.

In the FL phase, the spectral function $A_c(\omega)$ (blue curves in Figs.~\ref{fig:pseudogap_ferro}(a) and \ref{fig:pseudogap_anti}(a)) closely resembles that of a free Fermi liquid (dashed line), apart from a high-energy Hubbard gap of size $U=8t$.
At low energies, the self-energy exhibits the standard analytic structure $-\mathrm{Im}\Sigma(\omega)\sim \omega^2$ expected for a Fermi liquid, and the quasiparticle band remains continuously connected to the noninteracting dispersion.

By contrast, in the sFL phase the low-energy structure undergoes a qualitative reorganization.
A pronounced dip develops in $A_c(\omega)$ at positive energy, indicating the formation of a pseudogap.
This suppression of spectral weight is directly tied to the emergence of a pole-like structure in the self-energy, as highlighted by the red arrows in Figs.~\ref{fig:pseudogap_ferro}(b) and \ref{fig:pseudogap_anti}(b).
The self-energy in this regime can be approximated by
$\Sigma(\omega)\sim \frac{\Phi^2}{\omega+\mu_\psi}$ as suggested by the ancilla Fermi-liquid description discussed in Appendix.~\ref{app:ancilla}. 
The pole structure signals the appearance of an emergent fermionic mode that hybridizes with the physical electron. 

Correspondingly, the real part of the self-energy develops a kink at the pseudogap energy. 
Importantly, $-\mathrm{Re}\Sigma(0)$ shifts the effective chemical potential and therefore controls the Fermi surface volume.
Across the FL–sFL transition, $-\mathrm{Re}\Sigma(0)$ changes discontinuously, signaling a reconstruction of the Fermi surface volume, as already discussed in the previous subsection.

\begin{figure}
    \centering
    \includegraphics[width=0.99\linewidth]{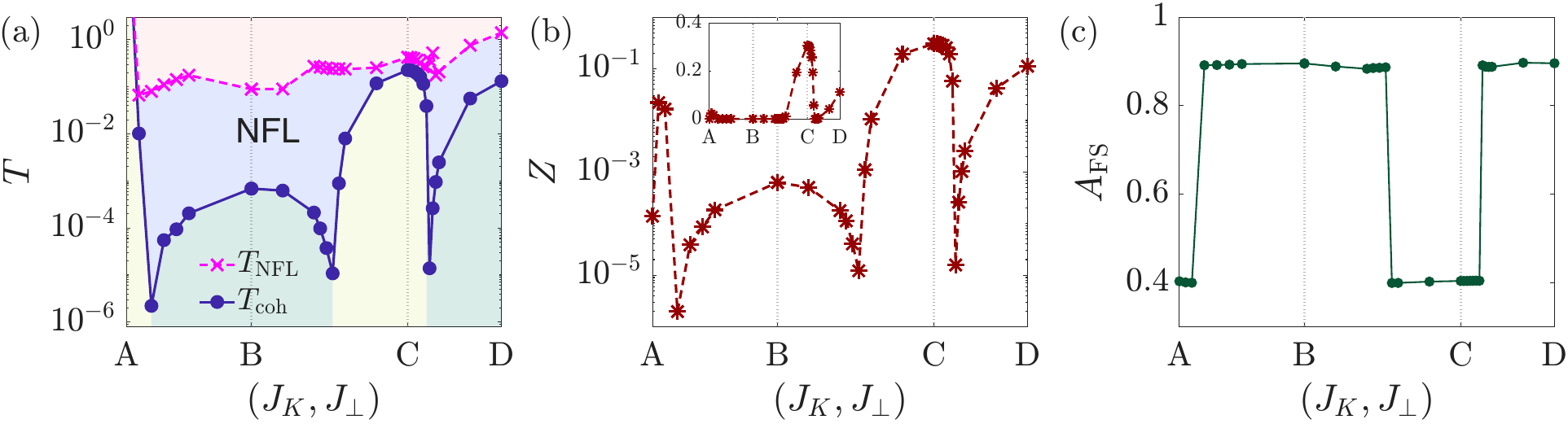}
    \caption{ (a) Upper bounds $T_{\mathrm{NFL}}$ and lower bounds $T_{\mathrm{coh}}$ of the NFL temperature along the line cut indicated in Fig.~\ref{fig:phase_diagram}(b), 
    passing through the four marked $(J_K, J_\perp)$ points, 
    A $(-10,0)$, B $(-10,2)$, C $(0,2)$, and D $(6,0)$. 
    (b) shows the quasiparticle spectral weight $Z = (1 - \partial_\omega \mathrm{Re}\Sigma(\omega) |_{\omega = 0})^{-1}$ evaluated along the same line cut. 
    $Z$ vanishes at each phase boundary between the FL and sFL phases. The linear-scale $Z$ is shown in the inset.
    (c) The Fermi surface volume $A_{\mathrm{FS}}$ evaluated along the same line cut.  
    }
    \label{fig:phase_diagram_JkJp}
\end{figure}

\begin{figure}
    \centering
    \includegraphics[width=0.7\linewidth]{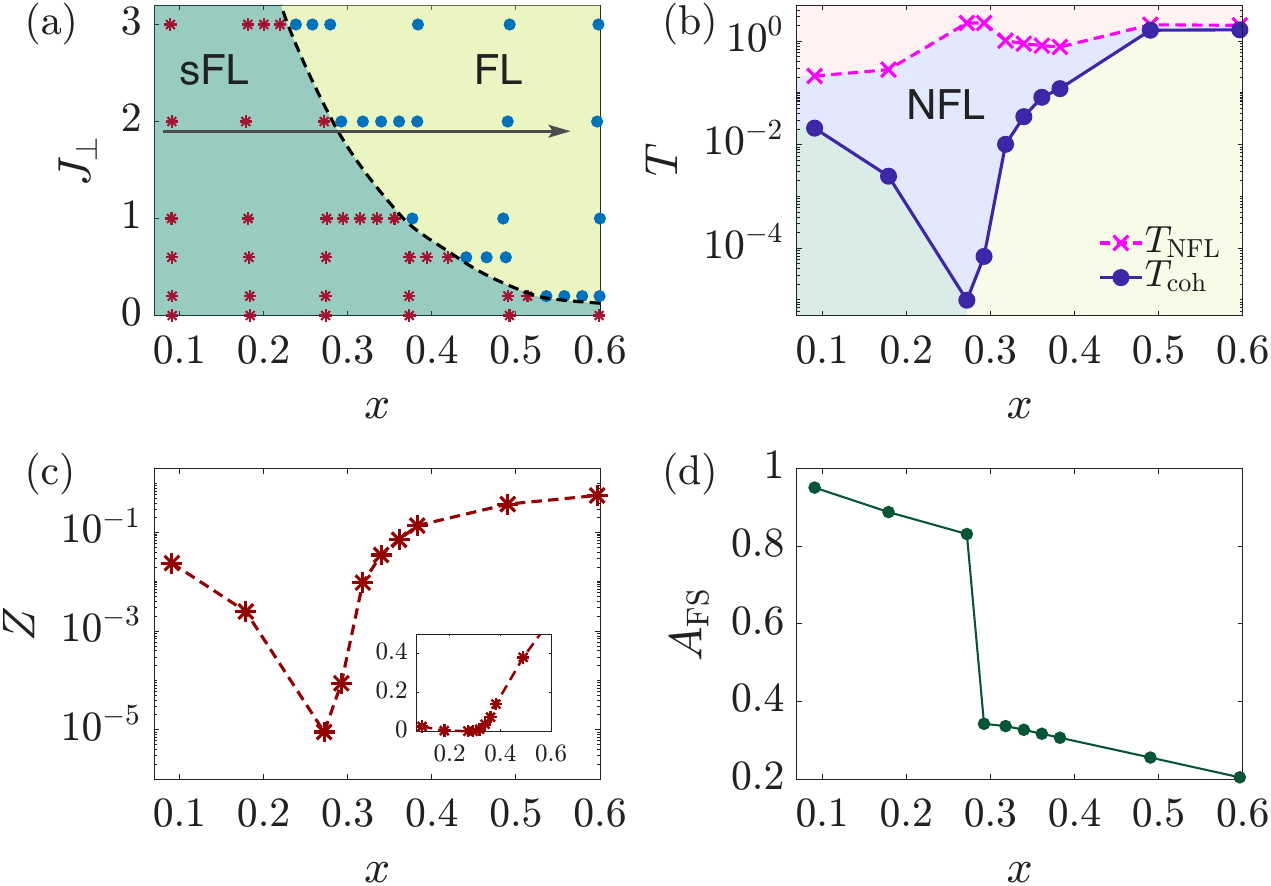}
    \caption{ (a) More phase diagram of the double kondo model Eq.~\eqref{eqn:double_kondo} by self-consistent DMFT calculation 
    as a function of doping $x$ and interlayer coupling $J_\perp $. 
    A fixed anti-Hund's coupling $J_K=2.0t$, Hubbard interaction $U = 8t$ and interlayer hoppint $t_\perp=0$ is used for all the calculations. 
    The blue dots and red stars indicate odd and even values of the $\mathbb{Z}_2$ charge $Q$, characterizing the FL phase (yellow region) and sFL phase (green region), respectively.
    (b) Upper and lower bounds of the NFL $T_{\mathrm{NFL}}$ and $T_{\mathrm{coh}}$ along the line cut indicated in (a). 
    (c) Quasiparticle spectral weight $Z=\left(1-\partial_\omega \mathrm{Re}\,\Sigma(\omega)\big|_{\omega=0}\right)^{-1}$ evaluated along the line cut in (a). 
    $Z$ vanishes at the phase boundary between the FL and sFL phases. The linear-scale $Z$ is shown in the inset
    (d) Fermi surface volume $A_{\mathrm{FS}}$ along the same cut, showing a jump $\Delta A_{\mathrm{FS}}=-1/2$ at the transition point.
    }
    \label{fig:phase_diagram_app}
\end{figure}

\begin{figure}
    \centering
    \includegraphics[width=0.7\linewidth]{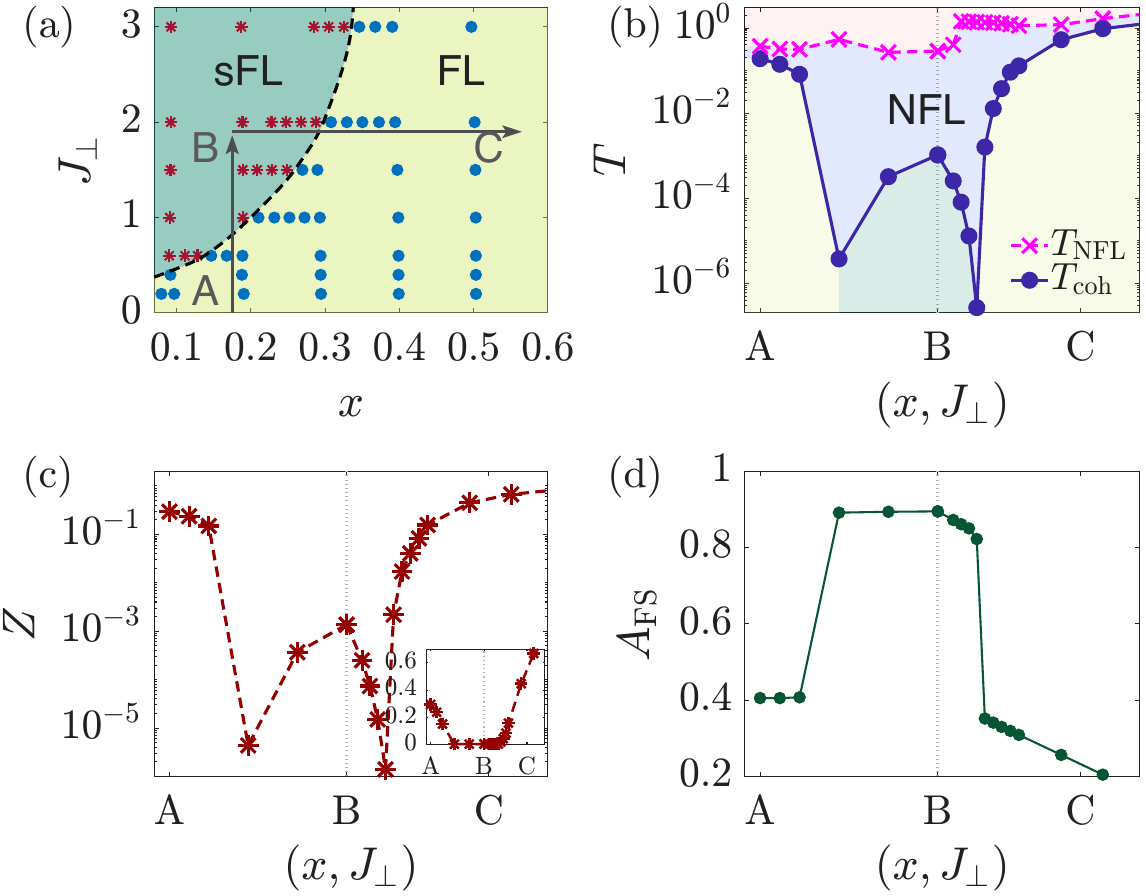}
    \caption{(a) Phase diagram of the double Hubbard model Eq.~\eqref{eqn:dtJ} as a function of doping $x$ and interlayer spin coupling $J_\perp$, with a fixed $U=8t$ and $V_\perp=0$. 
    The blue dots and red stars indicate odd and even values of the $\mathbb{Z}_2$ charge $Q$, characterizing the FL phase (yellow region) and sFL phase (green region), respectively.
    (b) Upper and lower bounds of the NFL temperature $T_{\mathrm{NFL}}$ and $T_{\mathrm{coh}}$ along the line cut indicated in (a). 
    (c) Quasiparticle spectral weight $Z=\left(1-\partial_\omega \mathrm{Re}\,\Sigma(\omega)\big|_{\omega=0}\right)^{-1}$ evaluated along the line cut in (a). 
    $Z$ vanishes at the phase boundary between the FL and sFL phases. The linear-scale $Z$ is shown in the inset
    (d) Fermi surface volume $A_{\mathrm{FS}}$ along the same cut, showing a jump $\Delta A_{\mathrm{FS}}=-1/2$ at the transition point.
    }
    \label{fig:phase_diagram_dtJ}
\end{figure}

\section{More phase diagrams}

\subsection{Phase diagram of $J_K$ and $J_\perp$}

In Fig.~\ref{fig:phase_diagram_JkJp}, 
we present additional details of the phase diagrams shown in Fig.~\ref{fig:phase_diagram}(b), 
obtained by fixing the doping $x=0.2$, the Hubbard interaction $U=8t$ and $t_\perp=0$. 
The two characteristic temperature scales $T_{\mathrm{NFL}}$ and $T_{\mathrm{coh}}$ along the line cut in Fig.~\ref{fig:phase_diagram}(b) are displayed. 
$T_{\mathrm{coh}}$ vanishes at every phase boundary between the FL and sFL phases, indicating that the QCP is universal. 
The QCP is further confirmed by the vanishing quasiparticle weight 
$Z=\left(1-\partial_\omega \mathrm{Re}\,\Sigma(\omega)\big|_{\omega=0}\right)^{-1}$. 
Across the transition, the QCP is always accompanied by a jump of the Fermi-surface volume $\Delta A_{\mathrm{FS}}\approx\pm 1/2$.

\subsection{The double kondo model with $J_K>0$.}

Here we present additional phase diagrams of the double Kondo model in Eq.~\eqref{eqn:double_kondo}, 
obtained by fixing the anti-Hund’s coupling at $J_K = 2.0t$, the Hubbard $U=8t$, the interlayer hopping $t_\perp=0$ and varying the doping $x$ and interlayer exchange $J_\perp$, as shown in Fig.~\ref{fig:phase_diagram_app}.
Once again, two distinct Fermi-liquid phases are observed, separated by a continuous quantum phase transition.
The transition is characterized by the simultaneous vanishing of the coherence temperature $T_{\mathrm{coh}}$ and the quasiparticle weight $Z$. 
In the anti-Hund’s case $J_K>0$, the phase labeled as sFL corresponds to the conventional heavy Fermi-liquid phase, while the FL phase represents the RKKY-dominated regime. 
We retain the present notation to maintain consistency with the Hund’s coupling side ($J_K<0$), where the same $\mathbb{Z}_2$ topological classification applies.

Similar to the Hund's coupling side $J_K<0$, the sFL phase on the $J_K>0$ side is also stabilized at lower doping $x$ or higher electron density, 
with an enhanced tendency toward Kondo screening. 
The role of the interlayer coupling $J_\perp$ is however opposite for different signs of $J_K$.
For $J_K<0$ (Hund’s coupling), increasing $J_\perp$ promotes the sFL phase. 
By contrast, for $J_K>0$ (anti-Hund’s coupling), a larger $J_\perp$ favors the FL (RKKY-like) phase, suggesting that interlayer singlet formation competes with Kondo hybridization.

\subsection{Double Hubbard model}

In this section, we show that a considerably simplified model already captures the same QCP.
Specifically, we consider a bilayer Hubbard model Eq.~\eqref{eqn:dtJ}, 
where the system consists of two layers of itinerant electrons $c_{i;\alpha}$.
The physics of this bilayer Hubbard model is expected to resemble that of the double Kondo model in the strong Hund’s coupling limit, 
where the conduction electron and the local moment effectively bind together to form a spin-1 composite object. 

The phase diagram obtained by tuning the doping $x$ and the interlayer spin interaction $J_\perp$ is shown in Fig.~\ref{fig:phase_diagram_dtJ}.
Two distinct types of Fermi liquids are identified through the topological indicator introduced in Appendix.~\ref{app:topoz2}. 
The overall structure of the phase diagram closely resembles that of the double Kondo model in the large $-J_K$ limit of Eq.~\eqref{eqn:double_kondo}, with the sFL phase stabilized at lower doping $x$ and larger interlayer coupling $J_\perp$.

The critical doping $x_c$ increases with $J_\perp$ and approaches $x_c \sim 0.32$ in the large-$J_\perp$ limit.
Measurements of the characteristic temperatures $T_{\mathrm{NFL}}$ and $T_{\mathrm{coh}}$ and quasiparticle weight $Z$ indicate that the transition is continuous and is accompanied by an extended temperature regime of NFL behavior.
The Fermi-surface volume also exhibits a jump across the transition that closely parallel those observed in the double Kondo model.

Finally, we note that this model is relevant to several physical systems, including multiorbital molecular solids such as $C_{60}$\cite{caponeC60_2009} and moiré materials such as twisted bilayer graphene \cite{zhaoTopoMott2025a, zhaoRVB2025}, 
where two different kinds of Fermi liquid, as well as an intermediate superconducting state are discussed.

\section{An ancilla fermion understanding of Pseudogap}\label{app:ancilla}

In this section, we provide an ancilla-fermion interpretation~\cite{zhang_pseudogap_2020} of the pseudogap behavior observed in the sFL phase. 
As shown in Figs.~\ref{fig:pseudogap_ferro}(b) and \ref{fig:pseudogap_anti}(b), a pronounced peak develops in the imaginary part of the self-energy $-\mathrm{Im}\Sigma(\omega)$. 
Correspondingly, a dip structure appears in the electron spectral function. 
In momentum space, this dip can be traced back to an additional flat band contribution, as illustrated in Figs.~\ref{fig:momentum}(b) and (c).

We show that these features can be qualitatively captured by a simple effective model at the mean-field level, 
\begin{equation} \label{eqn:ancilla_app}
    H_{\mathrm{AFT}} = \sum_{\mathbf{k},a,\sigma} 
    \left(\epsilon(\mathbf{k})-\mu\right)c^\dagger_{\mathbf{k};a;\sigma} c_{\mathbf{k};a;\sigma} 
    + \Phi \sum_{i,a,\sigma} c^\dagger_{i;a;\sigma}\psi_{i;a;\sigma} +\mathrm{h.c.} - \mu_\psi\sum_{i,a} n_{i;a}^\psi~,
\end{equation}
where $\psi_{i;a;\sigma}$ denotes an ancilla fermionic degree of freedom. 
The hybridization $\Phi$ between $c$ and $\psi$ acts as a variational parameter that controls the coupling between the physical electrons and the ancilla sector. 
The chemical potential $\mu_\psi$ is tuned so that the $\psi$ fermions remain half filled.

When $\Phi = 0$, the $c$ and $\psi$ fermions are decoupled, and the theory reduces to a conventional Fermi liquid with Fermi surface volume $A_{\mathrm{FS}}=(1-x)/2$. 
When $\Phi \neq 0$, the ancilla fermions hybridize with the physical electrons and effectively become part of the Fermi sea. 
As a result, the Fermi volume becomes $A_{\mathrm{FS}}=(1-x)/2+1/2 = -x/2\ \mathrm{mod}\ 1$. 
This change of Fermi volume is consistent with the sFL phase identified in the DMFT calculations.

Within the ancilla theory, the Green's functions can be obtained by solving
\begin{equation}\label{eqn:ancilla_green}
    \left(\begin{matrix}
    \omega + \mu -\epsilon_\alpha(\mathbf{k})  & -\Phi \\ -\Phi & \omega + \mu_\psi
    \end{matrix}\right) 
    \left(\begin{matrix}G_{\alpha}^{\mathrm{AFT}}(\omega,\mathbf{k}) \\ F_{\alpha}^{\mathrm{AFT}}(\omega,\mathbf{k}) \end{matrix}\right)
    = \left(\begin{matrix} 1 \\ 0 \end{matrix}\right)~,
\end{equation}
where $F^\mathrm{AFT}_\alpha (\omega,\mathbf{k})= \lla \psi_{\mathbf{k};\alpha};c^\dagger_{\mathbf{k};\alpha}\rra (\omega)$. 
Solving this equation gives the electron Green's function
\begin{equation}
    G^{\mathrm{AFT}}_\alpha(\omega,\mathbf{k}) = \frac{1}{\omega+\mu-\epsilon(\mathbf{k}) - \Phi^2/(\omega+\mu_\psi)}~. 
\end{equation}
We therefore get the form of the self-energy: 
\begin{equation}
    \Sigma^{\mathrm{AFT}}_\alpha(\omega) = \Phi^2/(\omega+\mu_\psi). 
\end{equation}
For $\omega+i0^+ $, the self energy would show a peak at $-\mu_\psi$ in the imaginary part and a kink at $-\mu_\psi$ in the real part, 
consistent with our earlier observation. 

We can also test the Luttinger theorem within the ancilla-fermion theory. 
For finite $\mu_\psi \neq 0$, $\Sigma^{\mathrm{AFT}}_\alpha(0)$ is not divergent so the Luttinger surface has zero volume. 
On the other hand, the Luttinger integral defined in Eq.~\eqref{eqn:luttinger_integral} can be evaluated explicitly as
\begin{equation}\label{eqn:luttinger_ancilla}
\begin{aligned}
    I^{\mathrm{AFT}}_{\alpha;L} =& \int\frac{\mathrm{d}^2k}{(2\pi)^2} \theta(-E_-(\mathbf{k})) \frac{\Phi^2}{(E_-(\mathbf{k})+\mu_\psi)(E_+(\mathbf{k})-E_-(\mathbf{k}))} + \theta(\mu_\psi)
    \\=& -n_{\psi;\alpha}=-1/2~. 
\end{aligned}
\end{equation}
Here $E_\pm(\mathbf{k}) = (\epsilon(\mathbf{k})-\mu-\mu_\psi)/2 
\pm \sqrt{\Phi^2+(\epsilon(\mathbf{k})-\mu-\mu_\psi)^2/4}$ 
denotes the hybridized band structure. 
For $\mu_\psi<0$, the Luttinger integral coincides with minus the $\psi$-fermion number, which is fixed to be $-n_{\psi;\alpha} = -1/2$.
The result is consistent with out numerical results as indicated in Fig.~\ref{fig:muResigma} (c) and (d). 

\begin{figure}[t]
    \centering
    \includegraphics[width=0.7\linewidth]{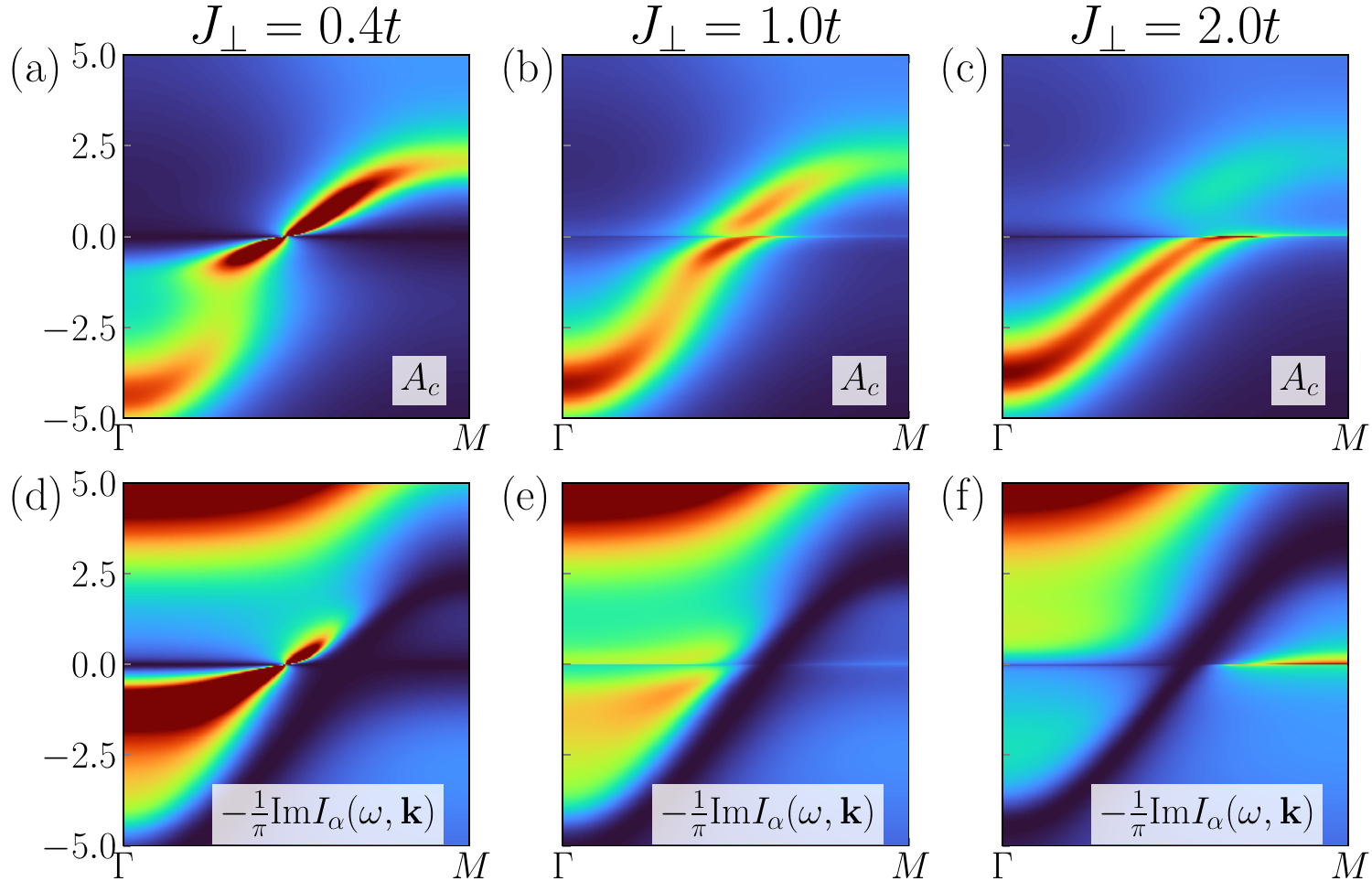}
    \caption{
    Momentum space spectrum for the double Hubbard model Eq.~\eqref{eqn:dtJ} with fixed $U=8t$ and $x=0.2$.  
    (a)-(c) Momentum space spectrum of the physical fermion $A_c$ along the line cut $(0,0)-(\pi,\pi)$ for $J_\perp=0.4t$, $J_\perp=1.0t$ and $J_\perp=2.0t$, respectively. 
    (d)-(f) The corresponding momentum space spectrum $-\frac{1}{\pi}\mathrm{Im}I_\alpha(\omega,\mathbf{k})$ of the trion like operator $[c_{k;\alpha};H_{\mathrm{int}}]$ extracted from Eq.~\eqref{eqn:spectrum_psi}. 
    }
    \label{fig:Apsi}
\end{figure}

\subsection{Ancilla fermion as a composite operator in double Hubbard model}

It was argued in Refs.~\cite{ledwithTrion2025,zhaoTopoMott2025a} that the ancilla fermion can be interpreted as a three-body trion operator in the Hubbard model and in twisted bilayer graphene systems.
Following the same spirit, here we identify the ancilla fermion as a spin-polaron like trion operator, and extract the corresponding spectral function. 

For concreteness, we consider the double Hubbard model introduced in Eq.~\eqref{eqn:dtJ}.
A natural way to identify composite fermionic modes generated by interactions is through the equation-of-motion hierarchy of the electron Green's function.
Carrying out two steps of the equation-of-motion using Eqs.~\eqref{eqn:eom000}\eqref{eqn:eom001} gives 
\begin{equation}\label{eqn:eom1_k}
    (\omega + \mu -\epsilon(\mathbf{k})) G_\alpha(\omega,\mathbf{k}) = 1 + F_\alpha(\omega,\mathbf{k})~,
\end{equation}
\begin{equation}\label{eqn:eom2_k}
    (\omega + \mu -\epsilon(\mathbf{k})) F_\alpha(\omega,\mathbf{k}) = \Sigma_{\mathbf{k};\alpha}^H + I_\alpha(\omega,\mathbf{k})~,
\end{equation}
where $F_\alpha(\omega,\mathbf{k}) = \lla [c_{\mathbf{k};\alpha}^{},H_{\mathrm{int}}];c^\dagger_{\mathbf{k};\alpha}\rra$, 
$I_\alpha(\omega,\mathbf{k}) = \lla [c_{\mathbf{k};\alpha}^{},H_{\mathrm{int}}];[c_{\mathbf{k};\alpha},H_{\mathrm{int}}]^\dagger\rra$, 
and $\Sigma^H_{\mathbf{k};\alpha} = \la \{[c_{\mathbf{k};\alpha},H_{\mathrm{int}}], c_{\mathbf{k};\alpha}^\dagger\}\ra $ is the Hartree contribution.  
These correlators naturally involve the composite operator $[c_{\mathbf{k};\alpha}^{},H_{\mathrm{int}}]$, which carries the same charge, spin and fermionic statistics as the original electron operator $c_{\mathbf{k};\alpha}$. 
Solving Eqs.~\eqref{eqn:eom1_k} and \eqref{eqn:eom2_k} yields 
\begin{equation}
    F_\alpha(\omega,\mathbf{k}) = \frac{\Sigma_\alpha(\omega)}{\omega+\mu-\epsilon(\mathbf{k})-\Sigma_\alpha(\omega)}~,
\end{equation}
\begin{equation} \label{eqn:spectrum_psi}
    I_\alpha(\omega,\mathbf{k}) = \frac{(\omega+\mu-\epsilon(\mathbf{k}))\Sigma_\alpha(\omega)}{\omega+\mu-\epsilon(\mathbf{k})-\Sigma_\alpha(\omega)} - \Sigma^H_{\mathbf{k};\alpha}~.
\end{equation}

The equation of motion structure suggests interpreting the composite operator $[c,H_{\mathrm{int}}]$ as a fermionic trion mode generated by the interaction.
To isolate this mode as an independent degree of freedom, we define
\begin{equation}
    \Psi_{\mathbf{k;\alpha}} = \frac{1}{\Phi}([c_{\mathbf{k};\alpha},H_{\mathrm{int}}]-\delta_{\mathbf{k};\alpha} c_{\mathbf{k};\alpha})~,
\end{equation}
where the subtraction $\delta_{\mathbf{k};\alpha}$ term removes the component overlapping with the original electron operator, and $\Phi$ is a normalization factor.
Specifically, we impose the orthogonality condition
$\langle\{\Psi_{\mathbf{k};\alpha},c^\dagger_{\mathbf{k};\beta}\}\rangle= 0$ and normalization condition $\la \{\Psi_{\mathbf{k};\alpha},\Psi^\dagger_{\mathbf{k};\beta}\}\ra = \delta_{\alpha\beta}$, 
from which one solves $\delta_{\mathbf{k};\alpha} = \la \{[c_{\mathbf{k};\alpha},H_{\mathrm{int}}], c_{\mathbf{k};\alpha}^\dagger\}\ra = \Sigma^H_{\mathbf{k};\alpha}$ and $\Phi = \sqrt{\la\{[c_{\mathbf{k};\alpha},H_{\mathrm{int}}],[c_{\mathbf{k};\alpha},H_{\mathrm{int}}]^\dagger\} - \delta_{\mathbf{k};\alpha}^2}$. 
With this definition, the equation of motion becomes
\begin{equation}
    (\omega + \mu+\delta_{\mathbf{k};\alpha} -\epsilon(\mathbf{k})) G_\alpha(\omega;\mathbf{k}) - \Phi \lla \Psi_{\mathbf{k};\alpha};c^\dagger_{\mathbf{k};\alpha}\rra = 1~.
\end{equation}
which has precisely the same structure as the first row of Eq.~\eqref{eqn:ancilla_green}.
This motivate us to identify the trion fermion $\Psi$ as the ancilla fermion $\psi$ in the ancilla Hamiltonian Eq.~\eqref{eqn:ancilla}. 

For the double Hubbard model Eq.~\eqref{eqn:dtJ} considered here, the commutator contains two contributions, 
\begin{equation}
\begin{aligned}
    [c_{i;t;\uparrow},H_{\mathrm{int}}] =& Uc_{i;t;\uparrow}n_{i;t\downarrow} + \frac{1}{2}J_\perp (c_{i;t;\uparrow}s_{i;b}^z + c_{i;t;\downarrow}s^-_{i;b})\\ 
    =& (\Phi_U \Psi^U_{i;t;\uparrow} + \delta_U c_{i;t;\uparrow}) + (\Phi_J \Psi^J_{i;t\uparrow} + \delta_Jc_{i;t;\uparrow})~,
\end{aligned}
\end{equation}
where $\delta_U = U(1-x)/2, \Phi_U = U\sqrt{1-x^2}/2$ and $\delta_J=0, \Phi_J = J_\perp\sqrt{3(1-x)/4-2\la\mathbf{s}_{i;t}\cdot \mathbf{s}_{i;b}\ra}/2$. 
The first term originates from the on-site Hubbard interaction and describes local charge fluctuations. 
It is not directly relevant to the FL–sFL quantum critical point studied here.
The second term instead corresponds to a spin-polaron operator. 
The corresponding ancilla fermions take the form
\begin{equation}
    \Psi^U_{i;t;\sigma}  = \frac{U}{\Phi_U}
    \left( c_{i;t;\sigma}n_{i;t;\bar\sigma} - \frac{1-x}{2} \right)~,
\end{equation}
and 
\begin{equation}
    \Psi^J_{i;t;\lambda} 
    = \frac{J_\perp}{2\Phi_J} \sum_\rho 
    c_{i;t;\rho}  (\boldsymbol{\sigma}\cdot \boldsymbol{s}_{i;b})_{\rho\lambda}~.
\end{equation}

In Fig.~\ref{fig:Apsi} we compare the spectral functions 
$A_c(\omega,\mathbf{k}) = - \frac{1}{\pi}\mathrm{Im}\,G_\alpha(\omega,\mathbf{k})$ 
and 
$-\frac{1}{\pi}\mathrm{Im}\,I_\alpha(\omega,\mathbf{k})$
on a linear energy scale for the double Hubbard model Eq.~\eqref{eqn:dtJ}. 
The latter is correlation function of $[c_{\mathbf{k};\alpha},H_{\mathrm{int}}]\sim \Psi^U+\Psi^J+\const$ and contains contributions from both the $\Psi^U_{i;\alpha}$ and $\Psi^J_{i;\alpha}$ fermions. 
While the underlying band structure remains the same in both $A_c(\omega,\mathbf{k})$ and 
$-\frac{1}{\pi}\mathrm{Im}\,I_\alpha(\omega,\mathbf{k})$, 
the spectral weight is redistributed differently between the two channels in the sFL phase. 
In the electron spectral function $A_c(\omega,\mathbf{k})$, the coherent quasiparticle weight is mainly concentrated on the dispersive band. 
By contrast, the spectral weight on the original dispersive band is strongly suppressed in 
$-\frac{1}{\pi}\mathrm{Im}\,I_\alpha(\omega,\mathbf{k})$. 
Instead, most of the spectral weight is transferred to a nearly flat band at low energy, which corresponds to the contribution from $\Psi^J_{i;\alpha}$, 
together with a broad high-energy feature associated with $\Psi^U_{i;\alpha}$.

\subsection{A variational wave function understanding. }

Although the self-energy obtained from the ancilla construction qualitatively reproduces the DMFT results, 
the ancilla fermion approach appears largely phenomenological and lacks a clear physical interpretation. 
To go beyond the mean-field description, we construct a variational wave function that provides a more direct physical understanding of the ancilla formulation.

We start from the simpler bilayer Hubbard model in Eq.~\eqref{eqn:dtJ} with $U=0$ and consider the effective ancilla Hamiltonian in Eq.~\eqref{eqn:ancilla}. 
The ground state of the ancilla Hamiltonian is a Slater determinant  $|\mathrm{Slater}[c,\psi]\rangle$ of the physical electrons $c$ and the ancilla fermions $\psi$. 
To recover the physical Hilbert space, we project the ancilla fermions on the two layers into an on-site spin singlet state 
$|s_i\rangle = \frac{1}{\sqrt{2}}(\psi_{i;t;\uparrow}^\dagger \psi_{i;b;\downarrow}^\dagger - \psi_{i;t;\downarrow}^\dagger \psi_{i;b;\uparrow}^\dagger)|0\rangle$, 
through the projection operator $P_s = \otimes_i |s_i\rangle\langle s_i|$.
After this projection, the wave function is mapped back to the physical Hilbert space 
\begin{equation}
    |\Psi_\mathrm{anc}\rangle = P_s |\mathrm{Slater}[c,\psi]\ra~,
\end{equation}
which is expected to capture the correct low-energy physics.
The wave function can be understood as a generalized version of Gutwillzer wavefunction. 

\textbf{Analytical result at half-filling} To justify this construction, we analyze the strong interlayer coupling limit $J_\perp \rightarrow \infty$. 
At half filling, the exact ground state is a trivial insulator where spins on the two layers form local singlets at each site,
\begin{equation}
    |\Psi\rangle_0 = \prod_i\frac{1}{\sqrt{2}} (c^\dagger_{i;1,\uparrow}c^\dagger_{i;2,\downarrow}
    -c^\dagger_{i;1,\downarrow}c^\dagger_{i;2,\uparrow})|0\rangle~.
\end{equation}
When a small hopping $t_{ij}$ is introduced, the ground state can be obtained perturbatively,
\begin{equation}
    |\Psi\rangle = \sum_{i,j} \left(1-\frac{2}{3J_\perp}t_{ij}c^\dagger_{i;a,\sigma}c_{j;a,\sigma}\right)|\Psi\rangle_0~.
\end{equation}

We now show that the same result can be reproduced within the ancilla construction with an appropriate choice of the hybridization parameter $\Phi$. 
First consider again the limit $J_\perp \rightarrow \infty$, where the hybridization should be taken as $\Phi \rightarrow \infty$. 
For each flavor $\alpha=(a,\sigma)$, the local $c$--$\psi$ hybridization is diagonalized by the two local hybridized orbitals
\begin{equation}\label{eqn:ancilla_gamma_pm}
    \gamma_{i;\alpha;\pm}^{\dagger}
    =
    \frac{1}{\sqrt{2}} \left(c^\dagger_{i;\alpha}\pm \psi^\dagger_{i;\alpha}\right)~.
\end{equation}
The index $\pm$ labels two hybridized modes of the $c$ and $\psi$; it is not a layer bonding/antibonding index. 
In the $\Phi\rightarrow\infty$ limit only the lower hybridized orbital $\gamma_-^\dagger$ is occupied, while the upper one is projected out. 
At half filling, every local lower hybridized orbital is occupied,
\begin{equation}
    |\mathrm{Slater}[c,\psi]\rangle = \prod_{i,\alpha}\gamma^\dagger_{i;\alpha;-}|0\rangle . 
\end{equation}

After projecting the two ancilla layers into spin singlet 
it is straightforward to verify that the projected state reproduces the exact ground state $|\Psi\rangle_0$.

When a finite hopping $t_{ij}$ is introduced in the ancilla Hamiltonian, 
perturbation theory shows that the projected wave function is now
\begin{equation}
    |\Psi_\mathrm{anc}\rangle =
    \sum_{i,j}
    \left( 1-\frac{1}{2\Phi}t_{ij}c^\dagger_{i;a;\sigma}c_{j;a;\sigma} \right) |\Psi\rangle_0 .
\end{equation}

Comparing the two results, we find that the two descriptions are equivalent provided that
\begin{equation}
    \Phi = \frac{3J_\perp}{4}.
\end{equation}
This establishes the validity of the ancilla construction, at least in this simplified model and in the strong $J_\perp$ limit at half-filling.  Even in this limit, we note that the wavefunction is not a simple Slater determinant and ancilla framework is useful to describe the opening of the gap from four-fermion interaction $J_\perp$.  Extension of this wavefunction by simply adding a chemical potential for $c$  describes the sFL phase upon doping!

\textbf{Small doping case} Similar construction can be extended to small hole doping in the $\Phi\rightarrow\infty$ limit. 
We assume Eq.~\eqref{eqn:ancilla_gamma_pm} remains unchanged for a small enough doping. 
The ground state before projection is obtained by removing a set $\mathcal{H}$ of electrons from the lower hybridized band:
\begin{equation}\label{eqn:ancilla_doped_slater}
    |\mathrm{Slater}[c,\psi]\rangle_x
    =
    \prod_{(\mathbf{k},\alpha)\notin \mathcal{H}}
    \gamma_{\mathbf{k};\alpha;-}^\dagger |0\rangle
    =
    \prod_{(\mathbf{k},\alpha)\in \mathcal{H}}
    \gamma_{\mathbf{k};\alpha;-}
    |\mathrm{Slater}[c,\psi]\rangle_{x=0}~, 
\end{equation}
where $\gamma_{\mathbf{k};\alpha;-}
    = \frac{1}{\sqrt{N}}
    \sum_i e^{-i\mathbf{k}\cdot \mathbf{x}_i}\gamma_{i;\alpha;-}~$. 
Let the holes in $\mathcal{H}$ be labeled by $h_m=(\mathbf{k}_m,\alpha_m)$ with a fixed ordering $m=1,\cdots,N_h$. 
The product of momentum-space hole operators can be expanded as
\begin{equation}\label{eqn:ancilla_hole_superposition}
    \prod_{m=1}^{N_h}\gamma_{\mathbf{k}_m;\alpha_m;-}
    =
    \sum_{\mathcal{C}} A_{\mathcal{H}}(\mathcal{C})
    \prod_{(i,\alpha)\in \mathcal{C}} \gamma_{i;\alpha;-}~,
\end{equation}
where $\mathcal{C}$ denotes a real-space hole configuration and
$A_{\mathcal{H}}(\mathcal{C})$ is the corresponding Slater-determinant amplitude built from the phases $e^{-i\mathbf{k}\cdot \mathbf{x}_i}$. 

For each configuration $\mathcal{C}$, let $\mathcal{H}_i(\mathcal{C})$ denote the subset of local flavors removed at site $i$, $
    \mathcal{H}_i(\mathcal{C}) \subset \{(t,\uparrow),(t,\downarrow),(b,\uparrow),(b,\downarrow)\}~$. 
To rewrite each term in a site-factorized form, one may need to permute the fermion operators into a fixed canonical ordering. 
The resulting permutation sign is configuration dependent, but it can be absorbed into a redefined coefficient $\widetilde A_{\mathcal{H}}(\mathcal{C})$. 
Therefore
\begin{equation}\label{eqn:ancilla_factorized_sites}
    P_s \prod_{m=1}^{N_h}\gamma_{\mathbf{k}_m;\alpha_m,-} |\mathrm{Slater}[c,\psi]\rangle_{x=0}
    =
    \sum_{\mathcal{C}} \widetilde A_{\mathcal{H}}(\mathcal{C})
    P_{s}
    \left(\prod_i\prod_{\alpha\in\mathcal{H}_i(\mathcal{C})}\gamma_{i;\alpha;-}\right)
    |\mathrm{Slater}[c,\psi]\rangle_{x=0}~,
\end{equation}

For every site $i$, it can be shown 
\begin{equation}\label{eqn:ancilla_local_identity_general}
    P_{s}
    \left(\prod_{\alpha\in\mathcal{H}_i}\gamma_{i;\alpha;-}\right)
    |\mathrm{Slater}[c,\psi]\rangle_{x=0}
    =
    \lambda_{|\mathcal{H}_i|}
    \left(\prod_{\alpha\in\mathcal{H}_i}c_{i;\alpha}\right)
    |\Psi_0\rangle~,
\end{equation}
where $|\mathcal{H}_i|$ is the dimension of set $\mathcal{H}_i$, and the coefficients are $\lambda_n = -\frac{2^{(n-1)/2}}{4}$. 
The formula can be directly verified for $|\mathcal{H}|_i=0$ and $|\mathcal{H}|_i=1$. 
For $|\mathcal{H}|_i>2$, both sides of the above formula vanish. 
For $|\mathcal{H}_i|=2$, both left and right side of Eq.~\eqref{eqn:ancilla_local_identity_general} remains nonzero only when ${\mathcal{H}}_i=\{(t,\uparrow),(b,\downarrow)\}$ or $\{(t,\downarrow),(b,\uparrow)\}$. 
The two nonvanishing cases can be evaluated explicitly as
\begin{equation}\label{eqn:ancilla_twohole_sign1}
    P_{s}\gamma_{i;t;\uparrow;-}\gamma_{i;b;\downarrow;-}|\mathrm{Slater}[c,\psi]\rangle_{x=0}
    =
    -\frac{1}{2\sqrt{2}}\,
    c_{i;t;\uparrow}c_{i;b;\downarrow}|\Psi_0\rangle~,
\end{equation}
\begin{equation}\label{eqn:ancilla_twohole_sign2}
    P_{s}\gamma_{i;t;\downarrow;-}\gamma_{i;b;\uparrow;-}|\mathrm{Slater}[c,\psi]\rangle_{x=0}
    =
    -\frac{1}{2\sqrt{2}}\,
    c_{i;t;\downarrow}c_{i;b;\uparrow}|\Psi_0\rangle~.
\end{equation}
Substituting Eq.~\eqref{eqn:ancilla_local_identity_general} into Eq.~\eqref{eqn:ancilla_factorized_sites}, we obtain
\begin{equation}\label{eqn:ancilla_preserve_amplitude}
    P_s \prod_{m=1}^{N_h}\gamma_{\mathbf{k}_m;\alpha_m,-} |\mathrm{Slater}[c,\psi]\rangle_{x=0}
    =
    \sum_{\mathcal{C}} \widetilde A_{\mathcal{H}}(\mathcal{C})
    \left[\prod_i \lambda_{|\mathcal{H}_i(\mathcal{C})|}\right]
    \prod_{(i,\alpha)\in \mathcal{C}} c_{i;\alpha}\,|\Psi_0\rangle~. 
\end{equation}
Let $N_0(\mathcal{C})$, $N_1(\mathcal{C})$, and $N_2(\mathcal{C})$ denote the numbers of sites with $|\mathcal{H}_i|=0,1,2$, respectively. 
Then
\[
    N_0(\mathcal{C})+N_1(\mathcal{C})+N_2(\mathcal{C})=N,
    \qquad
    N_1(\mathcal{C})+2N_2(\mathcal{C})=N_h~.
\]
Hence Eq.~\eqref{eqn:ancilla_preserve_amplitude} contains the factor
\[
    \prod_i \lambda_{|\mathcal{H}_i(\mathcal{C})|}
    =
    \lambda_0^{N_0(\mathcal{C})}
    \lambda_1^{N_1(\mathcal{C})}
    \lambda_2^{N_2(\mathcal{C})}
    =
    \left(-\frac{1}{4}\right)^{N_0+N_1+N_2}
    (\sqrt{2})^{N_1+2N_2}
    =
    \left(-\frac{1}{4}\right)^{N}
    (\sqrt{2})^{N_h},
\]
which is independent of how the $N_h$ holes are distributed in real space. 
Therefore projection does not generate any additional configuration-dependent sign or magnitude beyond the fixed fermionic ordering sign already contained in $\widetilde A_{\mathcal{H}}(\mathcal{C})$. 
Recombining the sum over $\mathcal{C}$ back into momentum space finally gives
\begin{equation}\label{eqn:ancilla_doped_projected}
    |\Psi_{\mathrm{anc}}(x)\rangle_{\Phi\rightarrow\infty}
    \propto
    \prod_{(\mathbf{k},\alpha)\in \mathcal{H}}
    c_{\mathbf{k};\alpha}\,
    |\Psi_0\rangle~.
\end{equation}
In other words, the $\Phi\rightarrow\infty$ state is an interlayer singlet background dressed by a Fermi sea of mobile physical holes. 
For a spin-symmetric and layer-symmetric state, the hole density is $x/2$ for each of the four flavors, so $\mathcal{H}$ forms four identical hole pockets with area $A_h=x/2$ centered near $(\pi,\pi)$, in agreement with the small-Fermi-surface volume of the sFL phase.

\begin{figure}[t]
    \centering
    \includegraphics[width=0.75\linewidth]{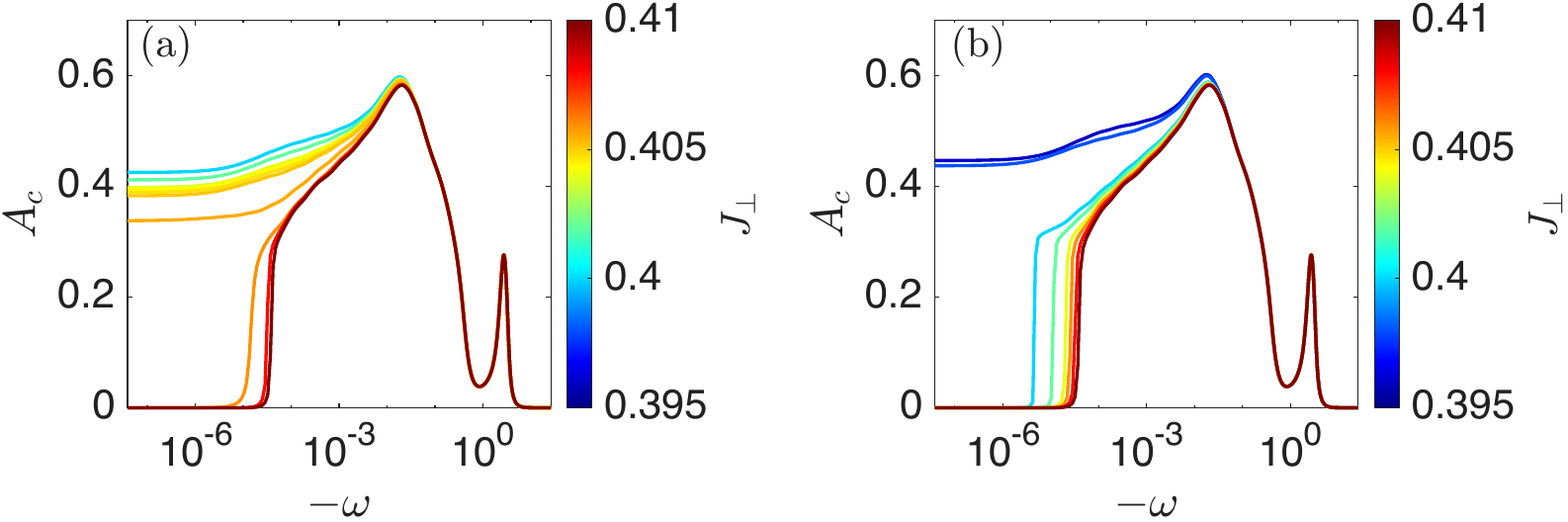}
    \caption{
    Hysteresis of the $J_\perp$-driven FL–sFL transition in the presence of isotropic interlayer hopping $t_{\perp;0}$ in the double Kondo model [Eq.~\eqref{eqn:double_kondo}]. 
    Other parameters are fixed at $J_K=-10t$, $U=8t$, $x=0.2$, and $t_\perp=0.1t$.
    The $\alpha$-band spectral function is shown for (a) a forward sweep of $J_\perp$ from the FL to the sFL side and (b) a backward sweep from the sFL to the FL side. 
    The resulting hysteresis indicates a weakly first-order transition, with coexisting FL and sFL solutions in the range $0.400t \lesssim J_\perp \lesssim 0.405t$.
    }
    \label{fig:first_tperp}
\end{figure}

\begin{figure}[t]
    \centering
    \includegraphics[width=0.75\linewidth]{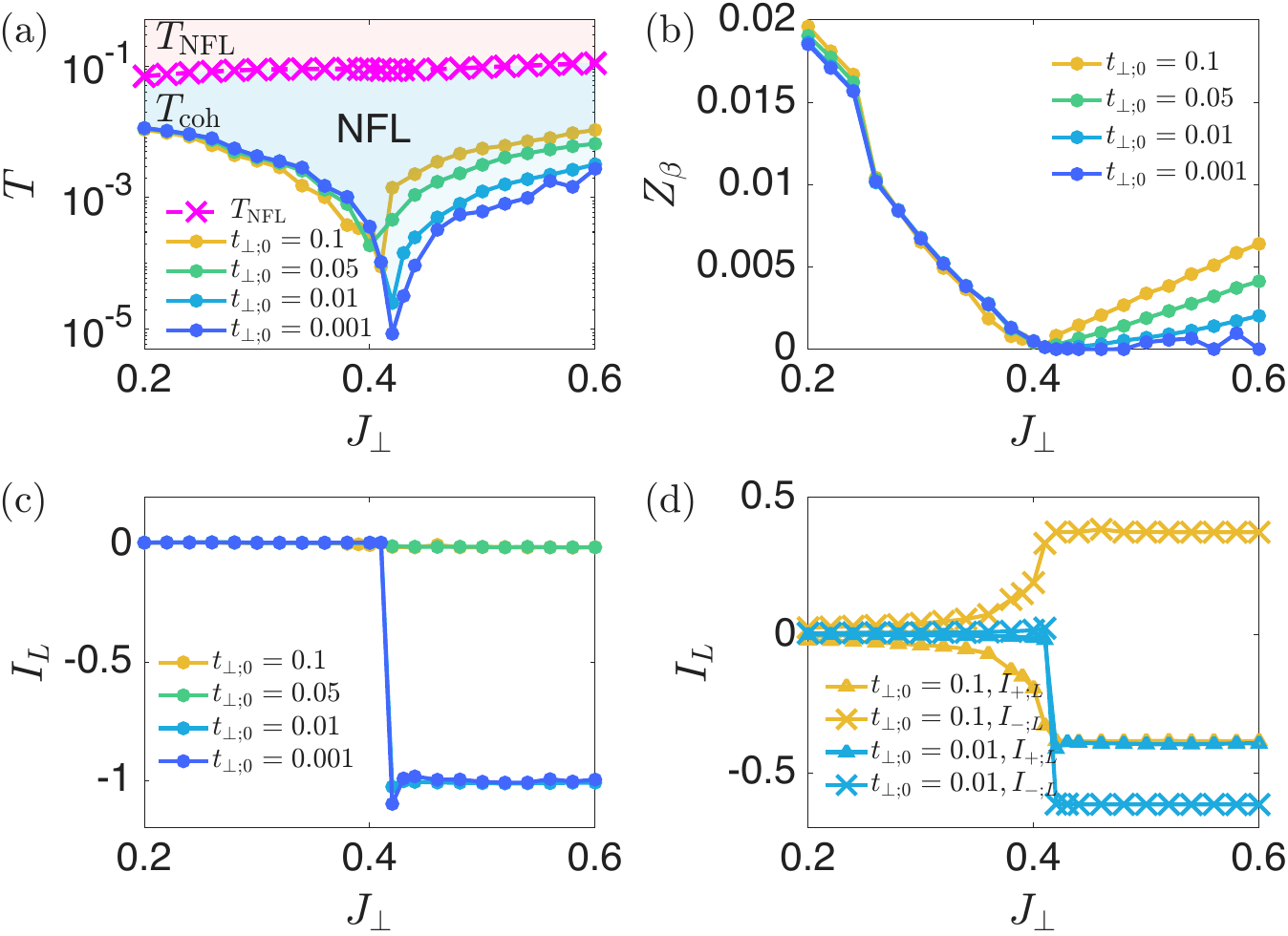}
    \caption{
    Effect of isotropic interlayer hopping $t_{\perp;0}$ on the QCP tuned by $J_\perp$ in the double Kondo model [Eq.~\eqref{eqn:double_kondo}], with Hund’s coupling fixed at $J_K = -10t$, Hubbard interaction $U = 8t$, and doping $x = 0.2$.
    (a) The NFL temperature regime as a function of $J_\perp$ for different values of the finite interlayer hopping $t_{\perp;0}$. 
    (b) Quasiparticle spectral weight of the $\beta$ band (corresponding to $c_+$) $Z_\beta=\left(1-\partial_\omega \mathrm{Re}\,\Sigma_\beta(\omega)\big|_{\omega=0}\right)^{-1}$. 
    $Z_\beta$ still vanishes at the phase boundary for finite $t_{\perp;0}$, but increases significantly in the sFL phase with increasing $t_{\perp;0}$.
    (c) The evolution of $I_{+;L} + I_{-;L}$ as a function of $J_\perp$ for different $t_{\perp;0}$. 
    (d) The evolution of separated $I_{+;L}$ and $I_{-;L}$ as a function of $J_\perp$ for $t_{\perp;0} = 0.1$ and $t_{\perp;0}=0.01$. 
    }
    \label{fig:tperp}
\end{figure}

\begin{figure}[t]
    \centering
    \includegraphics[width=0.99\linewidth]{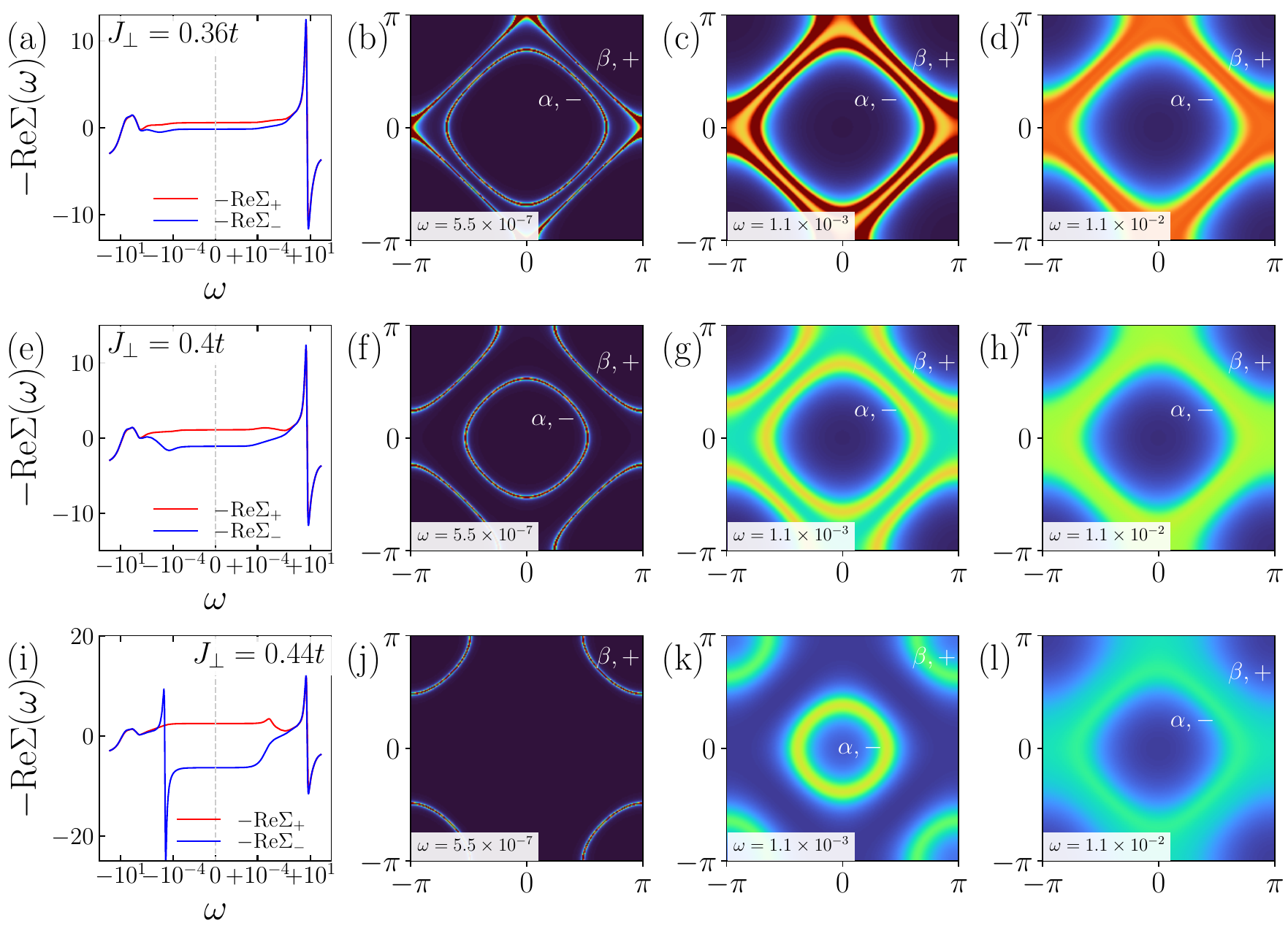}
    \caption{
    Self energy and const energy cuts on spectral function for the double Kondo model Eq.~\eqref{eqn:double_kondo} with $J_K=-10t$, $U=8t$, doping $x=0.2$ and $t_{\perp;0} = 0.1t$. 
    Panels (a)–(d), (e)-(h) and (i)–(l) correspond to $J_\perp = 0.36t$, $0.40t$ and $0.44t$, respectively. 
    In each row, the first panel shows the self energy $-\mathrm{Re}\Sigma_\pm (\omega)$, 
    while the remaining panels display constant-energy cuts of the spectral intensity $A(\omega,\mathbf{k})$ in momentum space at 
    $\omega = 5.5\times10^{-7}t$, $1.1\times10^{-3}t$, and $1.1\times10^{-2}t$. 
    A small imaginary part $\eta = 10^{-3}t$ is introduced via $\omega \rightarrow \omega + i\eta$ to broaden the spectrum.}
    \label{fig:FStperp}
\end{figure}

\begin{figure}[t]
    \centering
    \includegraphics[width=0.75\linewidth]{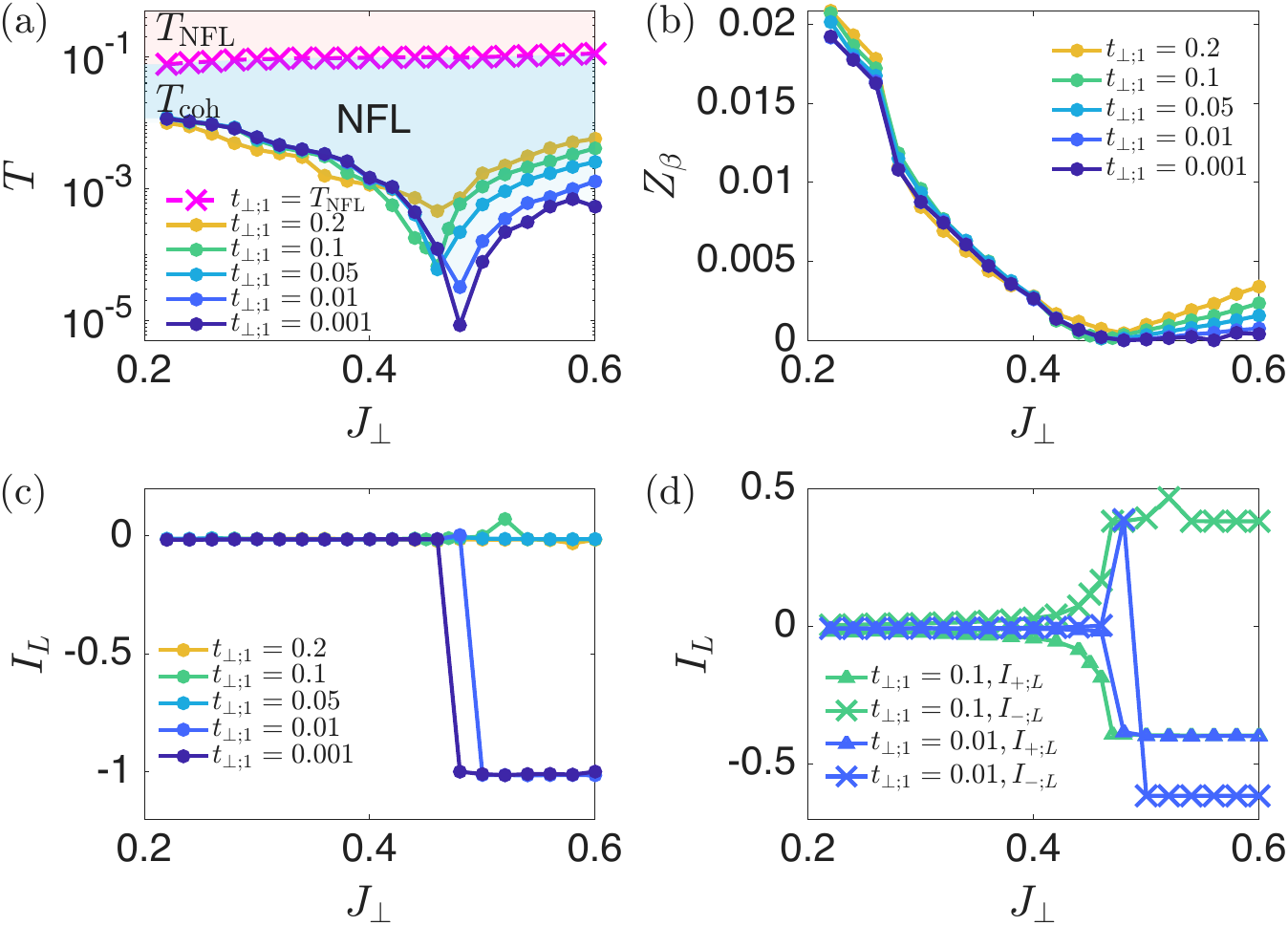}
    \caption{
    Effect of nodal interlayer hopping $t_{\perp;1}$ on the QCP tuned by $J_\perp$ in the double Kondo model [Eq.~\eqref{eqn:double_kondo}], with Hund’s coupling fixed at $J_K = -10t$, Hubbard interaction $U = 8t$, and doping $x = 0.2$.
    (a) The NFL temperature regime as a function of $J_\perp$ for different values of the finite interlayer hopping $t_{\perp;1}$. 
    The Hund's coupling is fixed at $J_K = -10t$, with $U = 8t$ and doping $x = 0.2$. 
    (b) Quasiparticle spectral weight of the $\beta$ band (corresponding to $c_+$) $Z_\beta=\left(1-\partial_\omega \mathrm{Re}\,\Sigma_\beta(\omega)\big|_{\omega=0}\right)^{-1}$. 
    $Z_\beta$ still vanishes at the phase boundary for finite $t_{\perp;1}$, but increases significantly in the sFL phase with increasing $t_{\perp;1}$. 
    (c) The evolution of $I_{+;L} + I_{-;L}$ as a function of $J_\perp$ for different $t_{\perp;1}$. 
    (d) The evolution of separated $I_{+;L}$ and $I_{-;L}$ as a function of $J_\perp$ for $t_{\perp;1} = 0.1$ and $t_{\perp;1}=0.01$. 
    }
    \label{fig:tperpd}
\end{figure}
\begin{figure}[t]
    \centering
    \includegraphics[width=0.99\linewidth]{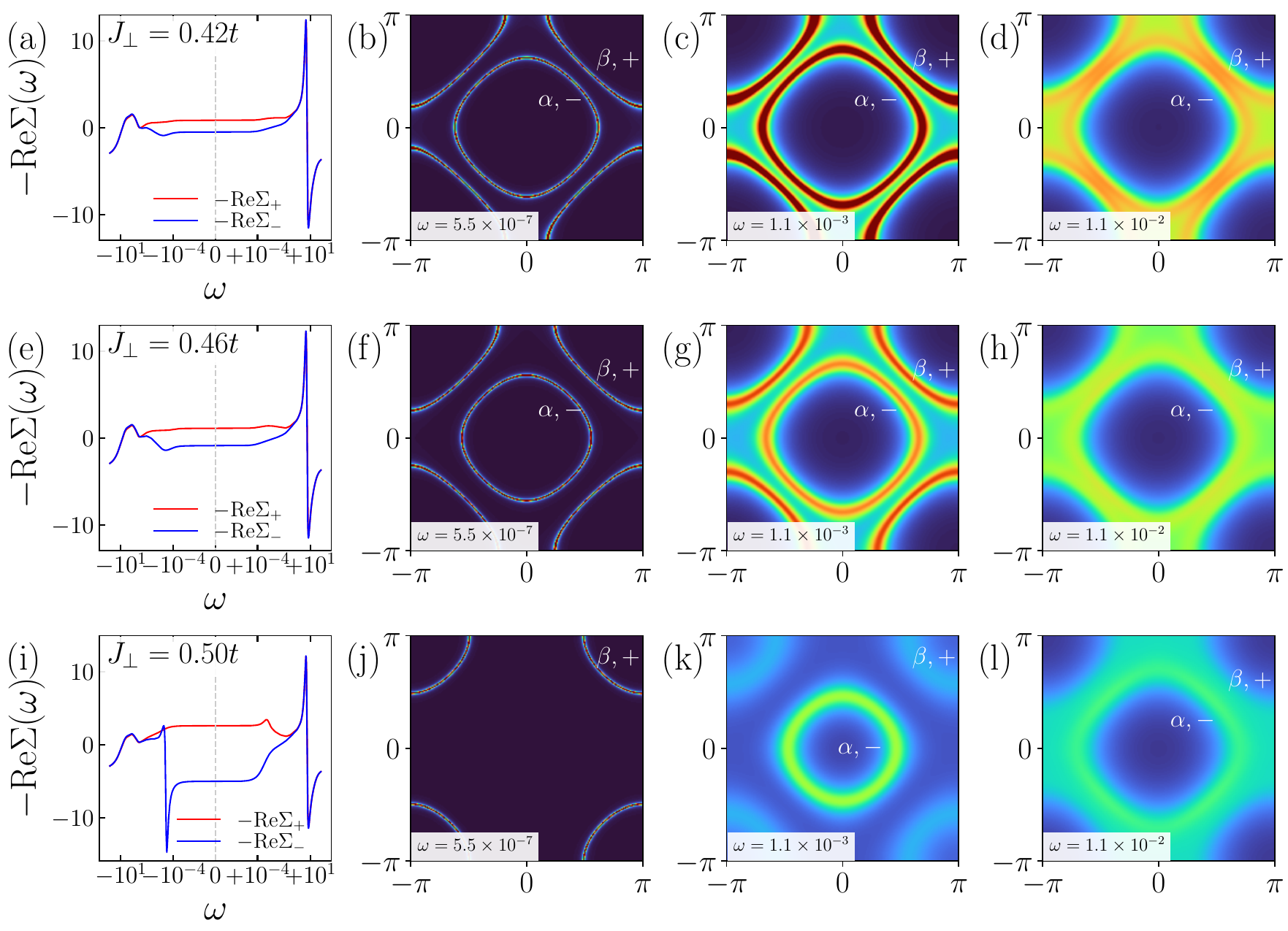}
    \caption{
    Self energy and const energy cuts on spectral function for the double Kondo model Eq.~\eqref{eqn:double_kondo} with $J_K=-10t$, $U=8t$, doping $x=0.2$ and $t_{\perp;1} = 0.2t$. 
    Panels (a)–(d), (e)–(h) and (i)-(l) correspond to $J_\perp = 0.42t$, $0.46t$, and $0.50t$, respectively. 
    In each row, the first panel shows the self energy $-\mathrm{Re}\Sigma_\pm (\omega)$, 
    while the remaining panels display constant-energy cuts of the spectral intensity in momentum space at 
    $\omega = 5.5\times10^{-7}t$, $1.1\times10^{-3}t$, and $1.1\times10^{-2}t$. 
    A small imaginary part $\eta = 10^{-3}t$ is introduced via $\omega \rightarrow \omega + i\eta$ to broaden the spectrum.
    }
    \label{fig:FStperpd}
\end{figure}

\begin{figure}[t]
    \centering
    \includegraphics[width=0.72\linewidth]{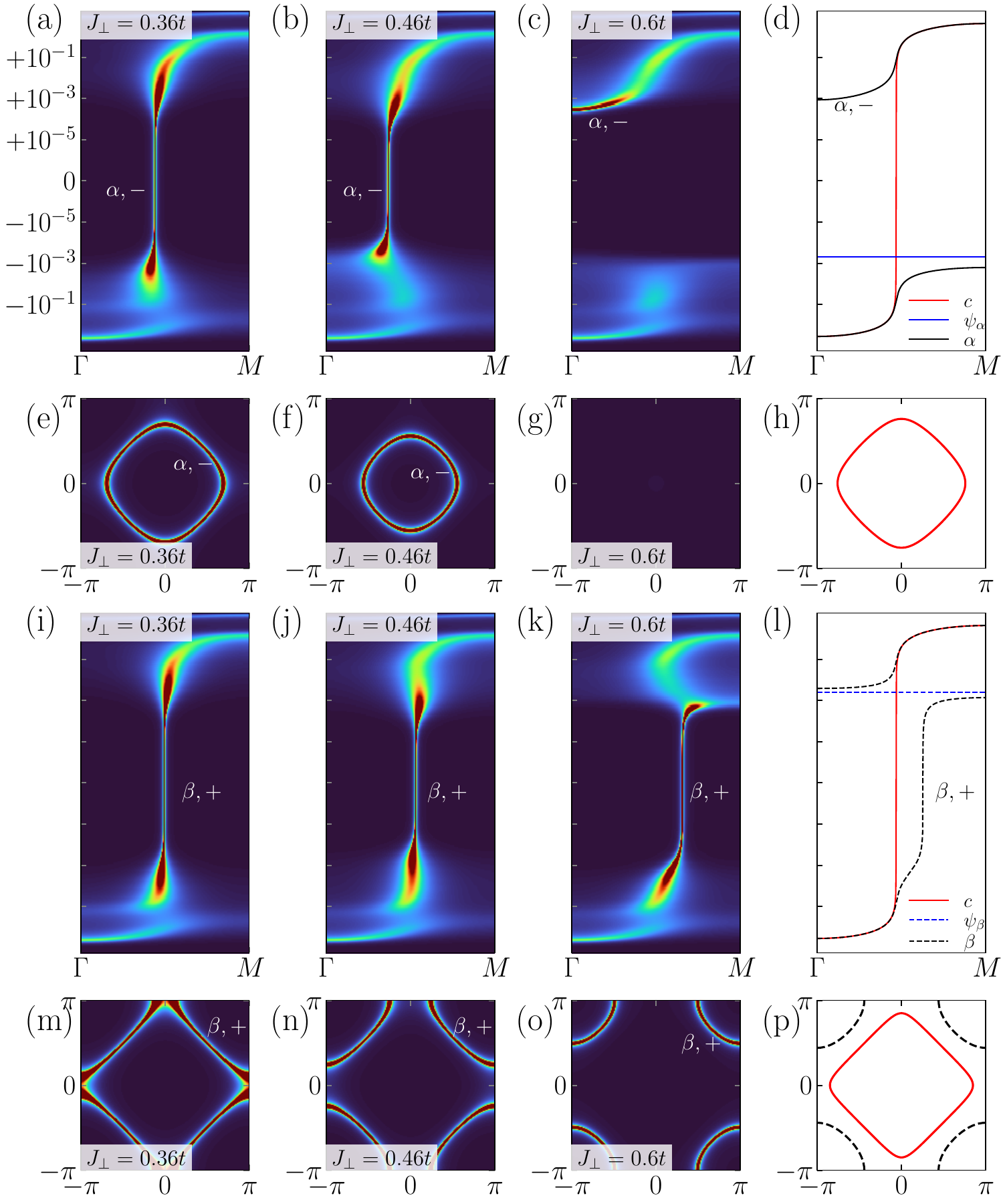}
    \caption{
    Momentum space spectrum of double Kondo model Eq.~\eqref{eqn:double_kondo} at fixed $J_K = -10t$, $U = 8t$ and $t_{\perp;1}=0.2t$. 
    (a)-(c) Reconstructed momentum-resolved $\alpha$ band spectral function $A_c(\omega,\mathbf{k})$ 
    along the momentum line cut $(0,0)$–$(\pi,\pi)$ for $J_\perp = 0.36t$, $0.46t$, and $0.6t$, respectively. 
    (e)-(g) Corresponding $\alpha$ band zero-frequency spectral function $A_c(\omega=0,\mathbf{k})$. 
    (i)-(k) Reconstructed momentum-resolved $\beta$ band spectral function $A_c(\omega,\mathbf{k})$ 
    along the momentum line cut $(0,0)$–$(\pi,\pi)$ for $J_\perp = 0.36t$, $0.46t$, and $0.6t$, respectively. 
    (m)-(o) Corresponding $\beta$ band zero-frequency spectral function $A_c(\omega=0,\mathbf{k})$. 
    (d), (l) Simulated band structure obtained from the ancilla-fermion description Eq.~\eqref{eqn:ancilla_tperp}, for $\alpha$ and $\beta$ bands with $\Phi=0.07t$ and $\mu_\psi=-0.001t\mp 0.0015t$. 
    The red, blue, and black lines represent the bare $c$ band, the $\psi$ band, and the hybridized bands, respectively. 
    For the $\alpha$ band, the ancilla band before hybridization is at energy $-\mu_\psi+\delta\mu_\psi<0$. 
    After hybridization with $c$, the lower band completely lies below $E=0$. The upper band is pushed to $E>0$ and gapped when $\Phi$ is large enough. 
    For the $\beta$ band,  the ancilla band is at $-\mu_\psi-\delta \mu_\psi > 0$.
    After hybridization with $c$, the upper band remains $E>0$, while the lower band always crosses $E=0$ and contributes to a small hole pocket dominated by $\psi$.
    (h), (p) Corresponding $\alpha$ and $\beta$ band Fermi surfaces in the ancilla-fermion theory. 
    The Fermi surfaces of the bare $c$ band and the hybridized bands are shown as red and black curves, respectively.
    }
    \label{fig:momentum_tperpd}
\end{figure}

\section{Effect of inter-layer coupling $t_\perp$} \label{app:tperp}

In the main-text calculations we mainly concentrate on the case where the interlayer hopping is zero, $t_\perp=0$, 
so that the $U(1)_{t,b}$ charge symmetries are preserved independently in the top and bottom layers. 
In realistic systems, however, a finite interlayer hopping generally exists and is believed to be responsible for the splitting of $\alpha$ and $\beta$ bands.  
The interlayer hopping explicitly breaks the separate $U(1)_t \times U(1)_b$ symmetry down to a single global $U(1)$ symmetry.
It is known that in the two-impurity problem the QCP becomes unstable once the separate charge symmetry is broken.  
It is therefore important to examine the stability of the NFL behavior under weak symmetry-breaking perturbations induced by a finite $t_\perp$.

Here we consider two types of interlayer hopping, denoted by $t_{\perp;0}$ and $t_{\perp;1}$. 
The former corresponds to an isotropic interlayer hopping
\begin{equation}
    t_{\perp;0} c^\dagger_{\mathbf{k};t;\sigma} c^{}_{\mathbf{k};b;\sigma} + \mathrm{h.c.}~,
\end{equation}
while the latter represents a hopping with a nodal line form factor
\begin{equation}
    \frac{t_{\perp;1}}{4}[\cos(k_x)-\cos(k_y)]^2 c^\dagger_{\mathbf{k};t;\sigma} c^{}_{\mathbf{k};b;\sigma} + \mathrm{h.c.}~.
\end{equation}

For either form of interlayer hopping the original $U(1)_{t,b}$ symmetry is explicitly broken and the Green's function and self-energy are no longer diagonal in the layer index. 
However, a $\mathbb{Z}_2$ mirror symmetry between the two layers still remains. 
It is convenient to introduce the bonding and antibonding fermions
\begin{equation}
    c_{i;\pm;\sigma} = \frac{1}{\sqrt{2}}(c_{i;t;\sigma}\pm c_{i;b;\sigma})~.
\end{equation}
Once a finite $t_{\perp}$ is introduced the $c_{\pm}$ band splits into an $\alpha$ band $c_-$ and a $\beta$ band $c_+$. 
Under the $\mathbb{Z}_2$ mirror operation, $c_{i;\pm;\sigma}\rightarrow \pm c_{i;\pm;\sigma}$. 
Therefore, the parity of each fermion $(-1)^{N_\pm}$ is still conserved given that the total charge is conserved. 
As a result, $\lla c_+;c_-^\dagger \rra=0$, 
both the Green's function $G$ and the self-energy $\Sigma$ remain diagonal in the $c_\pm$ basis.

\subsection{A weak first-order transition}

A careful hysteresis analysis reveals that the FL–sFL transition becomes weakly first-order when the isotropic interlayer hopping $t_{\perp;0}$ is sufficiently large.
By sweeping $J_\perp$ in both directions across the transition, we observe a small hysteresis in the $\alpha$-band spectral function, as shown in Fig.~\ref{fig:first_tperp}.
Deep in the sFL phase, the $\alpha$ band is fully gapped, whereas in the FL phase, it forms a Fermi surface with finite area.
The two Fermi-liquid solutions coexist within a narrow range, $0.400t \lesssim J_\perp \lesssim 0.405t$.
Given the extremely narrow coexistence region, we do not resolve the first-order transition in the following discussion and instead treat the rapid evolution between the two phases as a crossover.
However, at sufficiently high numerical resolution, the weakly first-order nature of the transition becomes apparent.

\subsection{Isotropic interlayer hopping}

In Fig.~\ref{fig:tperp} we calculate the phase diagram of the double Kondo model Eq.~\eqref{eqn:double_kondo} as a function of $J_\perp$ at fixed parameters $J_K=-10t$, $U=8t$, and doping $x=0.2$, for different values of $t_{\perp;0}$. 
As $t_{\perp;0}$ increases, the lowest achievable critical temperature becomes higher.

Different from the $t_\perp=0$ case, a topological $\mathbb{Z}_2$ indicator as in Eq.~\eqref{eqn:Z2topo_app} distinguishing the FL and sFL phases can no longer be defined. 
Instead we evaluate the Luttinger integer $I_L$, defined in Eq.~\eqref{eqn:luttinger_integral}, for the $c_\pm$ fermions. 
A jump of $I_{+;L}+I_{-;L}$ from $0$ to $-1$ is observed for small $t_{\perp;0}=0.001t$ and $t_{\perp;0}=0.01t$, whereas the jump disappears for larger values $t_{\perp;0}=0.05t$ and $t_{\perp;0}=0.1t$. 
This behavior can also be seen from the separately calculated $I_{\pm;L}$. 
For small $t_{\perp;0}=0.01t$ both $I_{+;L}$ and $I_{-;L}$ jump downward at the critical point. 
For larger $t_{\perp;0}=0.1t$ the same behavior evolves into a crossover: $I_{+;L}$ decreases while $I_{-;L}$ increases, compensating each other. 
These results suggest that the QCP is stable only for very small $t_{\perp;0}$. 

Nevertheless, although the zero-energy QCP is unstable with respect to finite $t_{\perp;0}$, the finite-energy physics remains strongly influenced by the nearby quantum criticality. 
As shown in Fig.~\ref{fig:tperp}(a), the lowest coherence temperature $T_{\mathrm{coh}}$ reachable at $t_{\perp;0}=0.1t$ remains as low as $\sim10^{-4}t$. 
On the sFL side slightly away from the QCP ($J_\perp=0.5t$), $T_{\mathrm{coh}}$ is still of order $\sim10^{-2}t$ at $t_{\perp;0}=0.1t$,  corresponding to a temperature scale of roughly $\sim10$ K. 
Therefore the NFL regime above $T_{\mathrm{coh}}$ naturally extends to temperatures comparable to the superconducting transition temperature $T_c$ observed in experiments.

In Fig.~\ref{fig:FStperp} we further show the Fermi surface evolution for different $J_\perp$ and energy cuts $\omega$ at fixed $t_{\perp;0}=0.1t$. 
Once a finite $t_{\perp;0}$ is introduced the original band splits into an $\alpha$ band around the $(0,0)$ point (corresponding to $c_-$) and a $\beta$ band around $(\pi,\pi)$ (corresponding to $c_+$). 
The splitting becomes increasingly pronounced as $J_\perp$ increases, and the $\alpha$ band eventually shrinks to zero in the sFL phase.

At first sight, the transition appears to be adiabatically connected to a trivial crossover when $t_{\perp;0}$ becomes sufficiently large. 
However, the underlying physics of the transition remains qualitatively different when one examines the self-energy and finite-temperature behavior. 
This can be seen from the different energy cuts of the spectrum function. 
When the energy scale is increased, the $\alpha$ band that disappears in the sFL phase reappears already at energies as low as $\sim10^{-3}t$ [Fig.~\ref{fig:FStperp}(k)]. 
At energies of order $\sim10^{-2}t$, the constant-energy momentum distribution becomes very similar in both the FL and sFL regimes, resembling the $t_{\perp;0}=0$ results shown in Fig.~\ref{fig:FS_T}.

\subsection{Nodal interlayer hopping}
We also study the case with nodal interlayer hopping $t_{\perp;1}$. 
The effect of $t_{\perp;1}$ enters the DMFT results in two ways. 
First, it modifies the density of states for the $c_+$ and $c_-$ bands. 
Second, it generates an overall isotropic contribution $t_{\perp;1}/4$ directly at the impurity site. 
Due to the absence of momentum dependence in the DMFT self-energy, the effect of the form factor is only partially captured in the present calculation. 

Qualitatively similar results are obtained for the case with $t_{\perp;0}$, as shown in Figs.~\ref{fig:tperpd} and \ref{fig:FStperpd}. 
The minimum $T_{\mathrm{coh}}$ increases with $t_{\perp;1}$, and the QCP evolves into a crossover for sufficiently large $t_{\perp;1}$. 
With regardless of absence of zero-temperature QCP, the NFL temperature remains large even for the largest $t_{\perp;1}=0.2t$ considered here. 
Note that the exact position of the QCP is slightly different from that of the $t_{\perp;0}$ case due to a different numerical method used to evaluate the momentum integral in Eq.~\eqref{eqn:greens}, but the overall physics is unaffected.

In Fig.~\ref{fig:FStperpd} we show the evolution of the spectrum function as a function of $J_\perp$ at different energy cuts. 
The spectrum looks similar in the FL and sFL phases for $\omega \sim 10^{-2}t$. 
Due to the $d$-wave form factor in $t_{\perp;1}$, the spectral function resembles the $\alpha$ and $\beta$ bands observed in experiments.

\subsection{Ancilla Fermion understanding. }

The ancilla-fermion description of the sFL can be naturally extended to the case with a finite $t_\perp$.
The main modification is the introduction of two ancilla bands separated by an energy splitting $\delta\mu_\psi$.
In general, both $\mu_\psi$ and $\delta\mu_\psi$ may exhibit momentum dependence.
However, within the DMFT self-consistent framework, the momentum dependence of $\Sigma(\omega)$ is neglected, and we therefore treat $\mu_\psi$ and $\delta\mu_\psi$ as constants.

The modified ancilla-fermion Hamiltonian reads
\begin{equation}\label{eqn:ancilla_tperp}
H_{\mathrm{AFT}}^{t_\perp} = \sum_{\mathbf{k},a,\sigma}
\left(\epsilon(\mathbf{k})-\mu\right)c^\dagger_{\mathbf{k};a;\sigma} c_{\mathbf{k};a;\sigma}
+ \sum_{\mathbf{k},\sigma}\left(t_\perp(\mathbf{k}) c^\dagger_{\mathbf{k};t;\sigma} c^{}_{\mathbf{k};b;\sigma}+\mathrm{h.c.}\right)
+ \Phi \sum_{i,a,\sigma} \left(c^\dagger_{i;a;\sigma}\psi_{i;a;\sigma}+\mathrm{h.c.}\right)
- \sum_{i,\pm}(\mu_\psi\pm\delta\mu_\psi) n_{i;\pm}^\psi~.
\end{equation}
In principle, the hybridization $\Phi$ should be a matrix; here, for simplicity, we take it to be proportional to the identity.

A comparison between the ancilla bands and the momentum-resolved spectral function is shown in Fig.~\ref{fig:momentum_tperpd}.
In the FL phase, the $\alpha$ and $\beta$ bands are only weakly split.
As the system approaches the sFL phase, the splitting becomes increasingly pronounced.
Deep in the sFL phase, the $\alpha$ band becomes fully gapped due to hybridization with the ancilla fermions.
A similar band structure is reproduced by the ancilla Hamiltonian, provided that $\delta\mu_\psi$ is sufficiently large.

\section{Relation to the QCP of 2-impurity Anderson model}

\begin{figure}
    \centering
    \includegraphics[width=0.7\linewidth]{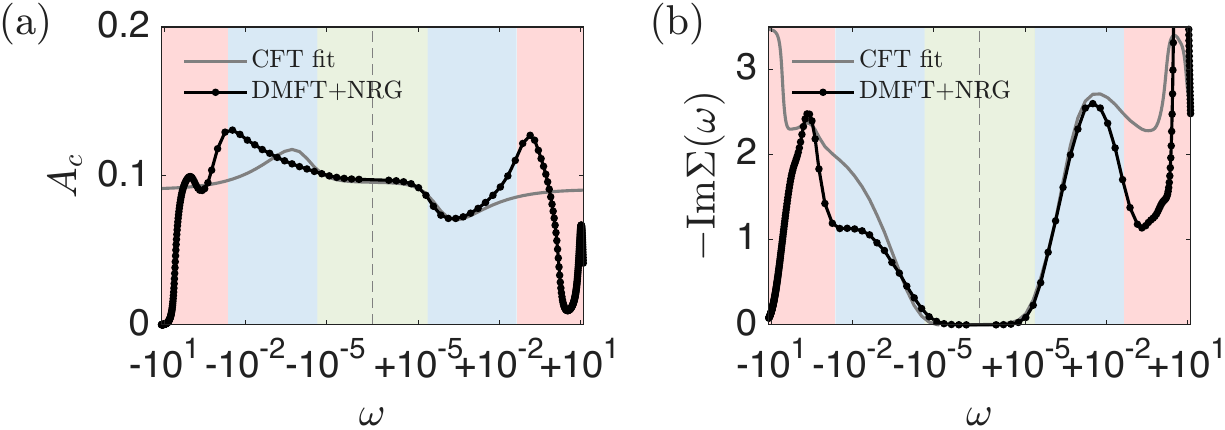}
    \caption{NFL behavior near the QCP, with parameters chosen as $J_\perp = 0.6t$, $J_K = -10t$, $x = 0.2$, and $U = 8t$.
    (a) and (b) display the spectral function $A_c(\omega)$ and the imaginary part of the fermionic self-energy, $-\mathrm{Im}\Sigma(\omega)$, near the QCP.
    A conformal field theory (CFT) fit is also shown, which shows qualitatively agreement with the DMFT+NRG numerical results.
    }
    \label{fig:correlators_app}
\end{figure}

The quantum critical point (QCP) discussed here is closely related to the well-known non-Fermi liquid QCP of the two-impurity Anderson model (2IAM). 
The 2IAM has been extensively studied using numerical renormalization group (NRG) techniques and conformal field theory (CFT) approaches \cite{Affleck1992,Affleck1995}. 
At the critical point, a residual entropy of $\log(2)/2$ is observed, which reflects the presence of a decoupled Majorana fermion degree of freedom.
In this appendix, we briefly review the key results from the CFT analysis of the 2IAM and apply them to fit the single-particle correlation function obtained from our DMFT calculations presented in the main text.

For a 2IAM, we consider a single impurity interacting with a conduction band. 
Assuming a linear dispersion (i.e., a constant density of states) near the Fermi energy, the 2IAM can be mapped onto a model of left-moving fermions scattering off an impurity located at $x=0$. The CFT analysis gives the following form of the Green’s function across the impurity:
\begin{equation}
    G^{\mathrm{imp}}_{\mathrm{CFT};\alpha\sigma}(\tau,x_1,x_2) = \frac{S_{\alpha\sigma}}{\tau+ix_2-ix_1}~,
\end{equation}
where $x_1 < 0$ and $x_2 > 0$ lie on opposite sides of the impurity, and $S_{\alpha\sigma}$ is the scattering matrix for spin $\sigma = \uparrow,\downarrow$ and channel $\alpha = L, R$. 
At the QCP of the 2IAM, it has been shown that $S_{\alpha\sigma} = 0$, corresponding to completely inelastic scattering.

In realistic systems, one is typically not exactly at the QCP, so it is important to understand how the Green’s function and the self-energy evolve as one moves away from criticality. 
By exploiting an emergent SO(7) symmetry at the QCP, it was shown in Refs.\cite{Sela2011,Mitchell2012} that the scattering $T$-matrix $T_{\alpha\sigma}(\omega)$ can be solved analytically. 
This matrix connects the free and full Green's functions via:
\begin{equation}\label{eqn:GT}
    G(\omega) = G_{0}(\omega) + G_{0}(\omega)T(\omega)G_{0}(\omega). 
\end{equation}
The $T$-matrix near the QCP, with a characteristic crossover scale $T^*$, takes the form:
\begin{equation}\label{eqn:T_cft}
    2\pi i\nu T_{\alpha\sigma}(\omega) = 1-S_{\alpha\sigma}\mathcal{G}(\omega/T^*)~,
\end{equation}
where $\nu$ is the density of states at the Fermi level, and $S_{\alpha\sigma} = 1$ for the Fermi liquid side of 2IAM. 
The function $\mathcal{G}(x) = \frac{2}{\pi} K(ix)$ involves the complete elliptic integral of the first kind $K(x)$.
The asymptotic behavior of $\mathcal{G}(x)$ is given by: For $x \ll 1$: $\mathcal{G}(x) = 1 + i x/4 - (3x/8)^2$;
For $x \gg 1$: $\mathcal{G}(x) \sim \frac{\sqrt{i}}{2\pi} \left[ \log(256 x^2) - i\pi \right]/\sqrt{x}$.
A detailed comparison between Eq.\eqref{eqn:T_cft} and NRG data can be found in Ref.\cite{Sela2011}, showing excellent agreement across a wide energy range.

Despite the success of the CFT approach for 2IAM, the situation in DMFT is more complex. 
In particular, the hybridization function $\Delta(\omega)$ is determined self-consistently and is not necessarily constant near $\omega = 0$. 
Nonetheless, we find that the CFT-inspired fitting still captures the qualitative behavior near the QCP.
In our DMFT calculations, we define the non-interacting Green’s function using the converged $\Delta(z)$ as:
\begin{equation}
    G_0(z)=\frac{1}{\omega+\mu-\Delta(z)}, 
\end{equation}
For the low-energy spectrum near QCP, $G_0(\omega)$ changes slowly with frequency $\omega$, and we choose the zero frequency value $G_0(\omega=0)$ as our starting point. 
Substituting into Eqs.~\eqref{eqn:GT} and\eqref{eqn:T_cft}, we choose the following fitted Green's function from CFT 
\begin{equation}\label{eqn:greens_CFT}
    G_{\mathrm{CFT}}(\omega) = G_{0}(0) + G_{0}(0)\frac{1-\mathcal{G}(\omega/T_\mathrm{coh})}{2\pi i\nu}G_{0}(0),  
\end{equation}
where $T_{\mathrm{coh}}$ is the lower bound of NFL regime. 
And $\nu$ is taken as the zero energy density of state measured by the Green's function defined in Eq.~\eqref{eqn:greens} as $\nu = -\mathrm{Im}G_{oo}(\omega=0)/\pi$.

Finally, with the above fitted Green's function, the self-energy can be estimated by 
\begin{equation}\label{eqn:selfenergy_CFT}
    \Sigma_{\mathrm{CFT}}(\omega) = G_0^{-1}(\omega) - G_{\mathrm{CFT}}^{-1}(\omega)~. 
\end{equation}
In Fig.~\ref{fig:correlators_app}, we show that such a simple fitting already gives good qualitative consistent with the numerical result for $J_K<0$. 
Considering only two parameters are used in the simulations, the qualitative agreement makes us believe that the QCP observed can already be captured by the CFT analysis  \cite{Affleck1992,Affleck1995,Sela2011,Mitchell2012} for the 2IAM critical point. 
Exactly at the QCP, $T_{\mathrm{coh}} \to 0$, so that $\mathcal{G}(\omega/T_{\mathrm{coh}})=0$, and $G_{\mathrm{CFT}}(\omega)$ becomes independent of $\omega$.
The self-energy $-\mathrm{Im}\Sigma_{\mathrm{CFT}}(\omega)$ is therefore expected to remain finite in the $\omega \to 0$ limit.

We note that there was report as a $\log(\omega)$ behavior of the self-energy on the frequency \cite{Gleis2024X}.
However, it is apparent that there cannot be any function that vanishes as $\log(\omega)$ for $\omega\to 0$. 
We therefore believe that the seemly $\log$ behavior is only an artifact of the asymptotical behavior of the elliptic function $K(x)$. 

\end{document}